\documentclass[twocolumn,showpacs,preprintnumbers,amsmath,amssymb]{revtex4}
\usepackage{epsfig}
\usepackage{dcolumn}% Align table columns on decimal point
\usepackage{bm}% bold math
\newcommand{\beq}{\begin{equation}} 
\newcommand{\eeq}{\end{equation}} 
\newcommand{\beqa}{\begin{eqnarray}} 
\newcommand{\eeqa}{\end{eqnarray}} 
\newcommand{\bea}{\begin{array}} 
\newcommand{\ea}{\end{array}} 

\newcommand{\dd}{{\rm d}}
\renewcommand{\pl}{\partial}
\newcommand{\inta}{\int_{-i\infty}^{+i\infty}} 
\newcommand{\lag}{\langle} 
\newcommand{\rag}{\rangle}
\newcommand{\law}{\stackrel{\rm law}{=}}
\newcommand{\ii}{{\rm i}}

\newcommand{\vu}{{\bf u}}
\newcommand{\vx}{{\bf x}}
\newcommand{\vk}{{\bf k}}
\newcommand{\vq}{{\bf q}}
\newcommand{\thetat}{{\tilde{\theta}}}
\newcommand{\vut}{{\tilde{\bf u}}}
\newcommand{\psit}{{\tilde{\psi}}}
\newcommand{\vxref}{{\bf x}_0}
\newcommand{\deltat}{{\tilde{\delta}}}
\newcommand{\cP}{{\cal P}}
\newcommand{\cL}{{\cal L}}
\newcommand{\vs}{{\bf s}}

\newcommand{\vQ}{{\bf Q}}
\newcommand{\vX}{{\bf X}}
\newcommand{\vU}{{\bf U}}
\newcommand{\vK}{{\bf K}}
\newcommand{\hx}{{\hat x}}
\newcommand{\hk}{{\hat k}}
\newcommand{\hu}{{\hat u}}
\newcommand{\thu}{{\tilde{{\hat u}}}}
\newcommand{\ut}{{\tilde{u}}}
\newcommand{\thpsi}{{\tilde{{\hat \psi}}}}
\newcommand{\hvk}{{\hat {\bf k}}}
\newcommand{\Nt}{N_{\rm tot}}
\newcommand{\Nv}{N_{\rm v}}
\newcommand{\Nf}{N_{\rm f}}
\newcommand{\hN}{{\hat N}}
\newcommand{\hP}{{\hat P}}

\newcommand{\rhob}{\overline{\rho}}

\newcommand{\Ai}{{\rm Ai}}

\begin{document}

\title{Density fields and halo mass functions in the Geometrical Adhesion toy Model}

\author{Patrick Valageas}
\affiliation{Institut de Physique Th\'eorique,\\
CEA, IPhT, F-91191 Gif-sur-Yvette, C\'edex, France\\
CNRS, URA 2306, F-91191 Gif-sur-Yvette, C\'edex, France}
\author{Francis Bernardeau}
\affiliation{Institut de Physique Th\'eorique,\\
CEA, IPhT, F-91191 Gif-sur-Yvette, C\'edex, France\\
CNRS, URA 2306, F-91191 Gif-sur-Yvette, C\'edex, France}
\vspace{.2 cm}

\date{\today}
\vspace{.2 cm}

\begin{abstract}

In dimension 2 and above, the Burgers dynamics, the so-called ``adhesion model'' in
cosmology, can actually give rise to several dynamics in the inviscid limit.
We investigate here the statistical properties of the density field when it is defined by
a ``geometrical model'' associated with this Burgers velocity field and where the matter 
distribution is fully determined, at each time step, by geometrical constructions. 
Our investigations are based on a set of numerical experiments that make use of an
improved algorithm, for which the geometrical constructions are efficient and robust.

In this work we focus on Gaussian initial conditions with power-law power spectra of
slope $n$ in the range $-3<n<1$, where a self-similar evolution develops, and we compute
the behavior of power spectra, density probability distributions and mass functions.
As expected for such dynamics, the density power spectra show universal high-$k$ tails
that are governed by the formation of pointlike masses. The two other statistical 
indicators however show  the same qualitative properties as those observed for 3D
gravitational clustering. In particular, the mass functions obey a Press-Schechter
like scaling up to a very good accuracy in 1D, and to a lesser extent in 2D. 

Our results suggest that the ``geometrical adhesion model'', whose solution is fully
known at all times, provides a precious tool to understand 
some of the statistical constructions frequently used to study the development of mass
halos in gravitational clustering.

\end{abstract}

 \vskip2pc

\maketitle

\section{Introduction}
\label{sec:intro}

The Burgers equation \cite{Burgersbook,Kida1979,Bec2007}, which describes 
the evolution
of a compressible pressureless fluid, with a nonzero viscosity, was first
introduced as a simplified model of fluid turbulence, as it shares the same
hydrodynamical (advective) nonlinearity and several conservation laws with the
Navier-Stokes equation. It also appears in many physical problems,
such as the propagation of nonlinear acoustic waves in
nondispersive media \cite{Gurbatov1991}, the study of disordered systems and
pinned manifolds \cite{LeDoussal2010}, or the formation of large-scale
structures in cosmology \cite{Gurbatov1989,Vergassola1994},
see \cite{Bec2007} for a recent review.
In the cosmological context, where one considers the inviscid limit without
external forcing, it is known as the ``adhesion model'' and it provides a
good description of the large-scale filamentary structure of the cosmic 
web \cite{Melott1994}.
In this context, one is interested in the statistical properties of the dynamics, 
as described by the density and velocity fields, starting with a random Gaussian
initial velocity \cite{Kida1979,Gurbatov1997} and a uniform density.
These initial conditions are expected for generic models of inflation of quantum fluctuations generated
in the primordial Universe and agree with the small Gaussian fluctuations
observed on the cosmic microwave background. In the hydrodynamical context,
this setup corresponds to ``decaying Burgers turbulence'' \cite{Gurbatov1997}.

This problem has led to many studies, especially in one dimension (1D) and  for
power-law initial energy
spectra (fractional Brownian motion) $E_0(k) \propto k^n$ of index\footnote{In the cosmological context
it is common to define the spectral index with respect to the density field spectrum $P(k) \propto k^{n_{\rho}}$. 
The two coincide in dimension 3 and in $d$ dimensions we have $n_{\rho}+d=n+3$.} $n$. The two 1D peculiar cases of white-noise initial velocity ($n=0$)
\cite{Burgersbook,Kida1979,She1992,Frachebourg2000,Valageas2009c}
and Brownian motion initial velocity ($n=-2$)
\cite{She1992,Sinai1992,Bertoin1998,Valageas2009a} have received much attention. 
%
%Indeed, in these two cases the initial velocity field is built from a white-noise 
%stochastic field (either directly or through one integration), which gives rise
%to Markovian processes and allows to derive many explicit analytical results.
For a more general $n$, it is not possible to obtain full explicit solutions,
but several properties of the dynamics are already known
\cite{Gurbatov1991,Gurbatov1997,Valageas2009b}. In particular,
for $-3<n<1$, the system shows a self-similar evolution as shocks merge
to form increasingly massive objects separated by a typical length, $L(t)$
- the integral scale of turbulence - that grows as $L(t) \sim t^{2/(n+3)}$,
while the shock mass function scales as $\ln[n(>m)] \sim -m^{n+3}$ at large
masses \cite{She1992,Molchan1997,Gurbatov1997,Noullez2005}.
In spite of these common scalings, the range $-3<n<1$ can be further split
into two classes, as shocks are dense for $-3<n<-1$ but isolated for $-1<n<1$ 
\cite{She1992}.

In higher dimensions, the situation gets more complicated as several prescriptions for the
matter distribution (again in the inviscid limit) can be associated to the same
velocity field, governed by the Burgers equation. 
They coincide over regular regions (i.e. outside of shocks)
but they can show significantly different behaviors on the shock manifold.
For instance, if one uses the standard continuity equation  mass clusters 
cannot fragment but they can leave shock nodes and travel along the
shock manifold \cite{Bogaevsky2004,Bogaevsky2006}.
By contrast, if one uses a modified continuity equation, associated with a
``geometrical model'' for the matter distribution, thus introducing the Geometrical Adhesion Model (GAM), 
mass clusters are always located
on shock nodes but they do not necessarily merge when they collide (in fact, collisions
can redistribute matter over a possibly different number of outgoing clusters, while
conserving the total momentum) \cite{Gurbatov1991,BernardeauVal2010b}.
The drawback of the prescription based on the standard continuity equation is that
the latter has to be numerically integrated over time depriving
the knowledge of the Hopf-Cole solution of the Burgers equation of much of its interest.
By contrast, the GAM extends the geometrical structure of the
Hopf-Cole solution to define an associated matter distribution 
\cite{Gurbatov1991,Saichev1996,Saichev1997,Vergassola1994,BernardeauVal2010b},
through Legendre transforms and convex hull constructions, so that both the
velocity and density fields can be derived at any time through geometrical
constructions.
This is a very convenient property, which allows faster numerical computations
\cite{Noullez1994,Vergassola1994} as well as greater analytical insights
\cite{Gurbatov1991,BernardeauVal2010b}. 

In all cases, beyond 1D and for generic initial conditions one has to rely on numerical
experiments to obtain quantitative results in those systems. 
The adhesion model has actually been studied in previous numerical works
\cite{Kofman1990,Kofman1992,Vergassola1994,Sahni1994},
in 1D, 2D and 3D, using the Hopf-Cole solution for the velocity field (however,
it was not always realized that one has to specifically complement the velocity field
construction to unambiguously define the density field).

The goal of this article is to revisit this problem, in 1D and 2D, with the use of a 
novel algorithm for the construction of the convex hull that is more efficient and more robust
for the construction of the convex hull triangulations (see Sec.~\ref{Two-dimensions}  and App.~\ref{Algorithms-for-the-2D-GAM} for details). We take
advantage of these simulations to investigate quantities
that have not been studied previously but that are of great interest in a
cosmological context. Thus, in addition to the mass function of shock nodes
(i.e. mass halos), we also consider the probability distributions of the density
contrast, within spherical and cubic cells, the low-order moments of the
density distribution and the density power spectrum, for which there exist specific
predictions for both the Geometrical Adhesion Model and the 3D gravitational dynamics.

In particular, the reduced cumulants of the smoothed density contrast can serve as both a test of the accuracy
of our numerical codes and a guide for comparison with the results obtained for 3D
gravitational clustering by $N$-body codes simulating the dynamics of a pressureless self-gravitating fluid.
They are defined as
\begin{equation}
S_{p}=\frac{\langle\delta_{R}^p\rangle_{c}}{\langle\delta_{R}^2\rangle^{p-1}} ,
\label{Spdef-delta}
\end{equation}
where $\delta_{R}$ is the filtered density field at scale $R$ (more precisely the filtered
density-contrast field, with $\delta(\vx)=(\rho(\vx)-\rhob)/\rhob$).
They were shown in \cite{1994ApJ...433....1B} to take a simple form for a top hat filter and were initially derived for $p=3,4$, and
then in \cite{1994A&A...291..697B} for all values of $p$, for the gravitational dynamics
in the large-scale limit.
In $d$ dimensions we have,
\begin{equation}
R\rightarrow\infty : \hspace{0.3cm} S_{3}^{\rm grav} \rightarrow
\frac{6}{7}\left(5+\frac{2}{d}\right)-\frac{3}{d}(n+3) .
\label{S3-gravity}
\end{equation}
In the context of the adhesion model, because prior to shell crossings the matter field follows the Zel'dovich approximation
the reduced cumulant values take the form (the result has been given for the 3D Zel'dovich approximation in
\cite{1994MNRAS.269..947B,1995ApJ...443..479B} and extended to the context of the adhesion model and to other dimensions in \cite{Valageas2009b}),
\begin{equation}
R\rightarrow\infty : \hspace{0.3cm} S_{3} \rightarrow \frac{3}{d}(d-n-2)
\label{S3-GAM}
\end{equation}
where $n$ is the energy spectrum index defined in Eq.(\ref{E0n}) below.
The result (\ref{S3-GAM}) only holds for $n\leq d-3$, since for larger $n$ shell crossing
keeps playing a role in the large-scale limit \cite{Valageas2009b}.
The behavior of those quantities at small scale is not fully understood. It has been argued that they should reach a constant value (at least for power-law spectra).
This is the case in the so-called ``hierarchical clustering models'' and it was partially checked in numerical simulations,  see \cite{1997MNRAS.287..241C} where an explicit description of the small-scale plateau is proposed. More
precise motivations from first principles have been put forward although it has never been proved
explicitly that such a family of solutions actually exists, and even less that it was relevant in a cosmological context.
The ``stable-clustering ansatz'' introduced in
\cite{Davis1977,Peebles1980} was such an attempt, based on the approximation that
once objects have formed they fully decouple from the dynamics and keep a
constant mass and physical size. This can also be set in a broader multifractal 
description \cite{Balian1989,Balian1989b}. More generally, some physical constraints
(such as the positivity of the density $\rho$) can be used to obtain some information
on the multifractal spectrum whence on the statistical properties of the density field
\cite{Valageas1999} (for instance, the coefficients $S_p$ can only grow or reach a
constant in a small-scale highly nonlinear regime).
However, there is no derivation of the precise form of the multifractal spectrum 
either from systematic approaches or well-controlled models.
As described in this article, the Geometrical Adhesion Model offers the opportunity
to check the validity of large-scale limits such as (\ref{S3-gravity})-(\ref{S3-GAM}),
while showing a nontrivial but well-understood small-scale limit.

Another focus of this paper is the Press-Schechter formalism \cite{Press1974},
which is widely used in cosmological large-scale studies.
Simulations of the formation of large-scale structures in cosmology have indeed shown
that for Gaussian initial conditions, such as those studied here,
the mass function of halos defined by a given density threshold (typically
$\rho/\rhob=200$)
is reasonably well described by the Press-Schechter formula.
%Since the Burgers dynamics is closely related to the gravitational dynamics
%\cite{Gurbatov1989,Vergassola1994} it is interesting to investigate this model
%in the present case.
This heuristic approach states that the fraction of matter, $F(>M)$, that is enclosed
within collapsed objects (infinitesimally thin shocks in the present adhesion model)
of mass larger than $M$
is given by the probability that, choosing a Lagrangian point $q$ at random,
the mass $M$ around this point has just collapsed to a point at the time of
interest if one assumes spherical collapse dynamics. %if particles follow the linear displacement field.
For Gaussian initial conditions this reads as
\beq
F_{\rm PS}(>M) = \int_{\nu(M)}^{\infty} \frac{\dd\nu'}{\nu'} \, f_{\rm PS}(\nu') 
\label{FM-PS}
\eeq
with (including the usual normalization factor $2$)
\beq
f_{\rm PS}(\nu) =  \sqrt{\frac{2}{\pi}} \, \nu \, e^{-\nu^2/2} .
\label{fPS}
\eeq
The value of $\nu(M)$ can be written as
\beq
\nu(M) = \frac{\delta_{c}}{\sigma(M)}, 
\label{nuM}
\eeq
where $\sigma(M)$ is the rms density fluctuation at scale $M$ and
$\delta_{c}$ is determined by the dynamical evolution of the density field.
It is $\delta_{c}\approx 1.69$ for 3D and $\delta_{c}\approx 1.47$ for 2D,
for the gravitational dynamics.
In the context of the Burgers equation,
prior to caustic formation particles follow the linear displacement field, which leads to
$\delta_{c}=d$ for $d$ dimensions.
%Here we used the fact that if particles follow the linear displacement field the density
%of a patch of matter evolves as $\rho=\rho_0/(1-\delta_L)$, as long as there is no
%shell crossing, where $\delta_L$ is the linear density contrast introduced in
%(\ref{linear}). Therefore, at the time of collapse we have $\delta_L=1$, whence the
%factor $1$ in the numerator of Eq.(\ref{nudef}).
The factor 2, which we have inserted in Eq.(\ref{fPS}) to ensure the correct normalization
to unity, is not accounted for
in this simple approach but can be using the random walk approach (as originally shown in \cite{1991ApJ...379..440B}) 
that (at least partially) takes into account
the cloud-in-cloud effects. It also implies that at late time all matter points are
comprised in a halo, which is expected to be the case for both the gravitational
dynamics and the adhesion model.
Although 3D realistic cosmological numerical experiments
\cite{Sheth1999,Jenkins2001,Reed2003} show deviations from
the simple Press-Schechter model (\ref{FM-PS})-(\ref{fPS}), the mass functions built by the
gravitational dynamics are still very well described by the scaling (\ref{FM-PS}), but
with a slightly different function $f(\nu)$ than (\ref{fPS}). It has been argued that 
those differences could be accounted for by various refinements (ellipsoidal collapse \cite{2001MNRAS.323....1S}, colored noise \cite{2010ApJ...711..907M}, etc.) that all could be similarly implemented in the GAM context.
One of the aims of this paper is therefore to test the mere validity of this 
Press-Schechter scaling, at the level of the one-point mass function, within the
Geometrical Adhesion Model, which is better controlled and provides a much larger
range of masses than 3D gravitational simulations.
This also allows us to check (especially in 1D) the predictions that can be obtained
for the large-mass tail of the halo mass function.

This article is organized as follows. In Sec.~\ref{Burgers-dynamics} we recall
the Burgers equation and its Hopf-Cole solution for the velocity field.
Then, we describe the associated ``Geometrical Adhesion Model,'' which defines the
matter distribution that we study here. We also present the power-law Gaussian
initial conditions that we focus on.
They correspond to the initial conditions that appear in the cosmological context,
for the formation of large-scale structures in the Universe, and they give rise
to self-similar dynamics. 
We briefly present in Sec.~\ref{1D-test-bench} our numerical results for the
1D case, where they can be checked with the help of the known analytical results.
The large dynamical range also allows a precise test of scaling laws and
asymptotic tails. We discuss in greater detail our results for the 2D case in
Sec.~\ref{Two-dimensions}.
After a brief description of our numerical algorithm, we study the shock mass
functions that we obtain and the dependence of the low-mass and high-mass tails
on the slope of the initial power spectrum. Then, we present our results for the
density probability distributions and the density power spectrum.
Next, we briefly discuss in Sec.~\ref{Separable-text} the case of separable initial
conditions in arbitrary dimensions, where exact results can be obtained.
Finally, we conclude in Sec.~\ref{Conclusion}. 

Note that this paper contains a few appendices that describe the
algorithms, compare them with previous numerical studies, and present
our detailed results for the 1D and separable cases.

The reader who is mostly interested in our results and the comparison with
behaviors observed for 3D gravitational clustering, may skip 
Sec.~\ref{Burgers-dynamics}, which is devoted to the definition of the dynamics,
and go directly to Sec.~\ref{1D-test-bench}.

\section{Burgers dynamics and Geometrical model}
\label{Burgers-dynamics}

\subsection{Equation of motion and Hopf-Cole solution for the velocity field}
\label{eq-motion}

We consider the $d$-dimensional Burgers equation \cite{Burgersbook} in the inviscid
limit (with $d\geq 1$),
\beq
\pl_t \vu + (\vu\cdot\nabla)\vu = \nu \Delta \vu , \hspace{1cm} \nu\rightarrow 0^+ ,
\label{Burgers}
\eeq
for the velocity field $\vu(\vx,t)$.
As is well known, for curlfree initial velocity fields the nonlinear Burgers equation
(\ref{Burgers}) can be solved
through the Hopf-Cole transformation \cite{Hopf1950,Cole1951}, by making the
change of variable $\psi(\vx,t)=2\nu\ln\Xi(\vx,t)$, where $\psi(\vx,t)$ is the
velocity potential defined by
\beq
\vu(\vx,t) = - \nabla \psi .
\label{u-psi}
\eeq
This yields the linear heat
equation for $\Xi(\vx,t)$, which leads to the solution
\beq
\psi(\vx,t) = 2\nu\ln\int\frac{\dd\vq}{(4\pi\nu t)^{d/2}} \, 
\exp\left[ \frac{\psi_0(\vq)}{2\nu}-\frac{|\vx-\vq|^2}{4\nu t}\right] .
\label{Hopf1}
\eeq
Then, in the inviscid limit $\nu\rightarrow 0^+$, a steepest-descent method
gives \cite{Burgersbook,Bec2007}
\beq
\psi(\vx,t) = \sup_{\vq}\left[\psi_0(\vq)-\frac{|\vx-\vq|^2}{2t}\right] .
\label{psixpsi0q}
\eeq
If there is no shock, the maximum in (\ref{psixpsi0q}) is reached at a unique
point $\vq(\vx,t)$, which is the Lagrangian coordinate of the particle that is
located at the Eulerian position $\vx$ at time $t$ \cite{Burgersbook,Bec2007}
(hereafter, we note by the letter $q$ the Lagrangian coordinates, i.e. the initial
positions at $t=0$ of particles, and by the letter $x$ the Eulerian coordinates
at any time $t>0$).
Moreover, this particle has kept its initial velocity and we have
\beq
\vu(\vx,t) = \vu_0[\vq(\vx,t)] = \frac{\vx-\vq(\vx,t)}{t} .
\label{vxv0q}
\eeq
If there are several degenerate solutions to (\ref{psixpsi0q}), we have a shock
at position $\vx$ and the velocity is discontinuous (as seen from 
Eq.(\ref{vxv0q}), as we move from one solution $\vq_-$ to another one $\vq_+$ 
when we go through $\vx$ from one side of the shock surface to the other side).

The solution (\ref{psixpsi0q}) has a nice 
geometrical interpretation in terms of paraboloids \cite{Burgersbook,Bec2007}.
Thus, let us consider the family of upward paraboloids $\cP_{\vx,c}(\vq)$ 
centered at $\vx$ and of height $c$, with a curvature radius $t$, 
\beq
\cP_{\vx,c}(\vq)=\frac{|\vq-\vx|^2}{2t}+c .
\label{Paraboladef}
\eeq
Then, moving down $\cP_{\vx,c}(\vq)$ from $c=+\infty$, where the paraboloid is
everywhere well above the initial potential $\psi_0(\vq)$ (this is possible
for the initial conditions (\ref{ndef}) below, since we have $|\psi_0(\vq)| \sim
q^{(1-n)/2}$, which grows more slowly than $q^2$ at large distances),
until it touches the surface defined by $\psi_0(\vq)$, the abscissa $\vq$ of this
first-contact point is the Lagrangian coordinate $\vq(\vx,t)$. If first-contact
occurs simultaneously at several points there is a shock at the Eulerian 
location $\vx$. One can build in this manner the inverse Lagrangian map,
$\vx\mapsto\vq(\vx,t)$.

\subsection{Geometrical Adhesion Model for the density field: Legendre conjugacy and convex hull}
\label{Lagrangian-potential}

To the velocity field $\vu(\vx,t)$, defined by the Burgers equation (\ref{Burgers}),
we associate a density field $\rho(\vx,t)$ generated by this dynamics, starting from
a uniform density $\rho_0$ at the initial time $t=0$.
The latter obeys the usual continuity equation outside of shocks.
However, along shock lines, where the inviscid velocity field is discontinuous,
it is possible to define several prescriptions for the evolution of the matter distribution
in dimensions higher than 1. 
In this article we use the prescription described in detail in
\cite{BernardeauVal2010b}, where both the velocity and density fields are first
defined for finite $\nu$ and the inviscid limit is taken on a par, in a fashion which
allows to derive the matter distribution from a geometrical construction in terms
of convex hulls. In terms of the continuity equation, this corresponds to adding
a specific diffusive term that is
proportional to $\nu$. Then, this term vanishes in the inviscid limit outside of shocks
but it has a nontrivial nonzero limit along shocks (just as the diffusive term in
Eq.(\ref{Burgers}) has a nontrivial inviscid limit, which prevents the formation
of multistreaming flows in $\vu(\vx,t$)).

We now describe how the matter distribution is obtained within this
``Geometrical Adhesion Model''.
Let us first recall that an alternative description of the Burgers dynamics to the
paraboloid interpretation (\ref{Paraboladef}) is provided by the Lagrangian potential
$\varphi(\vq)$ \cite{Vergassola1994,Bec2007,BernardeauVal2010b}.
Thus, let us define the ``linear'' Lagrangian potential $\varphi_L(\vq,t)$ by
\beq
\varphi_L(\vq,t) = \frac{|\vq|^2}{2} - t \psi_0(\vq) ,
\label{phiLdef}
\eeq
so that in the linear regime the Lagrangian map, $\vq \mapsto \vx$, is given by
\beq
\vx_L(\vq,t) =  \frac{\pl\varphi_L}{\pl\vq} = \vq + t \vu_0(\vq) .
\label{xL}
\eeq
Thus, we recover the linear displacement field, $\vx_L-\vq= t\vu_0(\vq)$,
which is valid before shocks appear, as seen in Eq.(\ref{vxv0q}) above.
Next, introducing the function
\beq
H(\vx,t) = \frac{|\vx|^2}{2} + t \psi(\vx,t) ,
\label{Hdef}
\eeq
the maximum (\ref{psixpsi0q}) can be written as the Legendre transform
\beq
H(\vx,t) = \sup_{\vq} \left[ \vx\cdot\vq - \frac{|\vq|^2}{2} + t \psi_0(\vq) \right]
= \cL_{\vx} [ \varphi_L(\vq,t) ] .
\label{Hxphiq}
\eeq
Here we used the standard definition of the Legendre-Fenchel conjugate $f^*(\vs)$
of a function $f(\vx)$,
\beq
f^*(\vs) \equiv \cL_{\vs} [ f(\vx) ] = \sup_{\vx} [ \vs\cdot\vx - f(\vx) ] .
\label{Legendredef}
\eeq
Therefore, Eulerian quantities, such as the velocity field $\vu(\vx,t)$, which can
be expressed in terms of the velocity potential $\psi(\vx,t)$, whence of $H(\vx,t)$,
can be computed through the Legendre transform (\ref{Hxphiq}).
In particular, this yields the inverse Lagrangian map, $\vx \mapsto \vq$,
$\vq(\vx,t)$ being the point where the maximum in Eq.(\ref{psixpsi0q}) or
(\ref{Hxphiq}) is reached. 

In 1D, one can derive the direct Lagrangian map, $\vq\mapsto\vx$, from this
inverse map, $\vx \mapsto \vq$, using the fact that both maps are monotonically
increasing as particles cannot cross. However, in higher dimensions this is no
longer the case and one must explicitly define the evolution of the matter distribution.
As explained in \cite{BernardeauVal2010b}, within an appropriate inviscid limit
for the density field it is possible to identify the ``Lagrangian-Eulerian'' mapping
$\vq\leftrightarrow\vx$ with the Legendre conjugacy associated with Eq.(\ref{Hxphiq}).
Thus, one obtains the direct map, $\vq\mapsto\vx$, by ``inverting'' Eq.(\ref{Hxphiq})
through a second Legendre transform,
\beq
\varphi(\vq,t) \equiv \cL_{\vq} [ H(\vx,t) ] = \sup_{\vx} \left[ \vq\cdot\vx 
-  H(\vx,t) \right] .
\label{varphicdef}
\eeq
From standard properties of the Legendre transform, this only gives back the
linear Lagrangian potential $\varphi_L(\vq,t)$ of Eq.(\ref{Hxphiq}) if the latter is
convex, and in the general case it gives its convex hull,
\beq
\varphi = {\rm conv}(\varphi_L) .
\label{phi-convex}
\eeq
Then, $\vq$ and $\vx$ are Legendre-conjugate coordinates and they are given by
\beq
\vq(\vx,t) = \frac{\pl H}{\pl\vx} , \;\;\; \vx(\vq,t) =  \frac{\pl\varphi}{\pl\vq} .
\label{qx-xq}
\eeq
Thus, both maps, $\vq(\vx)$ and $\vx(\vq)$, derive from a convex potential
and we can {\it define} the density field from these mappings by the
conservation of matter \cite{Gurbatov1991,Woyczynski2007,BernardeauVal2010b}, 
\beq
\rho(\vx,t) \dd\vx = \rho_0 \dd\vq ,
\eeq
which reads as
\beq
\frac{\rho(\vx)}{\rho_0} = \det\left(\frac{\pl\vq}{\pl\vx}\right) 
= \det\left(\frac{\pl\vx}{\pl\vq}\right)^{-1} .
\label{rhoJacob}
\eeq
Here we used the fact that both determinants are positive, thanks to the convexity
of $H(\vx)$ and $\varphi(\vq)$.
Thus, the ``Lagrangian-Eulerian'' mapping $\vq\leftrightarrow\vx$ (\ref{qx-xq})
and the density field (\ref{rhoJacob}) define the ``Geometrical Adhesion Model'' that we
study in this article.

Here we must note that it is possible to use other prescriptions for the evolution
of the distribution of matter. For instance, one can use the standard continuity
equation. However, in this case one needs to numerically integrate the continuity
equation over all previous times, so that one loses the advantages of the
Hopf-Cole solution, which allows to integrate the dynamics to obtain at once
the velocity field at any time through Eq.(\ref{psixpsi0q}).
By contrast, within the approach studied here, the density field at any time is fully
determined by the Legendre transform (\ref{varphicdef}), that is, by the convex hull
(\ref{phi-convex}). 
Therefore, the nonlinear dynamics has been reduced to a one-time geometrical
problem of convex analysis. In order to obtain the velocity and density fields at
time $t$ it is sufficient to compute the Legendre transforms (\ref{Hxphiq}) and
(\ref{varphicdef}), the latter being equivalent to the direct computation of the convex
hull (\ref{phi-convex}). In particular, there is no need to compute the evolution of the
system over previous times.

We must point out that, although different prescriptions coincide in regular
regions (outside of shocks), they can lead to very different behaviors on the shock
manifold. Thus, it has been shown that if one uses the standard continuity equation,
limit trajectories are unique \cite{Bogaevsky2004,Bogaevsky2006}, so that
trajectories that pass through a point at a given time coincide at all later times.
Then, halos cannot fragment (particles which have coalesced remain
together forever) but they can stop growing and leave shock nodes (while remaining
on the shock manifold).
In contrast, as described in details in \cite{BernardeauVal2010b}
(see also \cite{Gurbatov1991}), within the
approach (\ref{qx-xq}) studied in this article halos can fragment in dimension
$2$ or higher. More precisely, in 2D the matter within shock nodes is redistributed
through only two kinds of events, ``$(2\rightarrow 2)$ flips'' and ``$(3\rightarrow 1)$
mergings''. 
In the first case, 2 halos collide and give rise to 2 new halos, whereas in the second
case, 3 halos collide to form a single object. In each case there is a redistribution of
matter but the total momentum is conserved.
In 3D there are $(2\rightarrow 3), (3\rightarrow 2)$ and  $(4\rightarrow 1)$ events.

On the other hand, as noticed above, in 1D there are no ambiguities as the map
$q(x)$ is sufficient to build $x(q)$, and all prescriptions coincide (in the inviscid
limit).

The ``geometrical model'' defined by the Legendre conjugacy
(\ref{Hxphiq})-(\ref{varphicdef}) leads to specific tessellations of the
Lagrangian $\vq$-space and the Eulerian $\vx$-space
\cite{Gurbatov1991,Woyczynski2007,BernardeauVal2010b}.
More precisely, the Eulerian-space tessellation is fully defined by the Hopf-Cole
solution (\ref{psixpsi0q}), and for the power-law initial conditions (\ref{ndef})
that we consider in this article one obtains a Voronoi-like tessellation.
Eulerian cells correspond to empty regions (i.e. voids), which are associated to a
single Lagrangian coordinate $\vq$ as for all points $\vx$ in a cell the maximum
in (\ref{psixpsi0q}) is reached for the same value of $\vq$. The boundaries
of these cells correspond to shock lines in 2D (or shock surfaces in 3D) where the
velocity field is discontinuous, and are reminiscent of the filaments and sheets
observed in the 2D and 3D gravitational dynamics. However, for the power-law initial
conditions (\ref{ndef}) all the mass is contained within pointlike clusters
located at the summits of these Voronoi-like diagrams. Moreover, thanks to
the geometrical construction that underlies the ``geometrical model''
(\ref{qx-xq}), within this approach this Voronoi-like tessellation is associated
in a unique fashion to a dual Delaunay-like triangulation in Lagrangian
space. Thus, each shock node is associated with a triangle in 2D (a tetrahedron
in 3D) of this $\vq$-space triangulation, which gives the mass and the
initial location of the particles that make up this mass cluster.
Then, as time grows these tessellations evolve in a specific manner,
so that these dual constructions remain valid at all times. This implies
for instance in 2D that a collision between two shock nodes can only give
rise to two new shock nodes, and not to a single larger mass cluster, 
because two triangles cannot merge to build a single larger triangle
(this requires three triangles embedded in a larger one, which corresponds
to a three-body collision in Eulerian space) \cite{Gurbatov1991,BernardeauVal2010b}.

Here we may note that standard Voronoi tessellations have also been used in
cosmology to study the large-scale structures of the Universe, as they 
provide a model of these large-scale structures which can reproduce some properties
of the observed galaxy distribution \cite{Icke1987,Weygaert1989,Weygaert2007}.
The facts that the Burgers dynamics leads to generalized Voronoi cells as described
above, see also \cite{Woyczynski2007}, and that this model provides a good description
of gravitational clustering at large scales in cosmology
\cite{Gurbatov1989,Kofman1992,Vergassola1994}, provide
a further motivation for the use of Voronoi tessellations in this context.

The fact that within the approach defined by the ``geometrical model'' the system
can be integrated is obviously a great
simplification. This allows both to gain a better understanding of its properties,
taking advantage of this geometrical interpretation \cite{BernardeauVal2010b}, 
and to devise efficient numerical algorithms. This has already been investigated
in previous works, such as \cite{Vergassola1994}.
These nice properties are the main motivations for the use of the mapping
(\ref{qx-xq}), rather than alternative prescriptions which keep the standard continuity
equation even at shock locations but cannot be integrated in a similar fashion.
Moreover, as we shall describe in this article (see also 
\cite{Kofman1990,Kofman1992,Vergassola1994}) the density fields generated by
this  ``geometrical model'' 
show many properties that are similar to those observed in the large-scale
structures built by the gravitational dynamics that is relevant in cosmology.

\subsection{Initial conditions}
\label{Initial-conditions}

Since there is no external forcing in Eq.(\ref{Burgers}),
the stochasticity arises from the random initial velocity $\vu_0(\vx)$, which we 
take to be Gaussian and isotropic, whence $\lag\vu\rag=0$ by symmetry. Moreover, 
as is well known \cite{Bec2007}, if the initial velocity is potential,
$\vu_0=-\nabla\psi_0$, 
it remains so forever, so that the velocity field is fully defined by its 
potential $\psi(\vx,t)$, or by its divergence $\theta(\vx,t)$, through
Eq.(\ref{u-psi}) and
\beq
\theta = -\nabla\cdot\vu = \Delta \psi .
\label{thetadef}
\eeq
Normalizing Fourier transforms as
\beq
\theta(\vx) = \int\dd\vk \; e^{\ii\vk.\vx} \; \thetat(\vk) ,
\label{Fourier}
\eeq
the initial divergence $\theta_0$ is taken as Gaussian, homogeneous, and 
isotropic, so that it is fully described by its power spectrum $P_{\theta_0}(k)$ 
with
\beq
\lag\thetat_0\rag=0 , \;\;\; \lag\thetat_0(\vk_1)\thetat_0(\vk_2)\rag = 
\delta_D(\vk_1+\vk_2) P_{\theta_0}(k_1) ,
\label{Ptheta0def}
\eeq
where we note $\delta_D$ the Dirac distribution.
In this article we focus on the power-law initial power spectra,
\beq
P_{\theta_0}(k) = \frac{D}{(2\pi)^d} \, k^{n+3-d} \;\;\; \mbox{with} \;\; -3<n<1 ,
\label{ndef}
\eeq
which defines the normalization $D$ of the initial conditions.
Thus, the initial conditions obey the scaling laws
\beqa
\lambda>0 : \;\;\; \thetat_0(\lambda^{-1}\vk) & \law & \lambda^{d-(n+3)/2}
\; \thetat_0(\vk) , \label{scalingthetat0} \\
\theta_0(\lambda\vx) & \law & \lambda^{-(n+3)/2} \; \theta_0(\vx) ,
\label{scalingtheta0}
\eeqa
where ``$\law$'' means that both sides have the same statistical properties.
This means that there is no preferred scale in the system and for $-3<n<1$
the Burgers dynamics will generate a self-similar evolution 
\cite{Gurbatov1997,Valageas2009b}.
This is why we only consider the range $-3<n<1$ in this article.
For the initial velocity and potential this yields for any
$\lambda>0$,
\beq
\vu_0(\lambda\vx) \law \lambda^{-(n+1)/2} \; \vu_0(\vx) , \;\;\;
\psi_0(\lambda\vx) \law \lambda^{(1-n)/2} \; \psi_0(\vx) .
\label{scalingu0psi0}
\eeq
Since we have $\vut(\vk,t)=\ii(\vk/k^2)\thetat(\vk,t)$, the initial energy
spectrum is a power law,
\beq
\lag\vut_0(\vk_1)\cdot\vut_0(\vk_2)\rag = \delta_D(\vk_1+\vk_2) E_0(k_1) ,
\label{E0def}
\eeq
\beq
\mbox{with} \;\;\; E_0(k) = k^{-2} P_{\theta_0}(k) = \frac{D}{(2\pi)^d} \, k^{n+1-d} ,
\label{E0n}
\eeq
whereas the initial velocity potential power spectrum reads as
\beq
\lag\psit_0(\vk_1)\psit_0(\vk_2)\rag = \delta_D(\vk_1+\vk_2) P_{\psi_0}(k_1) ,
\label{Ppsi0def}
\eeq
\beq
\mbox{with} \;\;\; P_{\psi_0}(k) = \frac{D}{(2\pi)^d} \, k^{n-1-d} .
\label{Ppsi0n} 
\eeq

\subsubsection{``IR class'': $-3<n<-1$}
\label{IRclass}

For $-3<n<-1$ the initial velocity field is a continuous function but it is not
homogeneous and only shows homogeneous increments (if we do not add an
infrared cutoff). For instance, in the case $\{n=-2,d=1\}$ it is a Brownian motion.
Then, one may choose a reference point, such as the origin $\vx_0=0$, with
$\vu_0(\vxref)=0$, and define the initial velocity in real space as
\beq
\vu_0(\vx) = \int\dd\vk \left( e^{\ii\vk.\vx}-e^{\ii\vk.\vxref} \right) 
\vut_0(\vk) , \;\; \mbox{for} \;\; -3<n<-1 .
\label{uxukn}
\eeq
Alternatively, one may add an infrared cutoff and focus on much smaller scales
(i.e. push this cutoff to infinity in final results). In the numerical simulations below
we choose this second alternative as we always define our system on a finite box
with periodic boundary conditions.
In any case, the second-order velocity structure function $S_{u_0}$ does not suffer
from this IR divergence and it reads as
\beq
S_{u_0}(\vx_1,\vx_2) = \lag |\vu_0(\vx_1)-\vu_0(\vx_2)|^2\rag = D \, I_n \, x^{-n-1} ,
\label{Su0def}
\eeq
where $x=|\vx_2-\vx_1|$. Here we used Eq.(\ref{E0n}) and the factor $I_n$ is given
by
\beq
I_n = 2 (2\pi)^{-d/2} \int_0^{\infty} \dd k \, k^n \left( \frac{2^{1-d/2}}{\Gamma(d/2)}
- k^{1-d/2} J_{d/2-1}(k) \right).
\label{Indef}
\eeq
This reads as
\beqa
d=1 & : & I_n = \frac{2\sin(n\pi/2)}{\Gamma(-n)\sin[(n+1)\pi]} , \label{Ind1} \\
d=2 & : & I_n = \frac{2^{n+1}\sin(n\pi/2)}{\Gamma[(1-n)/2]^2\sin[(n+1)\pi]}
\label{Ind2} .
\eeqa
Note that $I_n$ is only defined for $-3<n<-1$ as it diverges for $n\geq -1$.

\subsubsection{``UV class'': $-1<n<1$}
\label{UVclass}

For $-1<n<1$ the initial velocity is homogeneous but it is no longer a continuous
function (if we do not add an ultraviolet cutoff). For instance, in the case
$\{n=0,d=1\}$ it is a white noise. Thus, the initial one-point velocity variance,
$\lag|\vu_0|^2\rag= \int\dd\vk E_0(k)$, shows an UV divergence. Then, it can
be convenient to consider the initial velocity potential $\psi_0$, which is continuous
(but not homogeneous), rather than the initial velocity $\vu_0$. Its initial
second-order structure function reads as
\beq
S_{\psi_0}(\vx_1,\vx_2) = \lag [\psi_0(\vx_1)-\psi_0(\vx_2)]^2\rag = D \, I_{n-2}
\, x^{-n+1} ,
\label{Spsi0def}
\eeq
where the coefficient $I_{n-2}$ is again given by Eqs.(\ref{Indef})-(\ref{Ind2}).

\subsection{Density contrast and linear mode}
\label{Linear-theory}

In order to follow the evolution of the matter distribution we define the density
contrast, $\delta(\vx,t)$, by
\beq
\delta(\vx,t) = \frac{\rho(\vx,t)-\rho_0}{\rho_0} .
\label{deltadef}
\eeq
Then, if we linearize the equation of motion (\ref{Burgers}) and the continuity
equation (which holds before the formation of shocks)
we obtain in the inviscid limit, $\nu\rightarrow 0^+$,
\beq
\thetat_L(\vk,t) = \thetat_0(\vk) , \;\; \deltat_L(\vk,t) = t \, \thetat_0(\vk)  ,
\label{linear}
\eeq
where the subscript $L$ stands for the ``linear'' mode.
Then, when we study the system at a finite time $t>0$, we can as well define the
initial conditions by the linear density field $\delta_L(\vx,t)$, which is
Gaussian, homogeneous, and isotropic, with a power spectrum
\beq
P_{\delta_L}(k,t) = t^2 \, P_{\theta_0}(k) = \frac{D}{(2\pi)^d} \,t^2 \, k^{n+3-d} ,
\label{PdeltaL}
\eeq
and an equal-time two-point correlation
\beqa
\lefteqn{C_{\delta_L}(\vx_1,\vx_2) 
= \lag \delta_L(\vx_1,t) \delta_L(\vx_2,t)\rag } \nonumber \\
&& = (2\pi)^{d/2} \int_0^{\infty}\dd k \, k^{d-1}
\, \frac{J_{d/2-1}(kx)}{(kx)^{d/2-1}} P_{\delta_L}(k)  \nonumber \\
&& \propto t^2 x^{-n-3} ,
\label{CdeltaL}
\eeqa
where $x=|\vx_2-\vx_1|$.
Note that for any $n>-3$ the initial density field is homogeneous, even though
the initial velocity only shows homogeneous increments when $-3<n<-1$.

Since we shall study the statistical properties of the density field smoothed
over arbitrary scales $x$, it is convenient to introduce the linear density
contrast $\delta_{Lr}$ smoothed over spherical cells of radius $r$,
\beq
\delta_{Lr} = \int_V \frac{\dd\vx}{V} \, \delta_L(\vx) 
= \int \dd\vk \, \deltat_L(\vk) W(kr) ,
\label{deltaLr}
\eeq
with
\beq
W(kr) = \int_V\frac{\dd\vx}{V} \, e^{\ii\vk\cdot\vx} = 2^{d/2} \, \Gamma(1+d/2)
\, \frac{J_{d/2}(kr)}{(kr)^{d/2}} .
\label{Wdef}
\eeq
Its variance is given by
\beq
\sigma^2(r) = \lag\delta_{Lr}^2\rag = \frac{2\pi^{d/2}}{\Gamma(d/2)}
\int_0^{\infty} \dd k \, k^{d-1} P_{\delta_L}(k) W(kr)^2 .
\label{sigmadef}
\eeq
Note that $\sigma^2$ is only finite over the range $-3<n<-1$ if $d=1$ and
over $-3<n<0$ if $d=2$. For higher $n$ it shows a UV divergence.

\subsection{Self-similarity}
\label{self-similarity}

For the initial conditions (\ref{ndef}) that we consider in this paper,
the rescaled initial velocity potential $\psi_0(\lambda\vq)$ has the same
probability distribution as $\lambda^{(1-n)/2}\psi_0(\vq)$ for any $\lambda>0$,
when we normalize by $\vu_0(0)=0$ and $\psi_0(0)=0$, as seen in 
Eq.(\ref{scalingu0psi0}).
Then, the explicit solution (\ref{psixpsi0q}) gives the scaling laws
\beqa
\psi(\vx,t) & \law & t^{\frac{1-n}{n+3}} \;
\psi\left(t^{\frac{-2}{n+3}}\vx,1\right) , \label{scale_psi} \\
\vu(\vx,t) & \law & t^{\frac{-n-1}{n+3}} \;
\vu\left(t^{\frac{-2}{n+3}}\vx,1\right) , \\
\vq(\vx,t) & \law & t^{\frac{2}{n+3}} \;
\vq\left(t^{\frac{-2}{n+3}}\vx,1\right) \label{scale_q} .
\eeqa
This means that the dynamics is self-similar: a rescaling of time is
statistically equivalent to a rescaling of distances, as
\beq
\lambda>0: \;\; t\rightarrow \lambda t, \;\; \vx \rightarrow 
\lambda^{\frac{2}{n+3}} \vx .
\label{selfsimilar}
\eeq
Thus, the system displays a hierarchical evolution as
increasingly larger scales turn nonlinear. More precisely, since in the
inviscid limit there is no preferred scale for the power-law initial conditions
(\ref{ndef}), the only characteristic scale at a given time $t$ is the
so-called integral scale of turbulence, $L(t)$, which is generated by the
Burgers dynamics and grows with time as in (\ref{selfsimilar}). Hereafter we choose
the normalization
\beq
L(t) =  (2Dt^2)^{1/(n+3)} ,
\label{Lt}
\eeq
where the constant $D$ was defined in Eq.(\ref{ndef}).
This scale measures the typical distance between shocks,
and it separates the large-scale quasi-linear regime, where the energy spectrum
and the density power spectrum keep their initial power-law forms, (\ref{E0n})
and (\ref{PdeltaL}), from the small-scale nonlinear regime, which is governed by shocks
and pointlike masses, where the density power spectrum reaches the universal
white-noise behavior (i.e. $P_{\delta}(k,t)$ has a finite limit for $k\gg 1/L(t)$).

This self-similar evolution only holds for $n<1$, so that $|\psi_0(\vq)|$ grows
at larger scales, see for instance Eq.(\ref{scalingu0psi0}), and $n>-3$,
so that $|\psi_0(\vq)|$ grows more slowly than $q^2$ and the solution
(\ref{psixpsi0q}) is well-defined \cite{Gurbatov1997}. 
This is the range that we consider in this paper.
The persistence of the initial power law at low $k$ for the energy spectrum,
$E(k,t) \propto k^{n+1-d}$, that holds in such cases, is also called the 
``principle of permanence of large eddies'' \cite{Gurbatov1997}.

In order to express the scaling law (\ref{selfsimilar}) it is convenient to
introduce the dimensionless scaling variables
\beq
\vQ= \frac{\vq}{L(t)} , \;\;\; \vX= \frac{\vx}{L(t)} , \;\;\; \vU= \frac{t\vu}{L(t)} ,
\;\;\; M= \frac{m}{\rho_0 L(t)^d} .
\label{QXU}
\eeq
Then, equal-time statistical quantities (such as correlations or probability distributions)
written in terms of these variables no longer depend on time and the scale $X=1$
is the characteristic scale of the system, associated with the transition from the 
linear to nonlinear regime.
In particular, the variance of the smoothed linear density contrast introduced in
Eq.(\ref{sigmadef}) writes as
\beq
d=1, \; -3<n<-1 : \;\; \sigma^2(X/2) = \frac{I_n}{2} \, X^{-n-3} ,
\label{sigma-1d}
\eeq
where $I_n$ was given in Eq.(\ref{Ind1}) and $X=2R$ (i.e. $X$ is the length of
the 1D interval and $R$ its radius), and
\beq
d=2, \; -3<n<0 : \;\; \sigma^2(R) = K_n \, R^{-n-3}
\label{sigma-2d}
\eeq
with
\beq
d=2, \; -3<n<0 : \;\; K_n = \frac{\Gamma(-n/2)\Gamma[(n+3)/2]}
{\pi^{3/2}(1-n)\Gamma[(1-n)/2]^2} .
\label{Kndef}
\eeq
On the other hand, in terms of the dimensionless scaling wavenumber $\vK$
and power spectra $P(K)$ defined as
\beq
\vK = L(t) \, \vk , \;\;\;  P(K) = L(t)^{-d} \, P(k,t) ,
\label{Kdef}
\eeq
the linear-regime density and velocity power spectra introduced in Eqs.(\ref{PdeltaL})
and (\ref{E0def}) read as
\beq
P_{\delta_L}(K) = \frac{1}{2(2\pi)^d} \, K^{n+3-d} , \;\;
E_0(K) = \frac{1}{2(2\pi)^d} \, K^{n+1-d} .
\label{PLE0}
\eeq

\section{The 1D case as a test bench}
\label{1D-test-bench}

Our numerical implementation follows the algorithms described in
appendix \ref{Algorithms-for-the-1D-Burgers-dynamics}. 
At any time $t$, the velocity field $u(x,t)$ and its potential $\psi(x,t)$ are obtained
from the Hopf-Cole solution (\ref{Hxphiq}). This also gives the inverse Lagrangian
map, $\vx\mapsto\vq$, which can be directly inverted to obtain the direct
Lagrangian map, $\vq\mapsto\vx$, because both mappings are monotonically
increasing. Then, the mapping $\vq\mapsto\vx$ fully determines the matter
distribution.
In order to compute the Legendre transforms associated with Eq.(\ref{Hxphiq})
we introduce the algorithm devised in \cite{Lucet1997}, which first builds
the convex envelope $\varphi$ before taking the Legendre transform.
This allows us to obtain the velocity and density fields with an
optimal running time that scales as $O(N)$, where $N$ is the number of grid
points used to set up the initial conditions. By contrast, previous works
\cite{Noullez1994,Vergassola1994} used a slower $O(N\ln N)$
algorithm.
We also take advantage of the fact that for the 1D case numerous exact results are
known allowing precise tests of the convergence properties of the codes.

\subsection{Shock mass function and large-mass tail}
\label{Shock-mass-function-text}

\begin{figure}
\begin{center}
\epsfxsize=7 cm \epsfysize=5 cm {\epsfbox{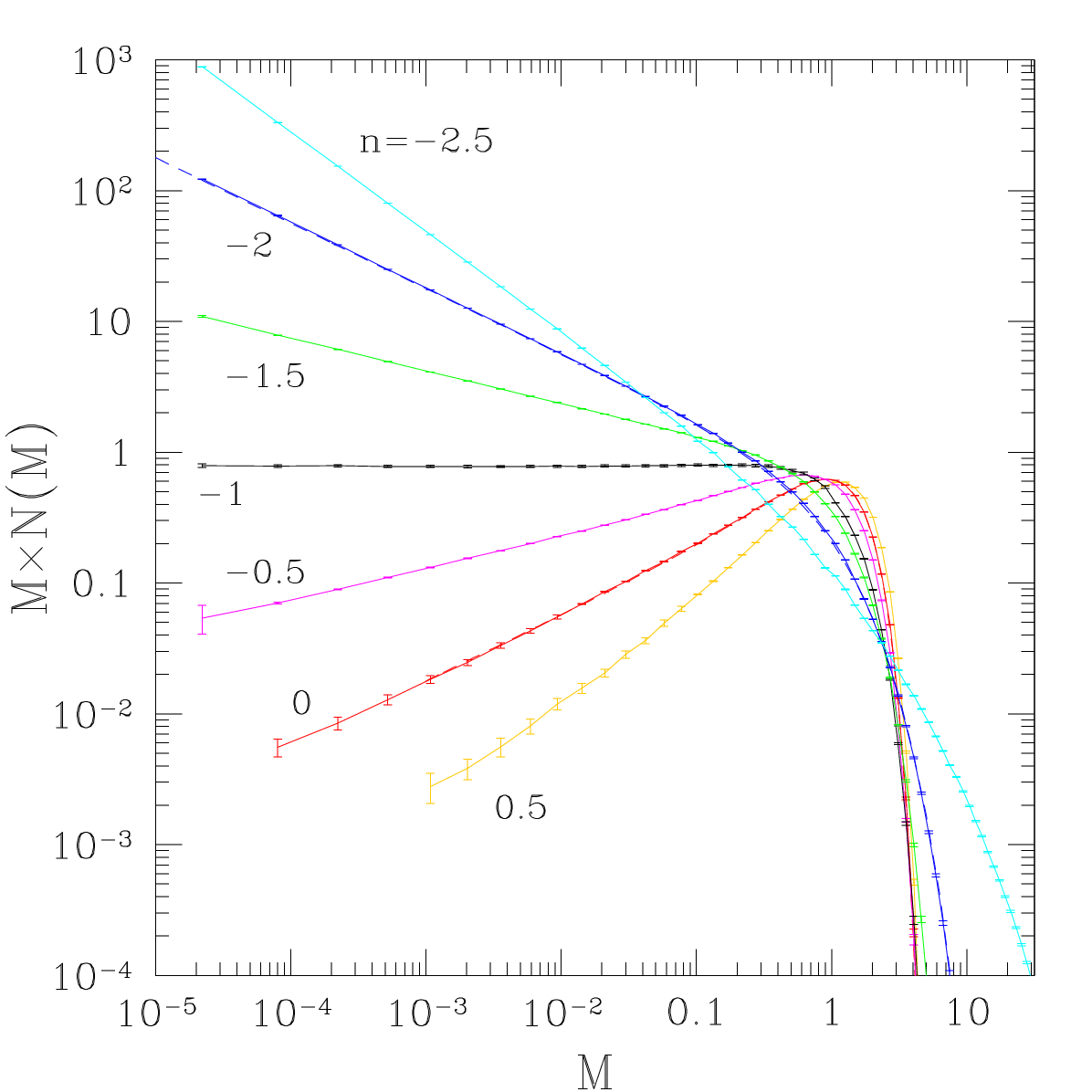}}
\end{center}
\caption{The shock mass functions $N(M)$ obtained for several indices $n$. We plot
the product $M\times N(M)$, in terms of the dimensionless scaling variables
(\ref{QXU}), to distinguish on the same plot both low-mass and high-mass regimes.
The small error bars show the statistical error (measured from the scatter between
different realizations). For $n=0$ and $n=-2$ the dashed lines (which can hardly be
distinguished from the numerical results) show the exact
analytical results (\ref{NMn0}) and (\ref{NMn-2}).}
\label{figmNm}
\end{figure}

In Fig. \ref{figmNm} we show the shock mass function obtained for different values of the
index $n$. Its shape is known  exactly for $n=0$ and $n=-2$ and the numerical results
are shown to be in exact agreement with the theoretical predictions (\ref{NMn0}) and
(\ref{NMn-2}).

\begin{figure}
\begin{center}
\epsfxsize=7 cm \epsfysize=5 cm {\epsfbox{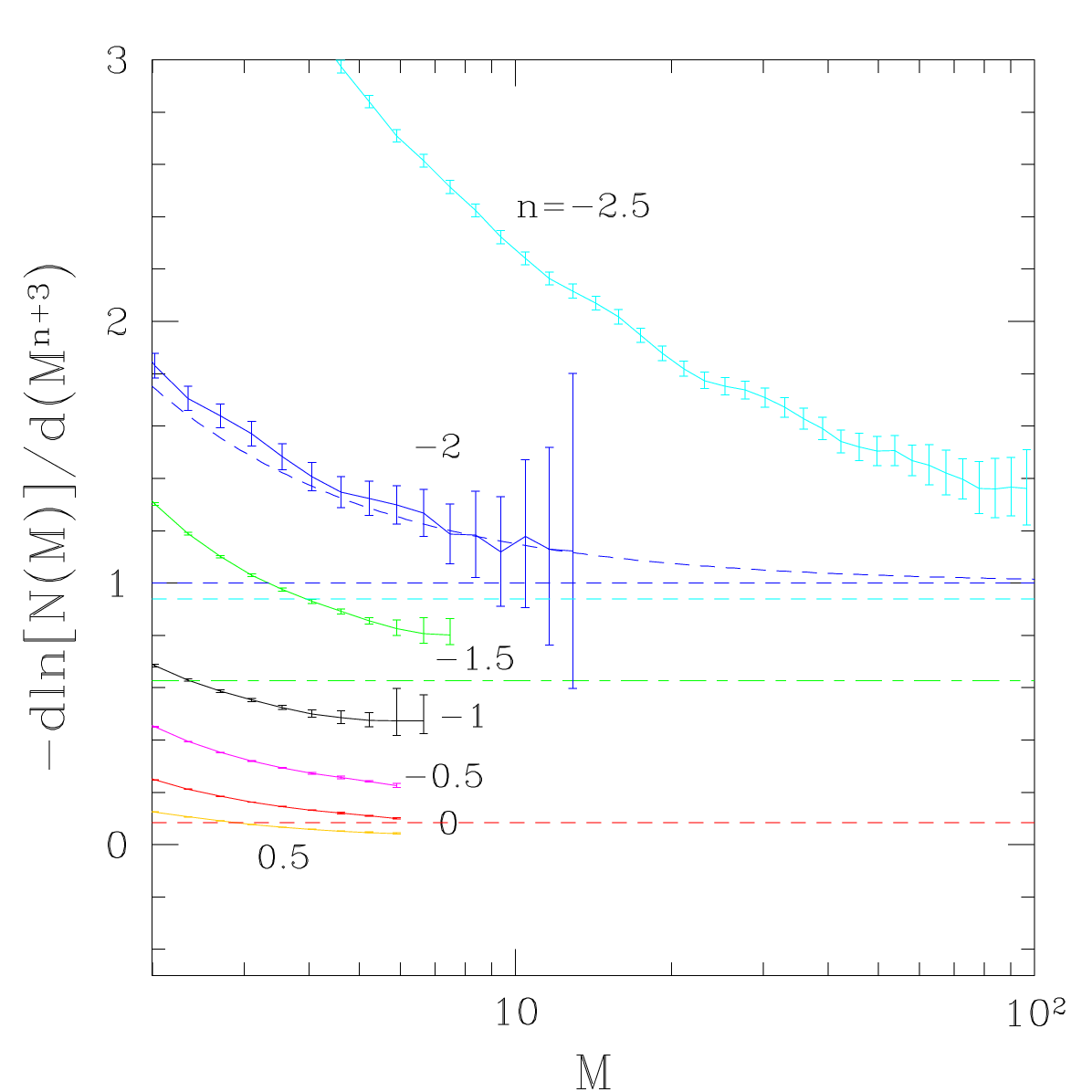}}
\end{center}
\caption{The derivative $-\dd\ln[N(M)]/\dd(M^{n+3})$ at high mass, for several values
of $n$. For $n=-2,-2.5$, and $0$, the horizontal dashed lines show the asymptotic
results (\ref{NMp-sd}) and (\ref{NMp-n0}). For $n=-2$ we also show the exact derivative
obtained from Eq.(\ref{NMn-2}) (curved dashed line). For $n=-1.5$ the dot-dashed line
is the value obtained from Eq.(\ref{NMp-sd}), which is only approximate in this case.}
\label{figlNmp}
\end{figure}

In order to measure the exponent that governs the high-mass tail, we show in
Fig.~\ref{figlNmp} the derivative $-\dd\ln[N(M)]/\dd(M^{n+3})$. As shown in 
\cite{Molchan1997,Ryan1998}, the shock mass function obeys the high-mass
asymptotic behavior 
\beq
-3<n<1 , \;\; M \rightarrow \infty : \;\; \ln N(M) \sim - M^{n+3} .
\label{NMp}
\eeq
For $-3<n\leq -2$ it is possible to obtain the numerical prefactor using a saddle-point
approach \cite{Valageas2009b}, which gives
\beq
-3 < n \leq -2 , \;\; M \rightarrow \infty : \;\; \ln N(M) \sim - \frac{M^{n+3}}{I_n} ,
\label{NMp-sd}
\eeq
where $I_n$ was defined in Eq.(\ref{Ind1}). One can check that this agrees with
the exact result (\ref{NMn-2}) obtained for $n=-2$ by different methods
\cite{Bertoin1998,Valageas2009a}.
For $n>-2$ the relevant saddle point develops shocks and makes the analysis more
complex, although for $n=0$ it is possible to obtain analytical results and to recover
the standard theoretical prediction (\ref{NMn0}) \cite{Frachebourg2000,Valageas2009c}
(using our normalizations),
\beq
n=0 , \;\; M \rightarrow \infty : \;\; \ln N(M) \sim - \frac{M^3}{12} .
\label{NMp-n0}
\eeq
We can check in Fig.~\ref{figlNmp} that each numerical curve reaches a constant
asymptote at high mass, in agreement with the scaling (\ref{NMp}).
Moreover, for $n=-2.5, -2$, and $0$, it is consistent with the analytical results
(\ref{NMp-sd}) and (\ref{NMp-n0}). 

Since these results can be recovered by a saddle-point approach \cite{Valageas2009b}, 
which also applies to the gravitational case, this suggests
that the large-mass tails of the halo mass functions can also be exactly obtained
for 3D gravitational clustering, as explained in \cite{Valageas2009d}.
However, as seen in Fig.~\ref{figlNmp}, the rate of convergence to the asymptotic regime
(\ref{NMp}) may be rather slow, especially for low $n$ (but note that the deviations from
the asymptotic behavior (\ref{NMp}) are magnified in Fig.~\ref{figlNmp} and would appear
much smaller in Fig.~\ref{figmNm}).

More precise comparisons regarding the high-mass or
low-mass tails can be found in the appendix \ref{One-dimension}.

\subsection{Press-Schechter like scaling}
\label{1D-PS}

With these results we are in a position to test the Press \& Schechter formalism 
\cite{Press1974} recalled in the introduction.
For the Gaussian initial conditions (\ref{ndef}) we have for $\nu(M)$,
\beq
-3<n<-1 : \;\; \nu(M) = \sqrt{\frac{2}{I_n}} \, M^{(n+3)/2}
\label{nuM1D}
\eeq
which leads to,
\beq
-3<n<-1 : \;\; N_{\rm PS}(M) = \frac{(n+3)}{\sqrt{\pi I_n}} \, M^{(n-1)/2} \,
e^{-M^{n+3}/I_n} .
\label{N-PS}
\eeq
As noticed in \cite{Vergassola1994}, it happens that Eq.(\ref{N-PS})
actually recovers both the known high-mass and low-mass exponents of Eqs.(\ref{NMp})
and (\ref{NMm}). In fact, a saddle-point approach \cite{Valageas2009b} shows that
it gives the exact high-mass asymptotic behavior (\ref{NMp-sd}) for $-3<n\leq -2$.
This is because in the inviscid limit i) particles move freely until shell-crossing in
the Burgers dynamics, and therefore follow the linear displacement field, and
ii) shell crossing only occurs for $n>-2$ in the saddle-point that governs the
high-mass tail (\ref{NMp}).
For $n=-2$, as noticed in \cite{Valageas2009a}, the Press-Schechter mass function
(\ref{N-PS}) is actually exact as shown by the comparison with the exact Eq.(\ref{NMn-2})
($I_{-2}=1$).
For $n>-2$ the factor $I_n$ in the exponential (\ref{N-PS}) no longer applies, as shocks
come into play (note that $I_n$ actually diverges for $n\geq -1$), but the exponent
is still valid as seen in (\ref{NMp}).

\begin{figure}
\begin{center}
\epsfxsize=7 cm \epsfysize=5 cm {\epsfbox{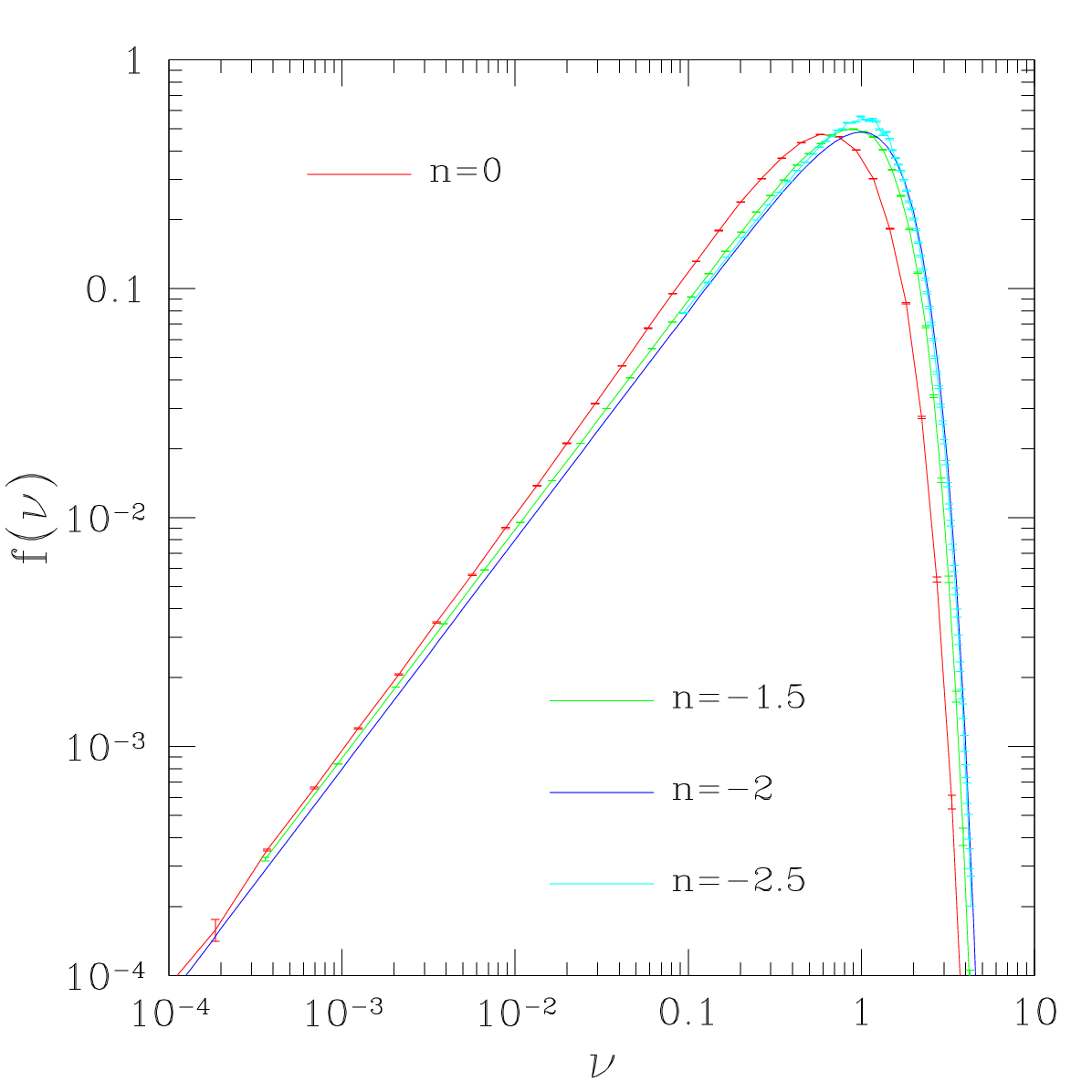}}
\end{center}
\caption{The shock mass function in terms of the scaling variable $\nu$ of
Eq.(\ref{nuM}), as defined by Eq.(\ref{fnudef}). For $n=-1.5, -2$, and $-2.5$, the
coefficient $I_n$ in Eq.(\ref{nuM}) is given by Eq.(\ref{Ind1}), but for $n=0$
(where $I_0$ would diverge) it is replaced by $I_0 \rightarrow 12$. For $n=-2$
the exact curve $f(\nu)$ happens to match the Press-Schechter model (\ref{fPS}).}
\label{figlfnu}
\end{figure}

In order to test whether the scaling with the reduced variable $\nu$ also works for the
Geometrical Adhesion Model, we plot in Fig.~\ref{figlfnu}
the function $f(\nu)$ defined from the shock mass function as
\beq
f(\nu) = M \, N(M) \, \frac{\dd M}{\dd\ln\nu} = \frac{2M^2}{n+3} \, N(M) ,
\label{fnudef}
\eeq
where we used Eq.(\ref{nuM1D}).

We can see that the curves obtained for $n=-1.5, -2$,
and $-2.5$, are almost identical, which shows that the scaling (\ref{fnudef}) is a
very good approximation over this range, $-2.5\leq n \leq -1.5$, even though for
$-2<n<-1$ the numerical factor in the exponential cutoff must show a weak
dependence on $n$ as explained above. In addition, since for $n=-2$ the function
$f(\nu)$ defined by Eq.(\ref{fnudef}) coincides with the Press-Schechter model (\ref{fPS}),
this implies that the latter is a very good approximation for this range of indices,
$-2.5\leq n \leq -1.5$. Since for $n \geq -1$ the factor $I_n$ diverges we can no
longer use Eq.(\ref{nuM}); this also shows that the scaling (\ref{fnudef}) can only be
approximate. However, we display in Fig.~\ref{figlfnu} the curve obtained for $n=0$
by making the change $I_0 \rightarrow 12$ in Eq.(\ref{nuM}), so as to recover the
exact high-mass tail (\ref{NMp-n0}). We can see that this gives a function $f(\nu)$ that
remains close to the Press-Schechter model (\ref{fPS}), as could be expected from the
fact that the latter agrees with both the exact low-mass slope (\ref{NMm}) and the
high-mass cutoff, although a noticeable deviation can be seen around the peak at
$\nu \sim 1$. Therefore, it appears that the scaling (\ref{fnudef}) provides a
reasonable description of the shock mass function for all $n$, provided one uses
the appropriate normalization in the $M\mapsto\nu$ relation (\ref{nuM}).

\subsection{Density field}
\label{1D-density-field}

In appendix  \ref{One-dimension}, we further show the behavior of the one-point probability 
distribution functions (PDF) of the smoothed density, and their low-order cumulants, and compare them with known 
results whenever possible. Those tests are successfully passed. It is to be noted that those quantities show behaviors 
in qualitative agreement with what is expected for the gravitational dynamics, in both the low 
variance regime and the large variance regime, although not necessarily for the very same reasons. 

In particular, we can check in Fig.~\ref{figSrho} that at large scale the reduced cumulants
$S_p$, defined by Eq.(\ref{Spdef}) (or the equivalent Eq.(\ref{Spdef-delta})), agree with
the analytical predictions (\ref{S3-d1})-(\ref{S34-QL}) (whence with Eq.(\ref{S3-GAM})
with $d=1$), for $n\leq -2$
(this upper boundary is due to strong shell-crossing
effects for $n>-1$, which are beyond the reach of perturbation theory).
This large-scale behavior can be analyzed in the same terms as for the 3D gravitational dynamics,
through perturbation theory or saddle-point approaches, and both systems are
similar in this respect. 

By contrast, at small scale
the universal flat plateau exhibited by the reduced cumulants is a direct consequence
of the formation of pointlike structures in the GAM. For the same reason,
the density power spectrum (and also poly-spectra, although we do not explicitly show it here) precisely exhibits
a universal $k^0$ tail in the high-$k$ limit, see Fig.~\ref{figPk}, 
which is characteristic of the fact that formed 
objects are pointlike. This is not expected for the gravitational dynamics as the small-scale
behavior of the matter spectrum, and of the reduced cumulants $S_p$,
depends on the matter profile within objects \cite{Cooray2002,Valageas1999}.
Then, the ratios $S_p$ do not seem to reach constant asymptotes
at small scale \cite{Colombi1996}, even though their scale-dependence is very weak.
Moreover, the ``stable-clustering ansatz''
introduced in \cite{Davis1977,Peebles1980}, which would predict constant asymptotes
(and fares reasonably well), is not based on such universal singularities but
on very different arguments on the decoupling of collapsed halos, so that the
high-$k$ slope of the power spectrum depends on $n$.
Thus, although both dynamics show partly similar behaviors at small scales, in this
regime the correspondence is not exact and can be due to different physical processes.
In spite of these limitations, some key statistical quantities still show
similar behaviors at small scales, such as the mass function
and the PDF of the smoothed density described in appendix  \ref{One-dimension}.
Therefore, with some care the Geometrical Adhesion Model could still prove to be a
useful tool to understand processes or to test approximation schemes encountered
within the 3D gravitational dynamics.

\section{Two-dimensional dynamics}
\label{Two-dimensions}

We now consider the two-dimensional case, $d=2$.
Our numerical implementation follows the numerical algorithms described in
appendix \ref{Algorithms-for-the-2D-GAM}.

As in the 1D case, in Eulerian space the velocity field $\vu(\vx,t)$ and its potential
$\psi(\vx,t)$ are again obtained from the Hopf-Cole solution (\ref{Hxphiq}).
Since 2D Legendre transforms can be obtained from two successive 1D partial
Legendre transforms, we again use the algorithm of Lucet \cite{Lucet1997},
and we obtain an optimal running time that is linear over the number of
initial grid points, $\Nt=N^2$.

The main difficulty that arises in 2D, and higher dimensions, is that it is no
longer possible to read the direct Lagrangian map, $\vq\mapsto\vx$, whence
the matter distribution, from the ``inverse'' Lagrangian map, $\vx\mapsto\vq$.
As recalled in Sec.~\ref{Lagrangian-potential}, within the ``Geometrical Adhesion
Model'' this nontrivial ``inversion'' is performed through the second Legendre
transform (\ref{varphicdef}), or equivalently through the convex hull
(\ref{phi-convex}).
Taking this Legendre transform on a grid, as in some previous works, would allow
us to obtain an approximation of the matter distribution on such a grid.
Then, using the same algorithm as for the first Legendre transform (\ref{Hxphiq})
we would reach an optimal running time $O(\Nt=N^2)$.
However, as explained in App.~\ref{compute-convex-hull}, this procedure
artificially splits large voids into smaller voids and introduces spurious 
matter concentrations.
Therefore, in this article we prefer to exactly compute the convex hull
(\ref{phi-convex}), without introducing any Eulerian grid at time $t$.
Thus, once the initial conditions are given on a grid, we {\it exactly} solve
the dynamics and we compute the exact Lagrangian and Eulerian space
tessellations, which have been discussed in details in  \cite{BernardeauVal2010b}.
In particular, the density peaks are not restricted to a predefined Eulerian grid.

Of course, the computation of the exact convex hull $\varphi$ is a much more 
difficult problem than the computation of the 2D Legendre transform on a grid.
Indeed, whereas the latter could be reduced to 1D problems, as explained above,
the former is a standard problem of 3D computational geometry (since $\varphi(\vq)$
is embedded in 3D). As described in App.~\ref{compute-convex-hull} and
\ref{Algorithms-for-the-2D-Burgers-dynamics}, we implement the 3D
divide-and-conquer algorithm devised by Chan \cite{Chan2003}.
This recursive algorithm allows us to reach an optimal running time $O(\Nt\ln\Nt)$.
(This is slower than the computation of the 2D Legendre transform on a grid,
because both problems do not have the same complexity, and the convex
hull contains more information.)

We discuss our numerical algorithms in greater detail in appendix
\ref{Algorithms-for-the-2D-GAM}, in particular we compare them with previous
numerical studies in appendix \ref{Numerical-comparison}.

\subsection{Shock mass function}
\label{Shock-mass-function-2D}

\begin{figure}
\begin{center}
\epsfxsize=7 cm \epsfysize=5 cm {\epsfbox{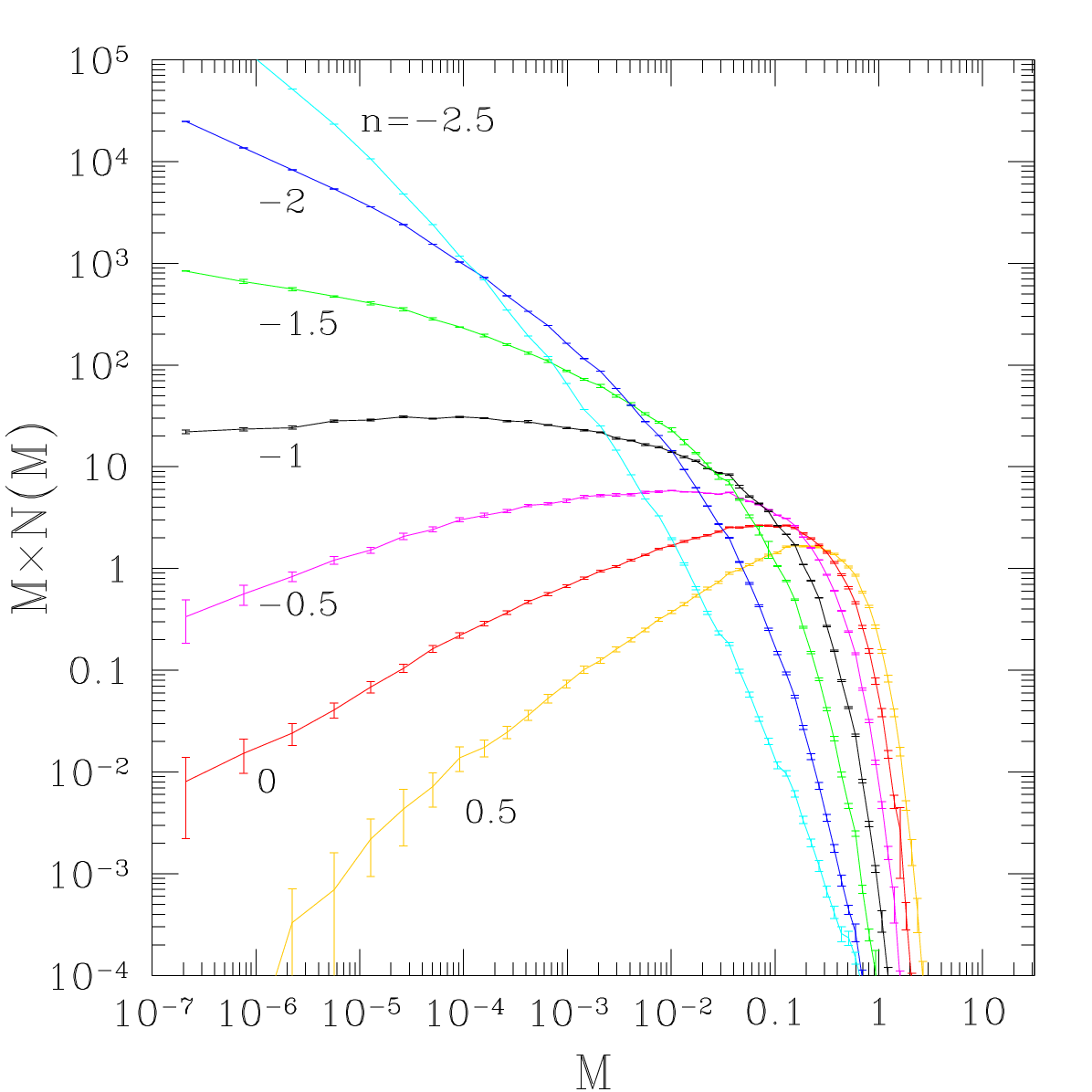}}
\end{center}
\caption{The product $M\times N(M)$, where $N(M)$ is the shock mass function in
the 2D case, for several $n$.}
\label{figmNm_2d}
\end{figure}

We show in Fig.~\ref{figmNm_2d} the shock mass functions obtained in the
2D case (more precisely we plot the product $M\times N(M)$). 
Thus, $N(M)\dd M$ is again the mean number of shocks of mass in the range
$[M,M+\dd M]$ within a unit volume.
As for the 1D case shown in Fig.~\ref{figmNm}, we can clearly see the power-law
tails at low mass and the exponential-like cutoffs at high mass, especially
for $n\geq -1.5$. For $n=-2$ and $n=-2.5$ where the low-mass tail grows faster
than $1/M$ it is more difficult to distinguish the low-mass asymptote from the
high-mass falloff.
Moreover, one can expect the convergence to the low and high-mass asymptotic
behaviors to be slower for lower $n$ if the scaling in terms of the Press-Schechter
variable $\nu$ defined as in Eq.(\ref{nuM}) is still a good approximation.
In agreement with previous works \cite{Vergassola1994},
we can check that, as for the 1D case, the shock mass function grows more slowly than
$1/M$ if $-1<n<1$ and faster than $1/M$ if $-3<n<-1$.
Therefore, the ``UV class'', $-1<n<1$, again corresponds to isolated shocks, which are
in finite number per unit volume, whereas the ``IR class'',  $-3<n<-1$, again
corresponds to dense shocks, which are in infinite number per unit volume.
This agrees with the study of the associated Lagrangian and Eulerian-space
tessellations described in \cite{BernardeauVal2010b}.

For $-3<n\leq -1$ it is again possible to obtain the high-mass cutoff of the
shock mass function \cite{Valageas2009b}, which now reads as
\beqa
\lefteqn{-3< n \leq -1 , \;\; M \rightarrow \infty : } \nonumber \\
&& \hspace{1.2cm} \ln N(M) \sim - \frac{2}{K_n} \pi^{-(n+3)/2} \, M^{(n+3)/2} ,
\label{NMp-sd-2D}
\eeqa
where the factor $K_n$ was defined in Eq.(\ref{Kndef}).
For higher $n$ the exponent $(n+3)/2$ is expected to remain valid, but shocks
should modify the numerical prefactor.
Note that the analytical result (\ref{NMp-sd-2D}) now extends up to $n=-1$,
instead of $n=-2$ in 1D. Indeed, in the general $d$-dimensional case, the overdense
saddle point associated with this high-mass tail is only affected by shocks for
$n>d-3$.
Unfortunately, the mass range of the numerical computations is too small to see the
convergence to the asymptotic behavior (\ref{NMp-sd-2D}), although they are
consistent with the scaling $M^{(n+3)/2}$.

\begin{figure}
\begin{center}
\epsfxsize=7 cm \epsfysize=5 cm {\epsfbox{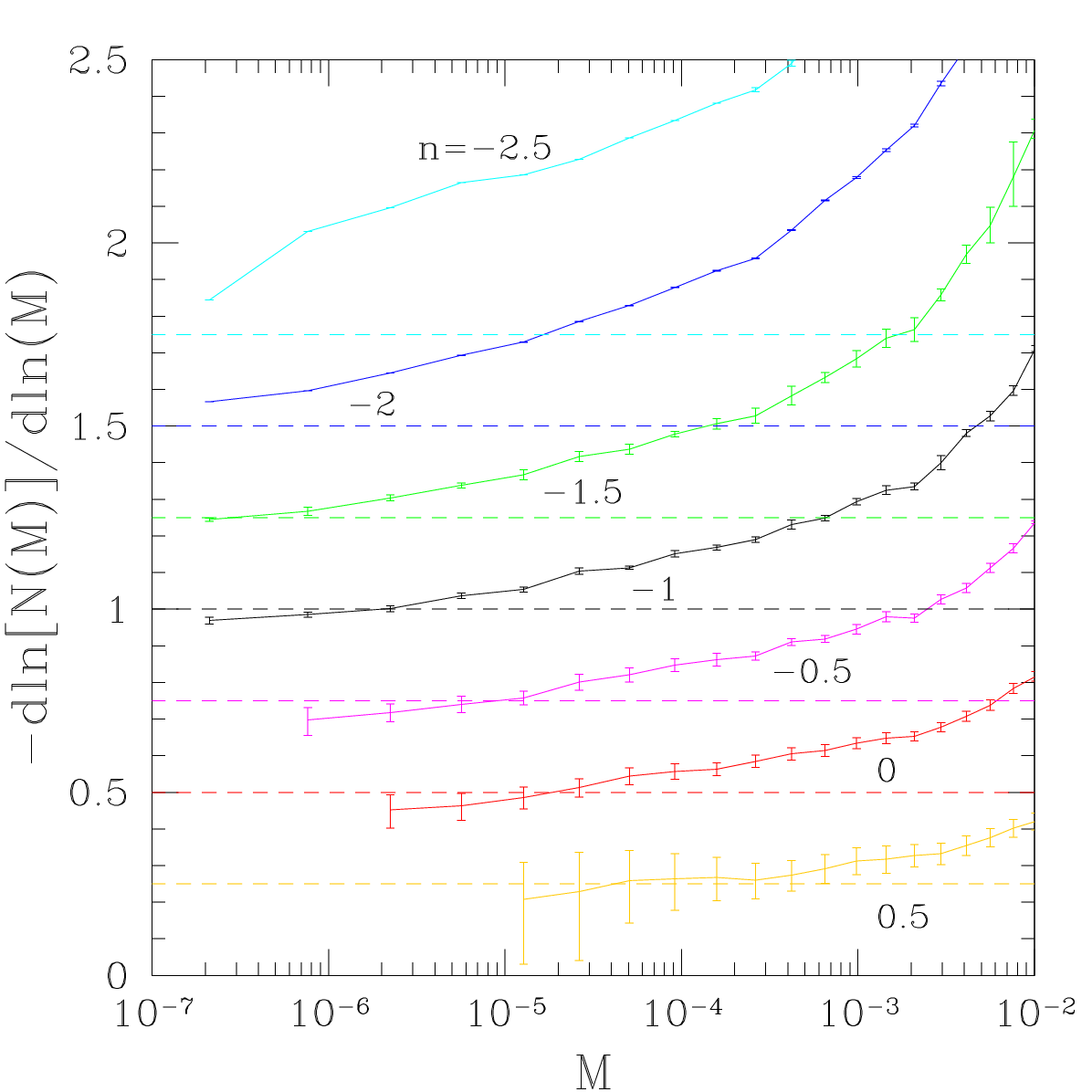}}
\end{center}
\caption{The derivative $-\dd\ln[N(M)]/\dd\ln(M)$ at low mass. The horizontal dashed
lines are the asymptotic behavior (\ref{NMm}).}
\label{figlNmm_2d}
\end{figure}

At low mass, previous numerical works and heuristic arguments \cite{Vergassola1994}
suggest that the 1D power-law tail (\ref{NMm}) remains valid, with the same
exponent $(n-1)/2$. As in Fig.~\ref{figlNmm}, we plot in Fig.~\ref{figlNmm_2d}
the derivative $-\dd\ln[N(M)]/\dd\ln(M)$. We can see that our numerical results
are consistent with Eq.(\ref{NMm}), although the numerical accuracy is not sufficient
to provide a precise measure of the exponent (we actually get slightly higher values
than $(1-n)/2$, but this might be due to logarithmic prefactors or to the fact that
the asymptotic regime is barely reached at these mass scales).
As already noticed in
Fig.~\ref{figmNm_2d}, the transition through the characteristic exponent
$N(M)\sim M^{-1}$, which marks the divide between dense and isolated shocks,
again appears to take place at $n=-1$.

\begin{figure}
\begin{center}
\epsfxsize=7 cm \epsfysize=5 cm {\epsfbox{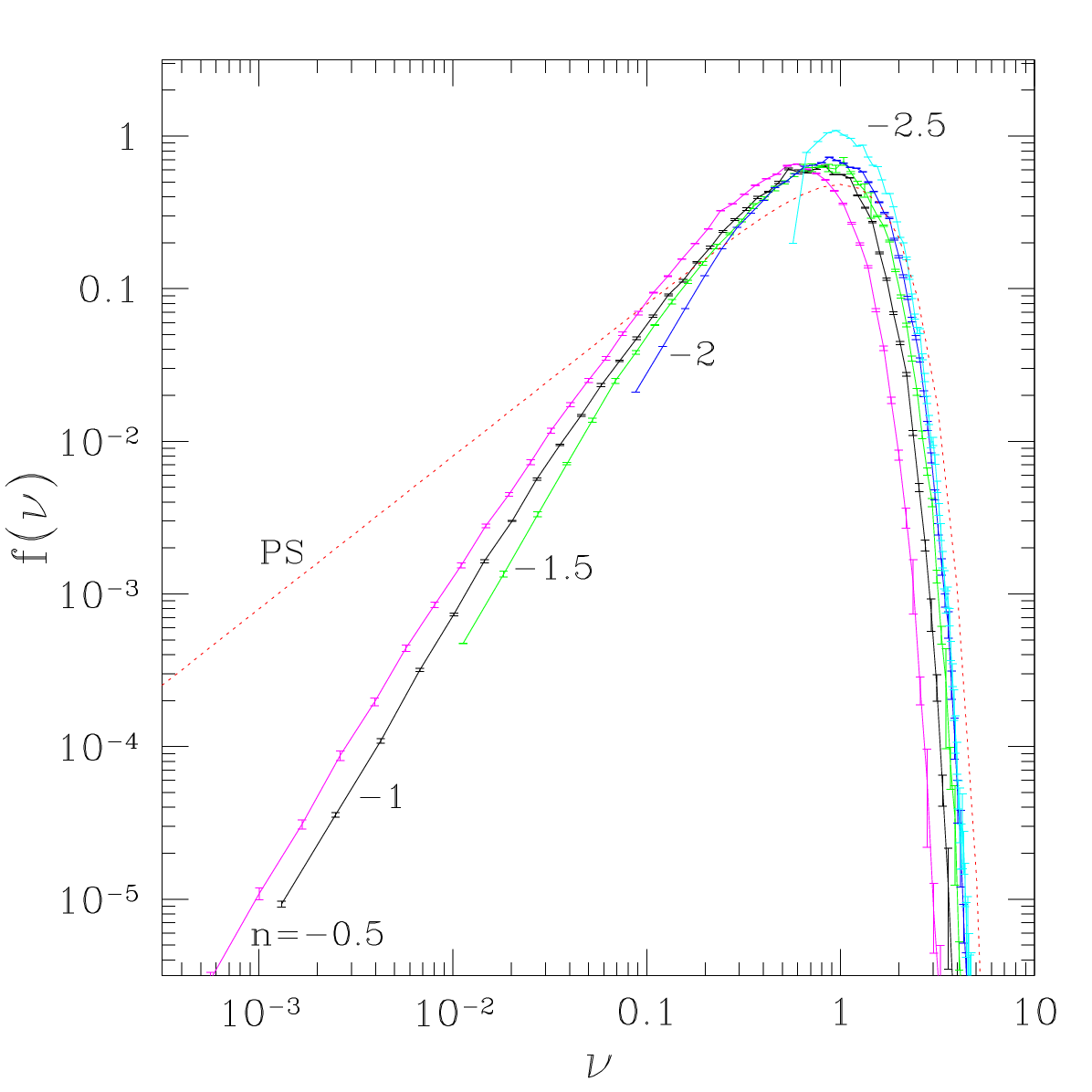}}
\end{center}
\caption{The shock mass function in terms of the scaling variable $\nu$ of
Eq.(\ref{nudef-2d}), as defined by Eq.(\ref{fnudef-2d}). The dotted line labeled
``PS'' is the Press-Schechter model (\ref{fPS}).}
\label{figlfnu_2d}
\end{figure}

Let us now investigate the Press-Schechter like scaling (\ref{FM-PS}).
In dimension $d$ the spherical collapse relates the linear density contrast
$\delta_L$ of a spherical region to its nonlinear density $\rho$ 
through \cite{Valageas2009b}
\beq
\rho = \rho_0 \left( 1 - \frac{\delta_L}{d} \right)^{-d} .
\label{FdeltaL-def}
\eeq
Therefore, in 2D complete collapse to a point is achieved at $\delta_L=2$ and
the variable $\nu$ of Eq.(\ref{nuM}) is now defined as
\beq
d=2 : \;\;\; \nu(M) = \frac{2}{\sigma(M)} .
\label{nudef-2d}
\eeq
Then, the Press-Schechter model \cite{Press1974} still reads as
Eqs.(\ref{FM-PS})-(\ref{fPS}). This now yields
\beq
d=2 , \;\; -3<n<0 : \;\; \nu = \frac{2}{\sqrt{K_n}} \, \pi^{-(n+3)/4} \, M^{(n+3)/4} ,
\label{nuM-2d}
\eeq
where $K_n$ was defined in (\ref{Kndef}), and
\beqa
-3 < n < 0 & : &  N_{\rm PS}(M) = \frac{n+3}{\sqrt{2K_n}} \, \pi^{-(n+5)/4}
\, M^{(n-5)/4} \nonumber \\
&& \hspace{1cm} \times \, e^{-2 M^{(n+3)/2}/(K_n\pi^{(n+3)/2})} .
\label{NPS-2d}
\eeqa
Thus, although the Press-Schechter prediction (\ref{NPS-2d}) recovers the
high-mass cutoff for $-3<n\leq -1$, for the same reasons as in 1D, it does
not reproduce the low-mass tail shown in Fig.~\ref{figlNmm_2d}, which
was consistent with Eq.(\ref{NMm}). This discrepancy was already noticed
in previous numerical works \cite{Vergassola1994}.
Nevertheless, it remains interesting to see whether a scaling of the form
(\ref{fnudef}) still provides a good approximation, albeit with a different function
$f(\nu)$ than the Press-Schechter model (\ref{fPS}). Thus, using Eq.(\ref{nuM-2d}),
we now define $f(\nu)$ as
\beq
f(\nu) = M \, N(M) \, \frac{\dd M}{\dd\ln\nu} = \frac{4M^2}{n+3} \, N(M) ,
\label{fnudef-2d}
\eeq
and we plot in Fig.~\ref{figlfnu_2d} the functions obtained for $n \leq -0.5$.
We can see that the scaling by the variable $\nu$ again provides a reasonable
description of the dependence on $n$ of the shock mass function, although
there remains a weak dependence on $n$. In particular, the low-mass exponent
$(n-1)/2$ of Eq.(\ref{NMm}) corresponds to the universal low-$\nu$ behavior
$f(\nu) \propto \nu^2$. Thus, even though the linear low-$\nu$ slope of
Eq.(\ref{fPS}) clearly fails, as seen in Fig.~\ref{figlfnu_2d}, a single quadratic
slope, $\propto \nu^2$, appears to match all mass functions, which was not
obvious a priori. However, its normalization shows a weak dependence on $n$.
At high mass the different curves are very close, in agreement with
(\ref{NMp-sd-2D}) (but prefactors are expected to depend on $n$), except for
the case $n=-0.5$ which falls somewhat below. This is expected from the constraint
in Eq.(\ref{NMp-sd-2D}), as for $n>-1$ shocks modify the normalization of this
high-mass asymptote.

Note that the range of masses shown in Fig.~\ref{figlfnu_2d} spans four orders
of magnitude in terms of the reduced variable $\nu$, whereas current cosmological
simulations of 3D gravitational clustering typically cover the range
$0.3<\nu<4.2$ \citep{Reed2003,Tinker2008}, that is only one order of magnitude.
This means that the asymptotic low-mass and high-mass tails are not really
probed by current 3D gravitational simulations. In particular, they cannot
measure the exponent of the low-$\nu$ tail.
For the Geometrical Adhesion Model, if the low-mass power-law tails
(\ref{NMm}) remain valid in higher dimensions, we actually obtain
$f(\nu) \sim \nu^d$ at low $\nu$ (with again no further dependence on $n$).
This agrees with the results described above
in 1D and 2D, as well as with the separable case in any dimension discussed in
Sec.~\ref{Separable-text} below, see Eq.(\ref{Md-Mm}).
Current cosmological simulations cannot discriminate between such behaviors in
3D, but it would be interesting to check in future works whether gravitational
clustering also gives rise to such a strong dependence on dimension, in terms
of the reduced variable $\nu$.
As seen in Fig.~\ref{figlfnu_2d} for the case of the Geometrical Adhesion
Model, such strong violations of the low-mass slope predicted by the simple
Press-Schechter prescription do not necessarily imply strong violations of the
Press-Schechter scaling itself.

This is another example of the benefits that can be obtained by studying
dynamics such as this ``Geometrical Adhesion Model'', which share many
properties with the gravitational dynamics and show complex nonlinear
behaviors while being simple enough to provide well controlled analytical
and numerical analysis. They provide nontrivial explicit examples that can
serve as a guide, to understand general properties or to confirm/rule out
simple expectations.

\subsection{Density distribution}
\label{Density-distribution-2d}

\begin{figure*}
\begin{center}
\epsfxsize=7 cm \epsfysize=5 cm {\epsfbox{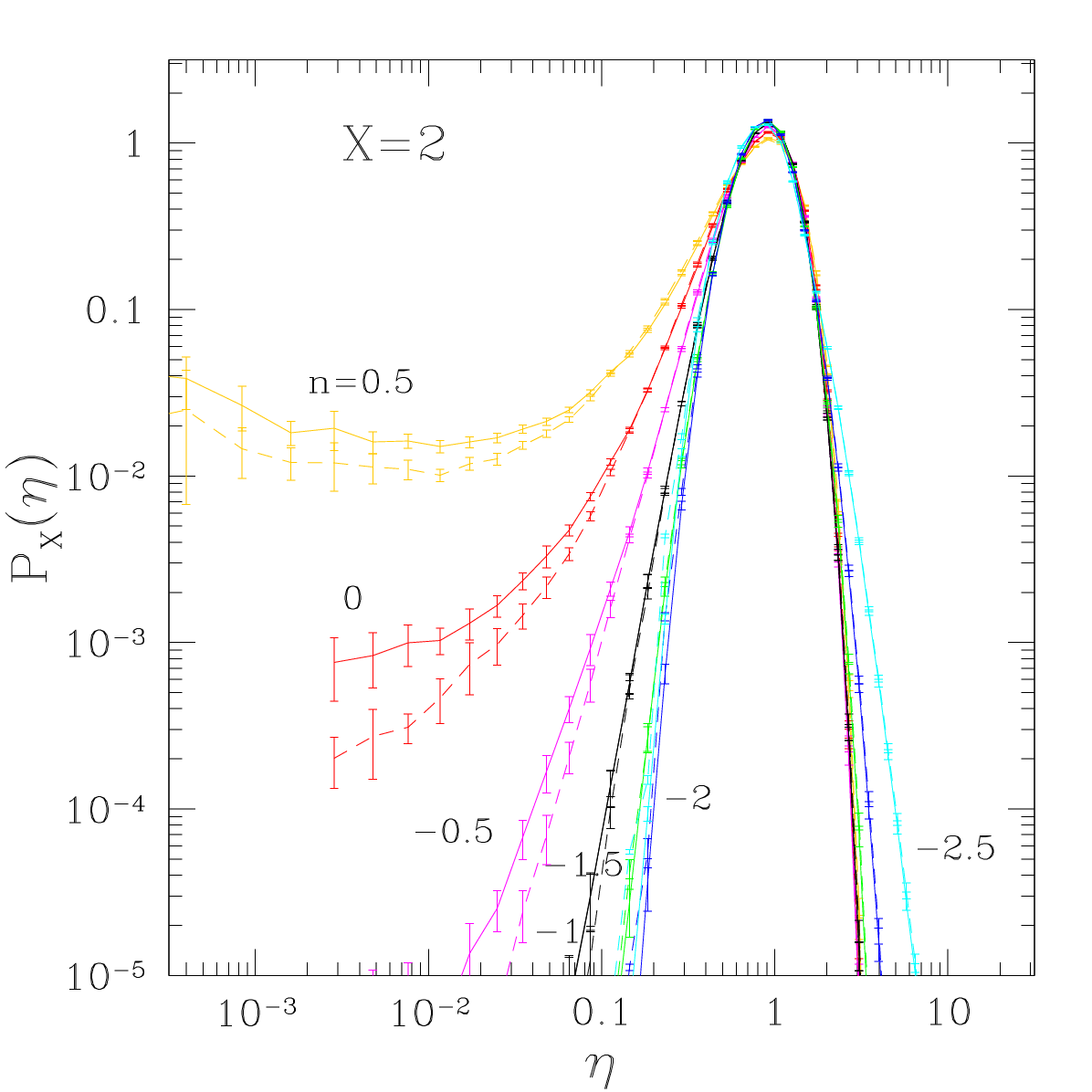}}
\epsfxsize=7 cm \epsfysize=5 cm {\epsfbox{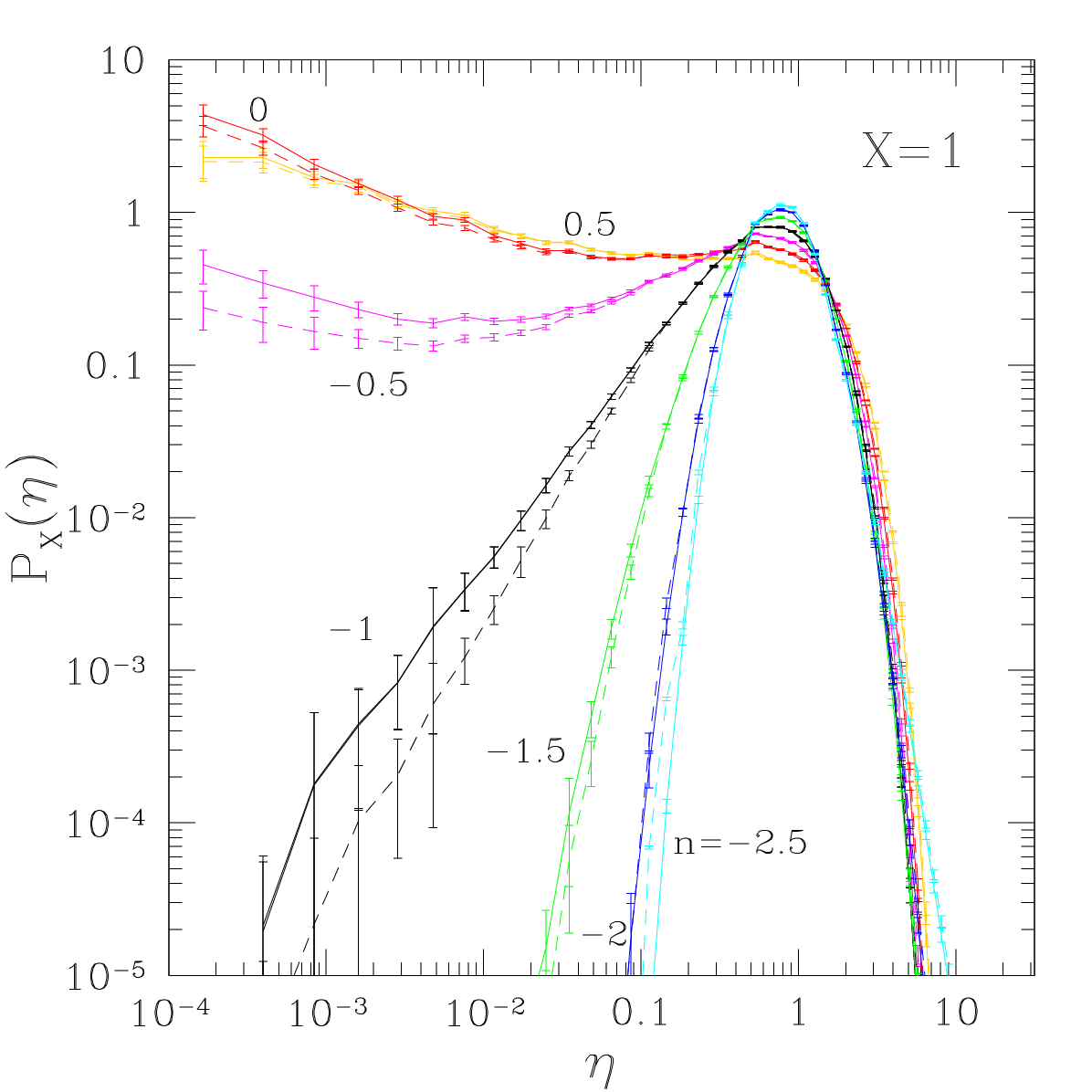}}
\epsfxsize=7 cm \epsfysize=5 cm {\epsfbox{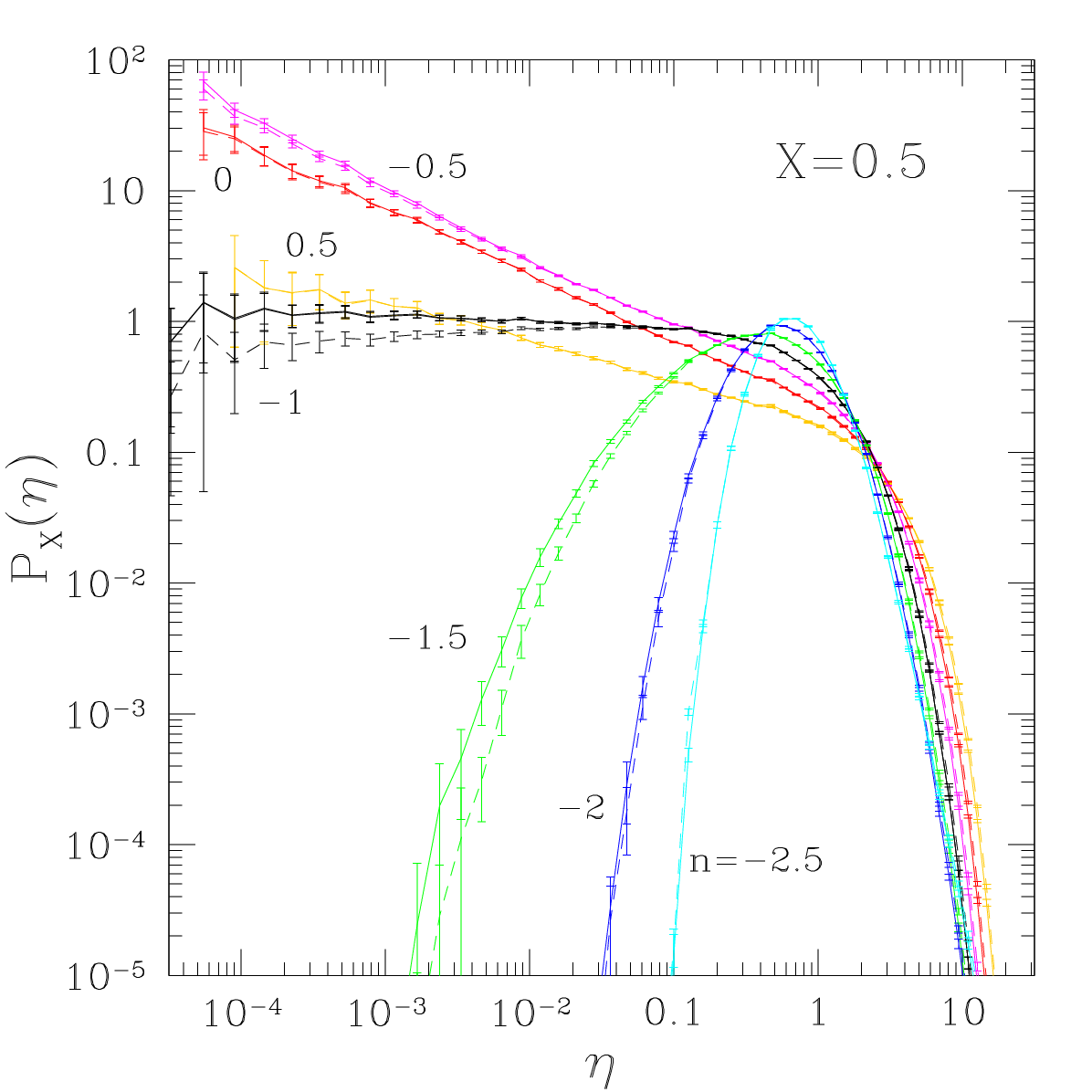}}
\epsfxsize=7 cm \epsfysize=5 cm {\epsfbox{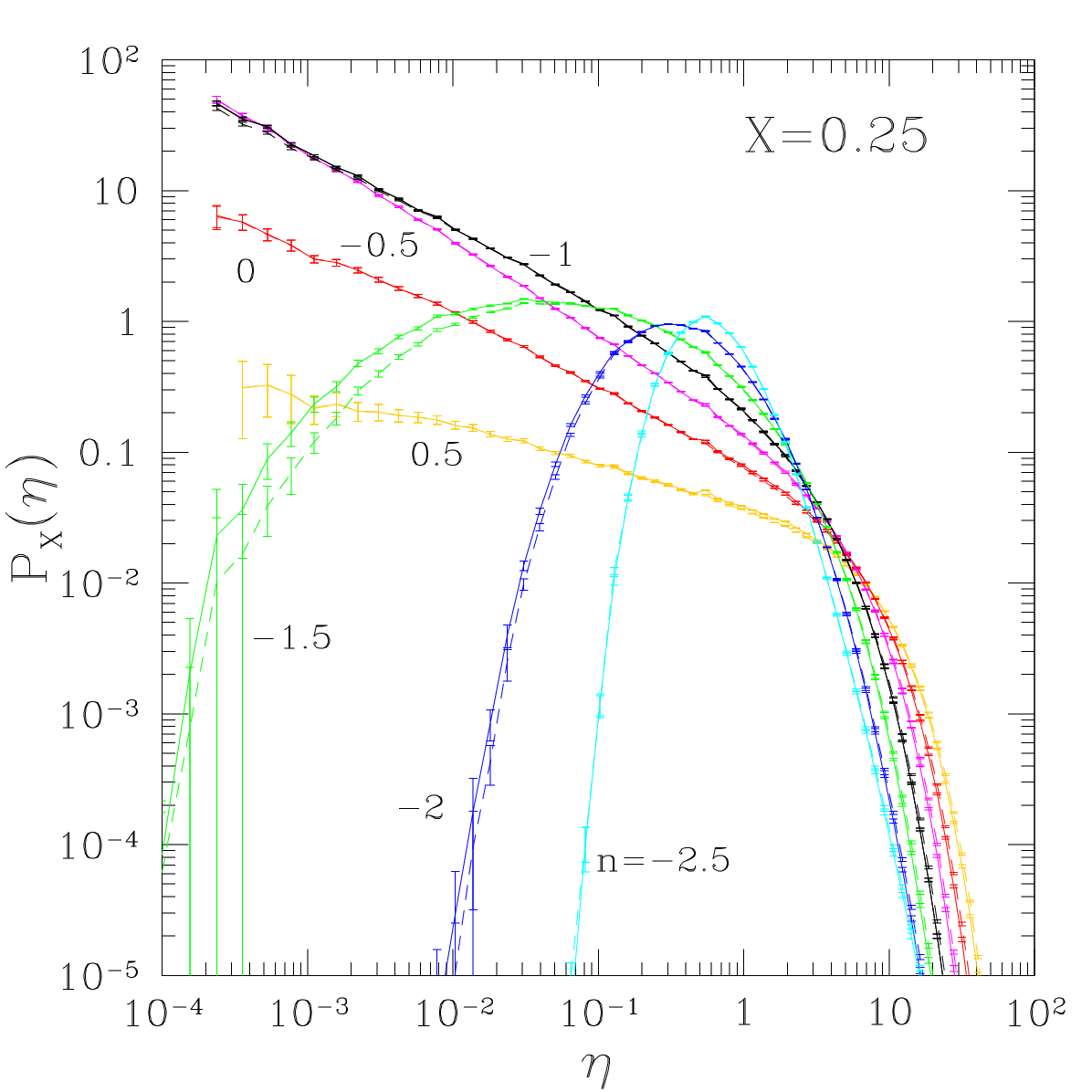}}
\end{center}
\caption{The probability distribution functions, $P_R(\eta)$ (solid lines) and
$P_X(\eta)$ (dashed lines), of the overdensity $\eta$ within spherical cells of radius
$R$ and within squares of size $X$. The radius $R$ is such that $\pi R^2=X^2$ (same
cell area), with $X$ given in each panel. For $-1<n<1$ there is an additional Dirac
contribution ($\propto \delta_D(\eta)$), associated with empty cells, which does not
appear in the figures.}
\label{figPxrho_2d}
\end{figure*}

We now consider the statistical properties of the smoothed density field.
In 2D we can study the probability distribution function, $P_R(\eta)$ or
$P_X(\eta)$, of the overdensity $\eta$ within circular cells of radius $r$ or
within squares of size $x$,
\beq
\eta= \frac{m}{\rho_0 \pi r^2} = \frac{M}{\pi R^2} \hspace{0.5cm} \mbox{or}
\hspace{0.5cm} \eta= \frac{m}{\rho_0 x^2}=\frac{M}{X^2} .
\label{eta-2d}
\eeq
We show both probability distributions in Fig.~\ref{figPxrho_2d} for cells
of the same area, that is, $\pi R^2= X^2$.
As expected, both distributions are close although we can distinguish modest
quantitative deviations, especially in the low-density tails.
We can see that we recover the qualitative features obtained in Fig.~\ref{figPxrho}
for the 1D case. For $-3<n<-1$, the probability distributions $P_R(\eta)$ and
$P_X(\eta)$ show both high-density and low-density exponential-like cutoffs,
whereas for $-1<n<1$ they show a low-density power-law tail.
Moreover, for $-1<n<1$ shock nodes are again isolated and in finite number, so that
there is an additional Dirac contribution of the form $P_R^0\delta_D(\eta)$ or
$P_X^0\delta_D(\eta)$ due to empty cells.

At large scales we recover for $-3<n\leq -1$ the Gaussian distribution associated
with the linear regime \cite{Valageas2009b},
\beqa
\lefteqn{-3 <n \leq -1 , \;\; R\rightarrow \infty : } \nonumber \\
&& \ln P_R(\eta) \sim 
- 2\left[\eta^{(n+1)/4}-\eta^{(n+3)/4}\right]^2/\sigma^2(R) \nonumber \\
&& \hspace{1.5cm} \sim - \frac{2R^{n+3}}{K_n} \, 
\left[\eta^{(n+1)/4}-\eta^{(n+3)/4}\right]^2 .
\label{PReta-QL-2d}
\eeqa
The asymptotic behavior (\ref{PReta-QL-2d}) holds for any finite $\eta$ if $-3<n\leq-2$,
and only above a low-density threshold $\eta_-$, with $0<\eta_-<1$, if $-2<n\leq -1$
(e.g., for $n=-1$ we have $\eta_-=1/4$).
Again, for $-3 <n \leq -1$ where typical density fluctuations are of order
$|\eta-1|\sim \sigma \propto R^{-(n+3)/2}$, we can expand the argument around
$\eta=1$ to recover the linear-regime Gaussian
\beqa
\lefteqn{-3 <n \leq -1 , \;\; R\rightarrow \infty , \;\; |\eta-1| \ll R^{-(n+3)/3} \; : }
\nonumber \\
&& \hspace{2cm} P_R(\eta) \sim e^{-(\eta-1)^2/(2\sigma^2(R))} .
\label{PReta-G}
\eeqa
For $-1<n<0$, where the linear variance (\ref{sigmadef}) is still finite, we expect
to recover the Gaussian (\ref{PReta-G}) at large scales, but the asymptotic behavior
(\ref{PReta-QL-2d}) no longer applies, since shocks modify the dependence on
$\eta$, whence the normalization of the cutoff as a function of $R$ for any
finite $\eta$.
For $0\leq n<1$, as for the cases $-1<n<1$ in 1D, where the linear variance
(\ref{sigmadef}) diverges, shocks play a key role at all scales and times and
the density probability distributions are always strongly non-Gaussian.

In agreement with Fig.~\ref{figPxrho_2d}, we can expect all these features to remain
valid for the probability distribution $P_X(\eta)$ within square cells, including
the exponents in (\ref{PReta-QL-2d}), but the numerical prefactors in
(\ref{PReta-QL-2d}) are modified.

At small scales, the density probability distributions are again governed by the
shock mass function, and the 1D scaling (\ref{PXeta-NL}) becomes
\beqa
\hspace{-0.2cm} -3<n<1, & & \hspace{-0.1cm} R\rightarrow 0 : 
\; P_R(\eta) \sim (\pi R^2)^2 \, N(\pi R^2\eta) , 
\label{PReta-NL-2d} \\
&& \hspace{-0.1cm} X\rightarrow 0 : \; P_X(\eta) \sim X^4 \, N(X^2\eta) .
\label{PXeta-NL-2d}
\eeqa
This implies in particular that the two distributions $P_R(\eta)$ and $P_X(\eta)$
coincide in the small-scale limit for equal-area cells, in agreement with
Fig.~\ref{figPxrho_2d}. However, at the scales shown in Fig.~\ref{figPxrho_2d}
this asymptotic regime has not been fully reached yet, hence we do not plot
the quantity $X^4 \, N(X^2\eta)$ of Eq.(\ref{PXeta-NL-2d}) to avoid overcrowding
the figure.

Using a saddle-point approach, the high-density tail of the probability distribution
$P_R(\eta)$ reads as
\beqa
\lefteqn{ -3 < n \leq -1 , \;\; \eta \rightarrow \infty : } \nonumber \\
&& \ln P_R(\eta) \sim - \frac{2\eta^{(n+3)/2}}{\sigma^2(R)} 
= - \frac{2R^{n+3}}{K_n} \, \eta^{(n+3)/2} ,
\label{PReta-large-eta-2d}
\eeqa
which also gives rise to the high-mass tail (\ref{NMp-sd-2D}) of the shock mass
function. For $-1<n<1$ shocks modify the asymptotic behavior, but they are expected
not to change the exponents.
Unfortunately, the range of our numerical computations is not sufficient to
check the tails (\ref{PReta-large-eta-2d}) to better than a factor 2, although they
are consistent with this scaling, $\ln P_R(\eta)\propto R^{n+3} \eta^{(n+3)/2}$.

\subsection{Density power spectrum}
\label{Power-spectra-2d}

\begin{figure}
\begin{center}
\epsfxsize=7 cm \epsfysize=5 cm {\epsfbox{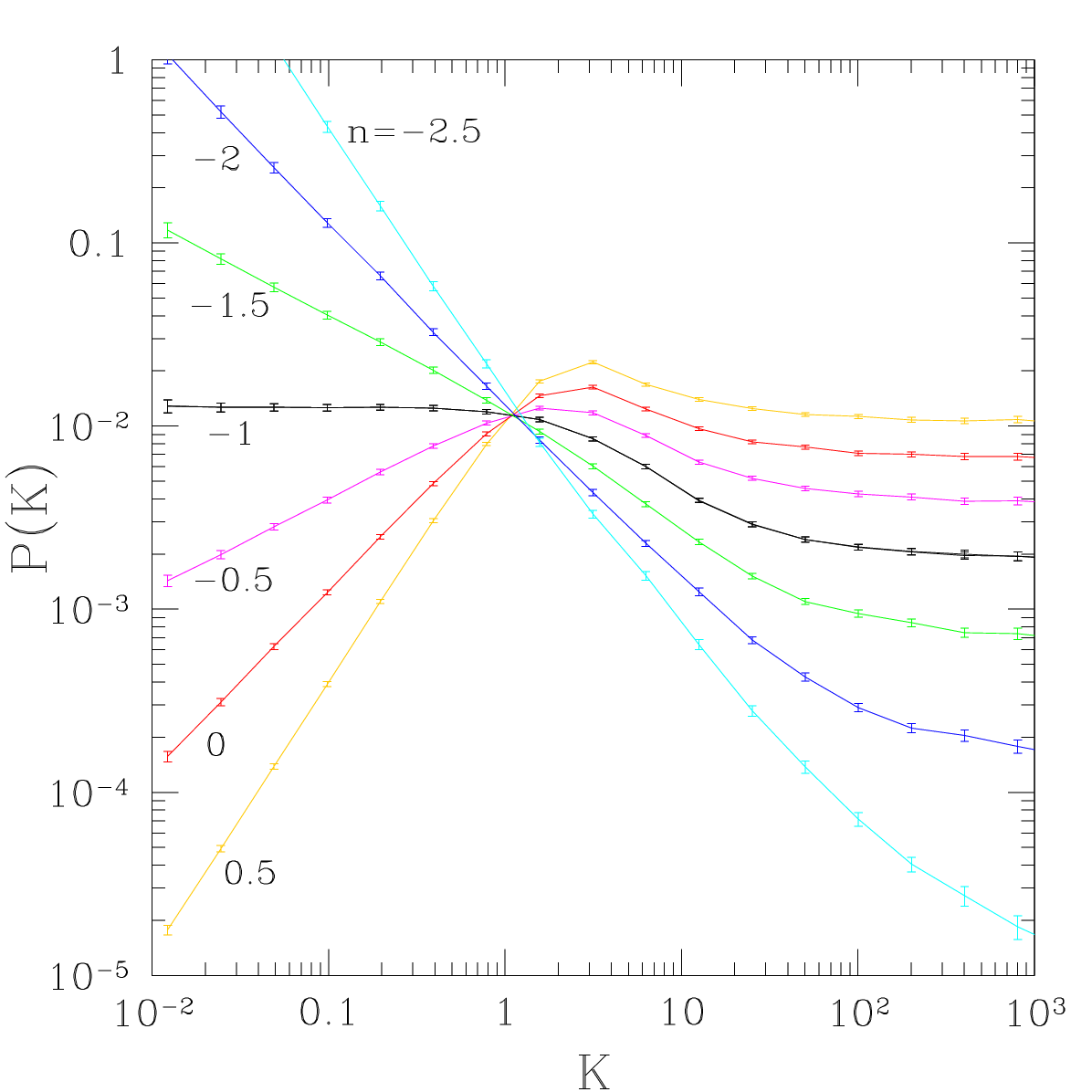}}
\end{center}
\caption{The density power spectrum $P(K)$.}
\label{figPk_2d}
\end{figure}

We show in Fig.~\ref{figPk_2d} the density power spectrum. At low $K$ we again
recover for all $-3<n<1$ the linear regime (\ref{PLE0}), whereas at high $K$ we
have the universal flat tail associated with shock nodes.
Indeed, for the power-law initial conditions (\ref{ndef}) that we consider in this
article, all the matter is located within pointlike shock nodes, which form
a Voronoi-like tessellation of the Eulerian space, while the boundaries of these
cells are massless shock lines and the interior of the cells is empty
\cite{BernardeauVal2010b}.
For $-1<n<1$, these Dirac density peaks appear to be isolated and in finite number
per unit volume, so that the constant high-$K$ tail is clearly seen in Fig.~\ref{figPk_2d}.
For $-3<n<-1$, shocks appear to be dense and there are no empty cells, in agreement
with the mass functions and the density distributions obtained in sections
\ref{Shock-mass-function-2D} and \ref{Density-distribution-2d}.
This makes the matter distribution closer to a continuous medium, so that the
constant high-$K$ tail is reached more slowly and at higher wavenumbers for lower $n$.

\section{Separable case in $d$ dimensions}
\label{Separable-text}

In dimensions two and higher there are no complete analytical results for the
statistical properties of the dynamics. However, it happens that the Burgers
dynamics, and the associated Geometrical Adhesion Model, exhibit exact
factorizable solutions in any dimension, for which we can obtain explicit results
(especially for the cases $n=0$ and $n=-2$).
This can be achieved for separable initial velocity potentials \cite{Gurbatov2003},
\beq
\psi_0(\vx) = \sum_{i=1}^d \psi_0^{(i)}(x_i) .
\label{psi0_sep}
\eeq
Then, this property remains true at all times, as can be seen at once from the
Hopf-Cole solution (\ref{Hopf1}), and each velocity component $u_i(\vx)$ only
depends on the coordinate $x_i$ along the same direction,
\beq
\psi(\vx,t) = \sum_{i=1}^d \psi^{(i)}(x_i,t) , \;\;\; u_i(\vx,t) = u^{(i)}(x_i,t) , 
\label{ux_sep}
\eeq
where the potentials $\psi^{(i)}(x,t)$ and the velocities $u^{(i)}(x,t)$ are the
solutions of the 1D Burgers dynamics defined by the initial conditions
$\psi_0^{(i)}(x)$. Thus, the dynamics is fully factorized into $d$ 1D Burgers dynamics.
In terms of the Legendre transforms, which fully determine the Eulerian and
Lagrangian fields as described in section~\ref{Burgers-dynamics},
this follows from the well-known property
\beq
f^*(\vs) = \sum_{i=1}^d f^*_i(s_i) \;\;\;  \mbox{for} \;\;\;
f(\vx) = \sum_{i=1}^d f_i(x_i) .
\label{Leg-d}
\eeq
This states that for any function $f(\vx)$ defined on $\mathbb{R}^d$ that is
separable (i.e., can be written as the second sum above) its Legendre transform
$f^*(\vs)$ is the sum of each 1D Legendre transform, as can be checked from the
definition (\ref{Legendredef}). This means that in $d$ dimension, if the initial
velocity potential is separable the Burgers dynamics can be fully factorized into
$d$ 1D Burgers dynamics.
This exact factorizability is specific to the Burgers dynamics, and it is not shared by
more complex dynamics such as the gravitational or Navier-Stockes dynamics.

As described in appendix \ref{Separable}, for such factorized initial conditions we
can obtain exact results for the shock mass function and the density probability
distributions. In particular, we obtain for the mass function the asymptotic
tails
\beq
M\rightarrow 0 : \;\; N(M) \sim \frac{(-\ln M)^{d-1}}{(d-1)!} \, M^{(n-1)/2} ,
\label{Md-Mm}
\eeq
and
\beq
M\rightarrow \infty : \;\; \ln N(M) \sim - M^{(n+3)/d} .
\label{Md-Mp}
\eeq
Therefore, we obtain the same asymptotic behaviors as those associated with the
isotropic 2D case studied in section~\ref{Shock-mass-function-2D}, but with a
logarithmic prefactor at low mass. This extends to any dimension $d$ for the
high-mass tail \cite{Valageas2009b}. 
Thus, keeping Gaussian initial conditions with the scaling (\ref{sigma-2d})
preserves the characteristic exponents of the shock mass function, even though
the isotropy of the system has been broken.
This is not surprising for the high-mass tail, which can be derived from a simple
saddle-point approach and as such mostly depends on the scaling (\ref{sigma-2d})
and the fact that the initial (linear) density field is Gaussian
\cite{Valageas2009b}.
The robustness of the low-mass power-law exponent is not so obvious a priori,
since it has not been derived in such a systematic fashion. From the analysis of
numerical computations, the scaling (\ref{NMm}) was advocated using simple
arguments that basically assume that the properties of the 2D and 3D convex hulls
are similar \cite{Vergassola1994}, that is, governed by the scaling (\ref{sigma-2d}). 
However, this also corresponds to assuming that the separable case studied
in this section and the isotropic case of section~\ref{Two-dimensions}
give the same low-mass exponents, which is not obvious.

\begin{figure}
\begin{center}
\epsfxsize=7 cm \epsfysize=5 cm {\epsfbox{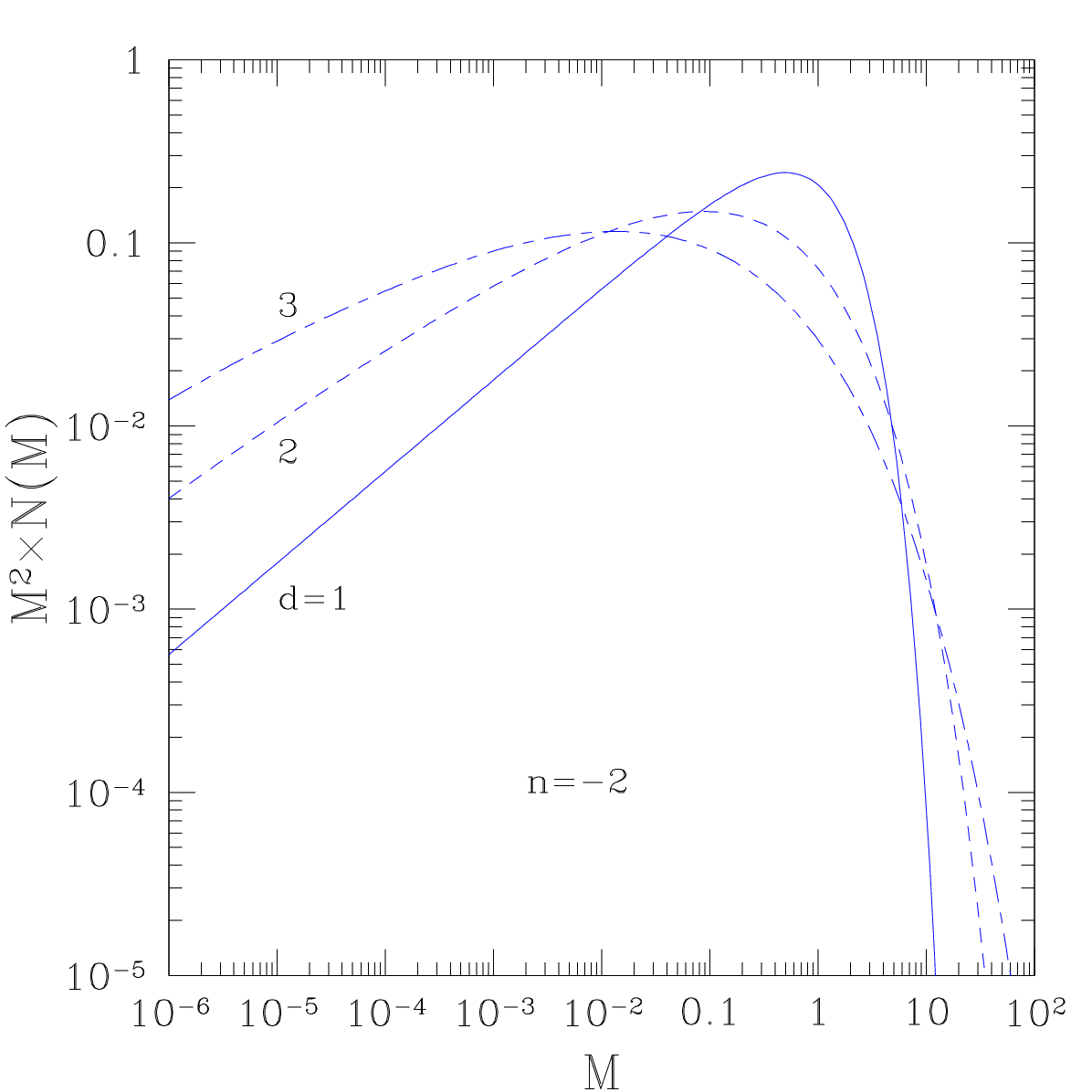}}
\end{center}
\caption{The product $M^2\times N(M)$, where $N(M)$ is the shock mass function in
the separable case for $n=-2$ and dimensions $d=1$ (solid line), $d=2$ (dashed line),
and $d=3$ (dot-dashed line).}
\label{figm2Nm_n-2}
\end{figure}

As pointed out in Sec.~\ref{Shock-mass-function-2D}, we can note that the
asymptotic tail (\ref{Md-Mm}) actually corresponds to a low-$\nu$ power-law tail
$f(\nu) \propto \nu^d$, in terms of the reduced variable $\nu$,
with no further dependence on $n$.
The large-mass tail (\ref{Md-Mp}) also corresponds to the usual falloff
$f(\nu)\sim e^{-\nu^2/2}$ at large $\nu$.
Therefore, the Press-Schechter scaling remains valid in any dimension,
at leading order for these separable cases, even though the low-$\nu$ exponent 
strongly depends on the dimension.
On the other hand, while the mass function appears more sharply peaked
at higher $d$ as a function of $\nu$, as seen in Fig.~\ref{figmNm_2d},
it flattens when it is drawn as a function of mass, as seen in
Fig.~\ref{figm2Nm_n-2} below for $n=-2$.

The same analysis can be applied to the probability distributions of the smoothed
density field, and we again recover the characteristic exponents
(\ref{PReta-large-eta-2d}) of the isotropic case.

For the index $n=-2$ we can obtain explicit expressions, which simplify in
2D as
\beqa
\lefteqn{\hspace{-0.6cm} n=-2, \;\; d=2 : \;\;\; N(M) = \frac{2}{\pi} \, M^{-3/2} \;
K_0\left(2\sqrt{M}\right) , } \label{NM-n-2-sep} \\
&& \hspace{-0.6cm} P_X(\eta) = \frac{2X}{\pi} \, e^{4X} \, \eta^{-3/2} \,
K_0\left(2X\sqrt{2+\eta+\frac{1}{\eta}}\right) .
\label{PXn-2-sep}
\eeqa
We show our results in Figs.~\ref{figm2Nm_n-2} 
for the shock mass function $N(M)$
in dimensions $d=1,2$ and $3$ and for the index $n=-2$.
In order to emphasize the low-mass power-law tails we plot the product
$M^2 N(M)$ in Fig.~\ref{figm2Nm_n-2}. 
In agreement with Eqs.(\ref{Md-Mm}) and (\ref{Md-Mp}), for higher $d$ the mass
function shows a smoother cutoff at high mass and a somewhat faster growth a low
mass due to logarithmic prefactors. This gives more weight to extreme events,
as is usually the case for multiplicative processes (since $M=\prod_i M_i$ the
shock mass can also be seen as the outcome of such a multiplicative process).
This also leads to a broadening of the peak of the product $M^2 N(M)$, which
gives the fraction of matter within shocks of mass $M$ per logarithmic interval
of mass. We can note that such a flattening can also be seen in the isotropic
case, by comparing Fig.~\ref{figmNm_2d} with Fig.~\ref{figmNm}. 
Similar results are obtained for the density distribution $P_X(\eta)$.

We can note that the separable solutions studied in this section might serve as a
basis for approximation schemes or perturbative expansions to describe the isotropic
case studied in Section~\ref{Two-dimensions}, however we shall not investigate
further this point in this article.

\section{Conclusion}
\label{Conclusion}

In this article we have presented a numerical analysis of density fields and
mass functions that can be generated by the Burgers dynamics in the inviscid limit, the ``adhesion model'' in cosmology,
when it is supplemented by a geometrical construction that explicitly defines the density field
in the shock manifolds. This leads to what we call the ``Geometrical Adhesion Model'' (GAM)
for the density field. 
Our analysis focused  on power-law Gaussian initial conditions, which are relevant within
the cosmological context, and we have considered the 1D and 2D cases.

We furthermore have taken advantage of new more efficient algorithms, which also make
use of the geometrical interpretation of the system, to measure mass functions
and density distributions over a large range of masses and scales. Our simulations
cover seven values of the slope $n$ of the initial density and velocity power spectra,
that span the range $-3<n<1$ where a self-similar dynamics develops.

In the 1D case, we have checked that our numerical results agree with the
complete analytical results that are known for the two cases $n=-2$ and
$n=0$. For general index $n$ we also obtain a good agreement with
the analytical results that apply to the tails of the mass function and of the
density probability distributions, and to the low-order moments of the
density contrast in the quasi-linear regime.
In particular, this confirms the validity of rare-event tails obtained by
steepest-descent methods.
Regarding the mass functions, we found that in 1D, they could be described with a good accuracy
with the reduced variable, $\nu=\delta_c/\sigma(M)$, although
there remains a small dependence with $n$. This is the basis for Press-Schechter like 
constructions commonly used in the cosmological context.
It also happens that the Press-Schechter prescription {\it per se} (e.g., derived from 
the 1D spherical collapse) provides a good
approximation for this scaling function $f(\nu)$ (and it actually gives the
exact mass function for $n=-2$).
The density probability distributions show the expected behaviors,
with a low-density tail for the ``UV'' class $-1<n<1$ and a sharp low-density
cutoff for the ``IR'' class $-3<n<-1$. Moreover, at small scales we have checked that
the density probability distributions reach their asymptotic form determined
by the shock  mass function.
For the density power spectrum we recover the universal constant
high-$k$ tail associated with shocks, which corresponds to pointlike masses
in the density field (and discontinuities in the velocity field).

In the 2D case we have performed a similar analysis, although the smaller
range of masses and scales does not allow to probe with a high accuracy the
rare-event tails. Nevertheless, our results are consistent with previous works
for the low-mass tails of the shock-node mass functions. We find that 
the scaling in terms of $\nu$ still captures most of the dependence on
$n$ of the mass functions, but deviations from this scaling law are slightly
larger than in 1D. Moreover, the scaling function is clearly different and it shows
a $\nu^2$-tail at low $\nu$ rather than the linear tail obtained in 1D.
In this regime, as noticed in previous works, the Press-Schechter prescription is no longer
a good approximation. The low-density tails of the density probability distributions
show the same behavior as in 1D, with again a qualitative difference between
the ``UV'' and ``IR'' classes. This is related to the low-mass exponents of the
mass functions, which are the same in 1D and 2D (in terms of $M$).
The density power spectra again reach the universal constant tail at high $k$
due to the formation of shock nodes, that is pointlike masses.

Finally, we have described how the mass functions and density probability
distributions can be obtained in any dimensions for separable initial conditions,
where the dynamics factorizes over $d$ one-dimensional dynamics.
%In particular, we give explicit analytical results for the case $n=-2$.

As compared with the collisionless gravitational dynamics,
the nonlinear behavior of this system thus appears as a whole much simpler to analyze as many
statistical properties can be derived from the mere fact that structures are all
pointlike objects. As we have just seen, 
this leads to the universal flat tail for $P(K)$ at high $K$. It also leads to
constant ratios $S_p$, defined in Eq.(\ref{Spdef-delta}), in the small-scale limit.
In contrast, in the gravitational case relevant for cosmology (or in the
Navier-Stokes dynamics relevant for hydrodynamics) characteristic structures 
are much more complex. Dedicated numerical simulations show the formation of
extended halos with nontrivial mean density profiles and some amount of
substructure \cite{Moore1999,Klypin1999,Gao2004,Navarro2010}.
This has prevented so far the derivation of simple universal laws
for the exponents associated with the density power spectrum and higher-order
correlations.

Despite these differences for the physical processes that take place at small scales,
we find that the matter distribution generated by the Burgers
dynamics, through the Geometrical Adhesion Model studied here, shares many
statistical properties with the one built by gravitational clustering in the
cosmological context. Moreover, this remains valid at small scales for several
quantities, such as the mass function and the probability distributions of the
smoothed density field.
We argue then that this system, 
because of the existence of an explicit geometrical solution that
can easily be implemented, provides a good tool for understanding 
the nonlinear processes that are common to both systems.
One example of this is to be found
in \cite{BernardeauVal2010a} where we explored the behavior of response functions (propagators)
within both the Eulerian and Lagrangian frameworks.
Another line of investigation which remains to be explored 
is the use the Burgers dynamics as the basis of new approximation schemes, for
instance through perturbative methods, for the 3D gravitational dynamics itself.

\appendix

\section{Algorithms for the 1D Burgers dynamics}
\label{Algorithms-for-the-1D-Burgers-dynamics}

\subsection{Set up of the initial conditions}
\label{algorithm-1d}

The system is discretized on a regular grid of $N$ points with a unit step,
$x_i= i$ with $i=0,..,N-1$, with periodic boundary conditions.
%In other words, the system extends on the whole real line and it is periodic
%of period $N$ (this symmetry being conserved by the dynamics).
As a consequence, the analysis is restricted to scales $x$ such that $1\ll x \ll N$,
and times $t$ such that $L(t)\ll N$ to avoid finite-size effects.
In order to simplify numerical computations (e.g., for Fast Fourier Transforms)
we choose $N$ to be a power of $2$, typically $N=2^{23}=8 \, 388 \, 608$.
We can also take $\rho_0=1$ so that each initial ``particle'' $i$ (i.e. initial grid
point) carries a unit mass.

To implement the initial conditions (\ref{ndef}) we define the rescaled coordinates
\beq
\hx= \frac{2\pi}{N} x , \;\;\; \hk= \frac{N}{2\pi} k ,
\label{hx}
\eeq
and we use the discrete Fourier transform
\beq
u_0(x) = \hu_0(\hx) = \sum_{\hk=-N/2+1}^{N/2-1} \thu_{0,\hk} \, e^{\ii \hk \hx} ,
\label{hu0}
\eeq
where the random complex Fourier coefficients $\thu_{0,\hk}$ are independent
Gaussian variables (except for $\thu_{0,-\hk}=\thu_{0,\hk}^*$) with a variance
\beq
\lag | \thu_{0,\hk} |^2 \rag = \frac{D}{2\pi} \, \left(\frac{2\pi}{N}\right)^{n+1} \, \hk^n .
\label{varthuk}
\eeq
We simultaneously obtain the initial potential $\psi_0(x)$, using
$\ut_0(k)=-\ii k \psit_0(k)$ from Eq.(\ref{thetadef}) (and taking $\ut_{0,k}=0$ and
$\psit(k)=0$ for $k=0$). Of course, with this prescription both the initial velocity
field $u_0(x)$ and potential $\psi_0(x)$ are homogeneous, for all $n$
(the discretization has introduced both UV and IR cutoffs, for $n=-1$ 
see \footnote{When 
we consider the case $n=-1$, which marks the transition between
the ``IR'' and ``UV'' classes we actually compute the two cases
$n=-1.001$ and $n=-0.999$ so as to fall within either category, and we check that for
quantities that remain finite for both ``IR'' and ``UV'' classes both results agree
 (up to the numerical resolution).
In order to take advantage of the accumulated statistics however (i.e. to reduce statistical error
bars) we display the mean of both results as the $"n=-1"$ curve.}).

\subsection{Computation of the 1D velocity and density fields}
\label{computation-1D}

For any time $t$, the velocity field $u(x,t)$ and its potential $\psi(x,t)$ are obtained
from the Hopf-Cole solution (\ref{Hxphiq}) using the algorithm described in
Sec.\ref{1D-Legende-2D-convex-hull} below.

% 
%Thus, the great simplification of the
%Burgers dynamics, as compared with other systems such as Navier-Stokes or
%gravitational dynamics, is that one does not need to solve the equation of motion
%forward in time, since one obtains directly the solution
%at the time of interest from Eq.(\ref{Hxphiq}).
%We describe in appendix~\ref{Algorithms-for-the-1D-Burgers-dynamics} the algorithm
%devised in \cite{Lucet1997} that we use to compute the Legendre transform
%(\ref{Hxphiq}).
%It takes advantage of the fact that we are given $\varphi_L(q)$ over an ordered grid,
%$q_j<q_{j+1}$, to achieve an optimal running time that scales as $O(N)$.
%It proceeds in two steps, first one computes the convex hull $\varphi$ with
%an incremental method, see appendix~\ref{2D-convex-hull}, and second one derives
%from $\varphi(q)$ the Legendre transform $H(x)$, see
%appendix~\ref{1D-Legendre-transform}.

For the 1D case, we do not need
to use the second Legendre transform (\ref{varphicdef}) or (\ref{qx-xq}) to
compute the Lagrangian map $x(q)$ since, as both functions $x(q)$ and $q(x)$ are
monotonically increasing,
$x(q)$ can be obtained simply by spanning $q(x)$.
As a consequence, our algorithm to obtain the velocity and density fields has an
optimal running time that scales as $O(N)$. By contrast, previous works
\cite{Noullez1994,Vergassola1994} used a slower $O(N\ln N)$
algorithm.

\subsection{Computation of a 1D Legendre transform by building
a 2D convex hull}
\label{1D-Legende-2D-convex-hull}

To compute the Legendre transform (\ref{Hxphiq}) we use the algorithm devised
in \cite{Lucet1997},
which scales linearly with $N$, taking advantage of the fact that we are given
$\varphi_L(q)$ over an ordered grid, $q_j<q_{j+1}$.
Thus, we first compute the convex envelope $\varphi$ of the linear Lagrangian
potential $\varphi_L$. Then, we obtain $H(x)$ as 
$H(x)=\cL_x[\varphi_L(q)]=\cL_x[\varphi(q)]$, using the property that the $\varphi_L$
and its convex envelope $\varphi$ have the same Legendre transform.
Moreover, thanks to the periodicity of $\psi_0(q)$
the Lagrangian coordinate, $q(x=0)$, of the particle that is located at the origin
at the time $t$ of interest obeys $-N/2\leq q \leq N/2-1$. Then, since particles do not
cross each other, so that $q(x)$ is monotonically increasing, to construct
$H(x)$ over the grid $x_i=0,1,..,N-1$ we simply need to span $\varphi_L(q)$ over
the set of points $\{q(0), q(0)+1,.., q(0)+N-1\}$.

\subsubsection{2D convex hull}
\label{2D-convex-hull}

\begin{figure}
\begin{center}
\epsfxsize=7 cm \epsfysize=4.5 cm {\epsfbox{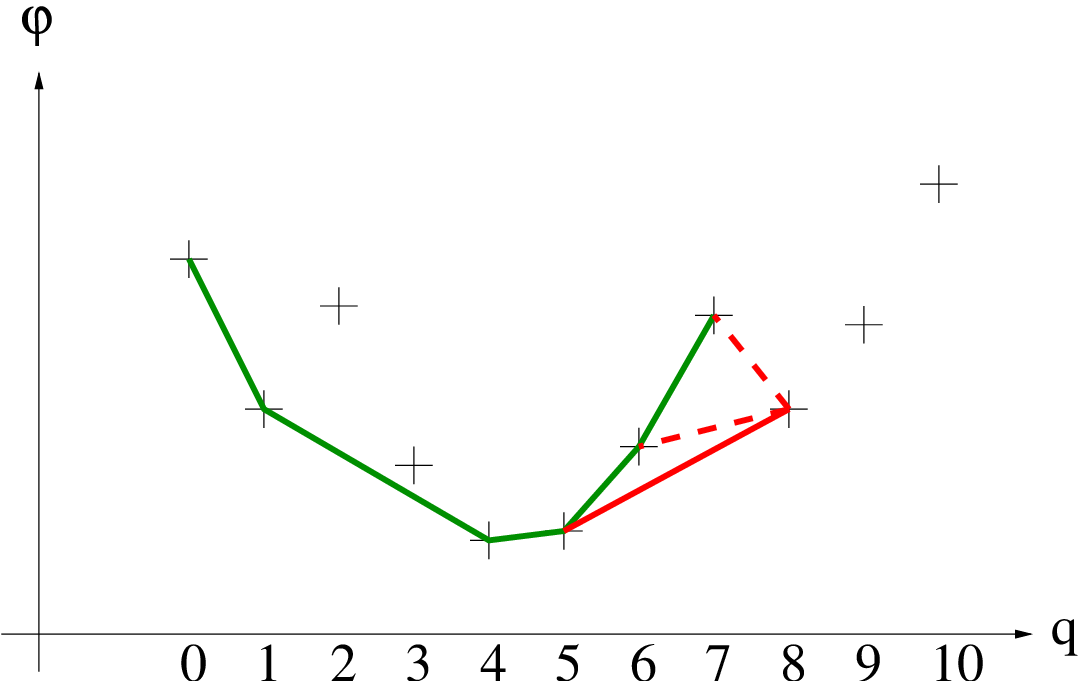}}
\end{center}
\caption{Construction of the convex envelope $\varphi(q)$ from the linear
potential $\varphi_L(q)$ given on a regular grid. Moving to the right in the
$(q,\varphi)$ plane we update the step-$n$ 2D convex hull $\varphi^{(n)}(q)$
as we add a new data point.}
\label{figconv1D}
\end{figure}

We first obtain the convex hull $\varphi$ through the following sequential
procedure.
Let us assume that at step $(n)$, with $n\geq 2$, we have built the convex hull
$\varphi^{(n)}(q)$ of $\varphi_L(q)$ over the first $n$ points of this set
$\{q(0), .., q(0)+n-1\}$. At this stage, $\varphi^{(n)}(q)$ is made of $p$ points
with $2\leq p \leq n$ (because of the discretization both $\varphi_L$ and its convex
hull $\varphi$ are defined by a finite number of points).
Moving to the next step $(n+1)$, we add the next point
$(q(0)+n,\varphi_L[q(0)+n])$ and going backward we remove if necessary the points
$p, p-1,.., p'+1$ of the previous convex hull $\varphi^{(n)}$ until its last two
vertices, $p'-1$ and $p'$, and the new point $n+1$ turn counterclockwise in the
$(q,\varphi_L)$ plane. This yields the new convex hull $\varphi^{(n+1)}$.
Iterating from $n=2$ up to $N$ we
obtain the $p$ vertices of the convex hull $\varphi$.

This algorithm is shown in Fig.~\ref{figconv1D} at step $(9)$. We have already built
the convex hull associated with the $8$ points $\{0,1,..,7\}$ and we are adding
the point $8$. Moving backward we see that we must remove the vertices $7$ and
$6$ and the new convex hull $\varphi^{(9)}$ is made of the list $\{0,1,4,5,8\}$.

\subsubsection{1D Legendre transform}
\label{1D-Legendre-transform}

\begin{figure}
\begin{center}
\epsfxsize=7 cm \epsfysize=4. cm {\epsfbox{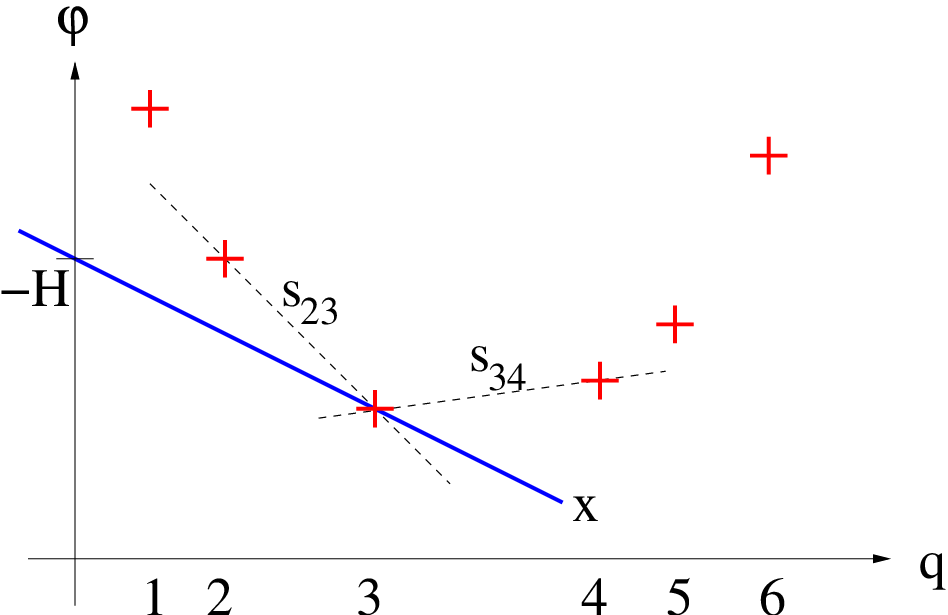}}
\end{center}
\caption{Computation of the Legendre transform $H(x)$ from the convex
envelope $\varphi(q)$. All lines of slope $x$ in-between the slopes $s_{23}$ and
$s_{34}$ of segments $(23)$ and $(34)$ in the $(q,\varphi)$ plane make first
contact from below with the vertex $3$. Hence $H(x)=x q_3-\varphi_3$ for
$s_{23}<x<s_{34}$.}
\label{figLeg1D}
\end{figure}

Second, as in \cite{Lucet1997},
spanning the vertices $j=1,..,p$, and computing the slope $s_{j,j+1}$ associated
with the segment $[j,j+1]$ of $\varphi(q)$, we note that all $x$ such that
$s_{j-1,j}<x<s_{j,j+1}$ have the Lagrangian coordinate $q(x)=q_j$, which also yields
$H(x)=x q_j-\varphi(q_j)$. Thus, by reading the discrete slopes $s_{j,j+1}$ of the
piecewise affine convex hull $\varphi(q)$ from left to right (whence $s_{j,j+1}$
is monotonically increasing since $\varphi$ is convex), we gradually ``fill in'' the
values $H(x_i)$ on the grid $x_i=i$ with $i=0,..,N$, in order of increasing $i$.
Clearly, both steps (computing the convex hull $\varphi$ and next the Legendre
transform $H$) scale linearly with the number
of points on the grid and are thus optimal \cite{Lucet1997}.
We illustrate in Fig.~\ref{figLeg1D} this second step, computing $H(x)$ on the grid
from the vertices of $\varphi(q)$.

\section{Results for the 1D case}%{One-dimensional dynamics}
\label{One-dimension}

We present here the results obtained for the 1D case, using the algorithm presented in
the previous appendix \ref{Algorithms-for-the-1D-Burgers-dynamics}
and for initial conditions as described in the text.

\subsection{Distribution of shocks}
\label{Distribution-of-shocks}

\begin{figure}
\begin{center}
\epsfxsize=7 cm \epsfysize=5 cm {\epsfbox{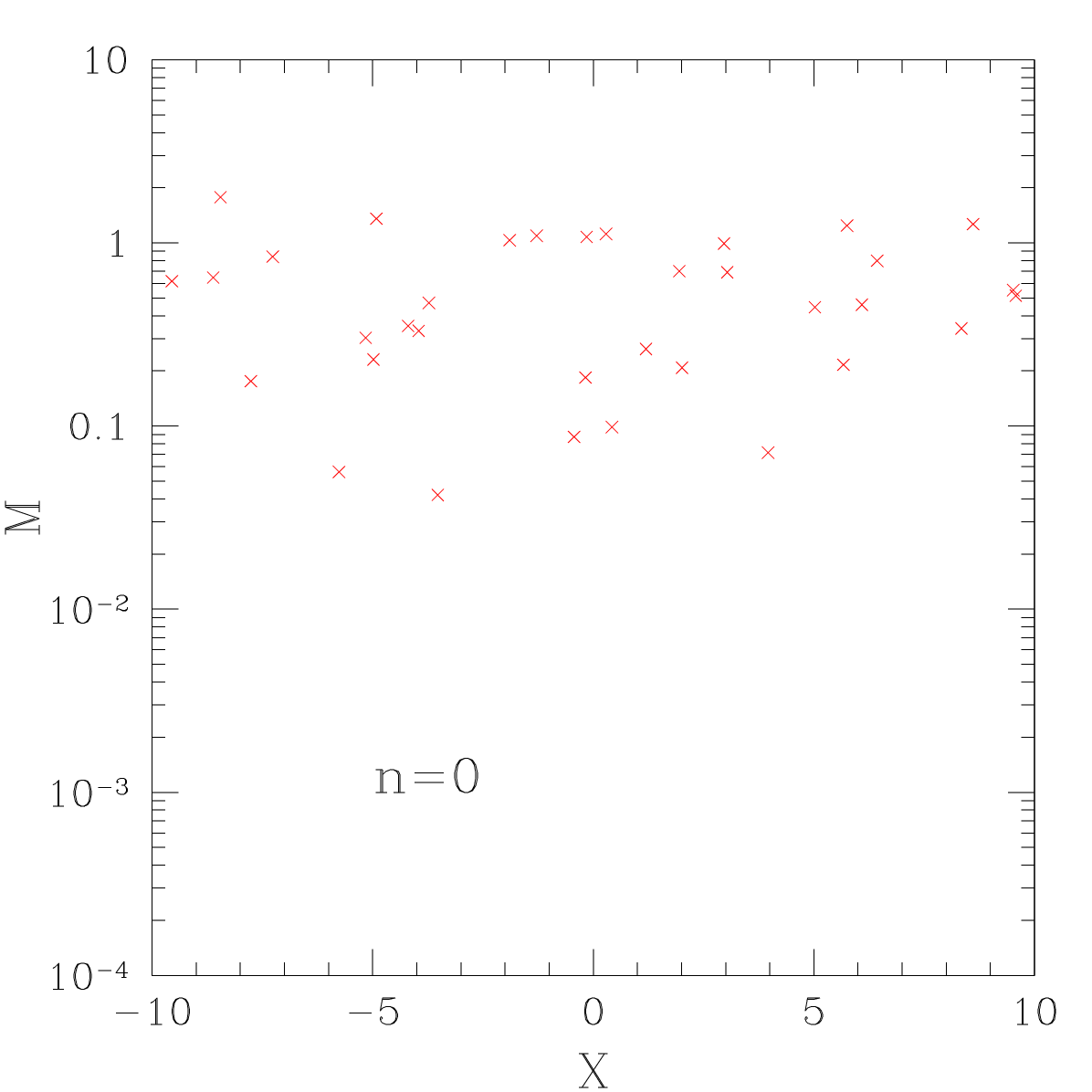}}
\epsfxsize=7 cm \epsfysize=5 cm {\epsfbox{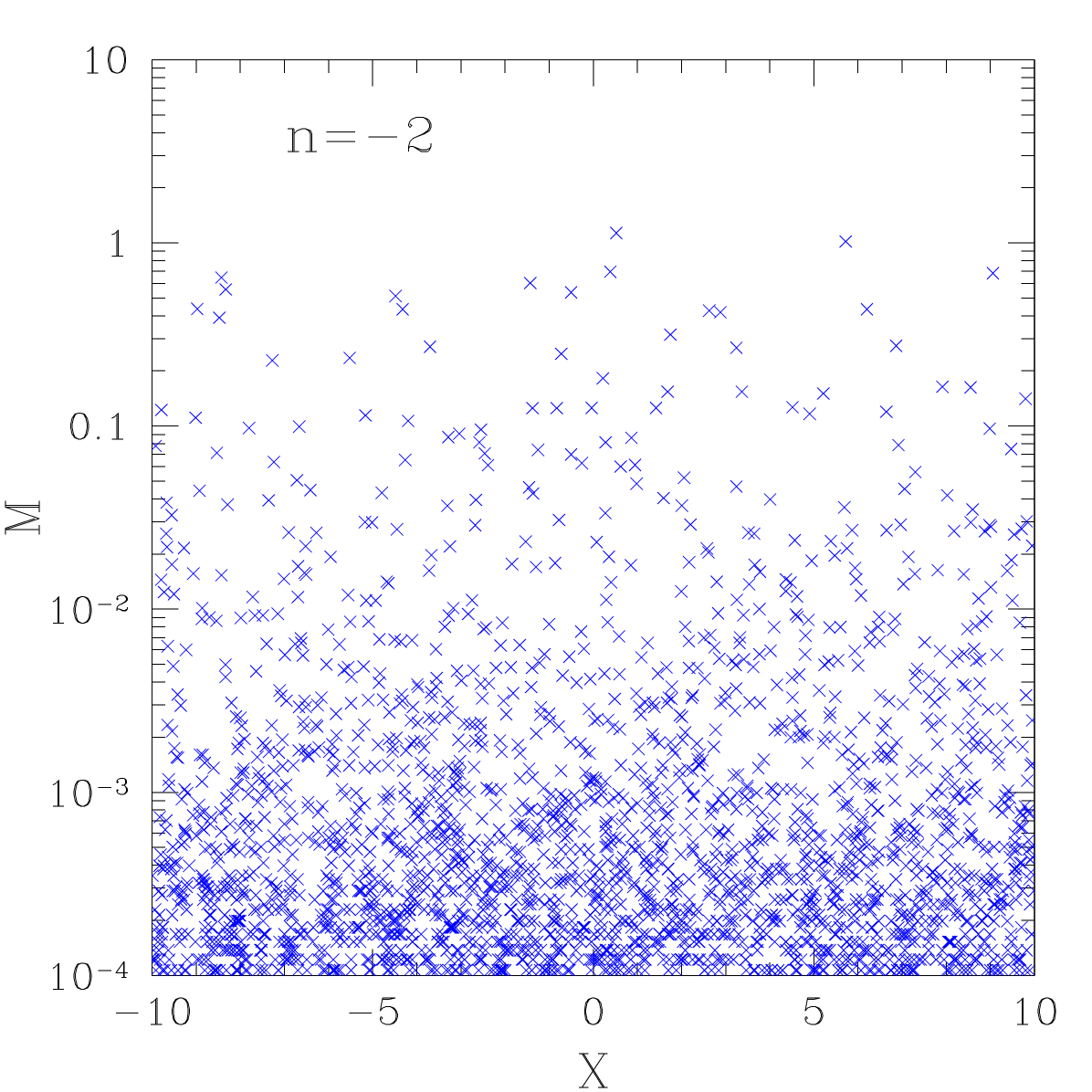}}
\end{center}
\caption{The distribution of shocks in the position-mass plane. Each cross corresponds
to a shock, observed at a given time for one realization of the Gaussian initial
conditions (\ref{ndef}) with $n=0$ (upper panel) and $n=-2$ (lower panel).
The position $X$ and the mass $M$ are the scaling variables (\ref{QXU}).}
\label{figldensity}
\end{figure}

We show in Fig.~\ref{figldensity} the resulting distribution of shocks in the position-mass plane
that we obtain for one realization of the Gaussian initial conditions (\ref{ndef}) at a
given time. Since we use the scaling variables (\ref{QXU}) the statistical properties
of the output do not depend on this time. In particular, the typical masses (at the
onset of the exponential cutoff of the mass function) and their typical length-scale
(e.g. the nearest-neighbor distance) are of order unity.
In agreement with previous works \cite{She1992,Vergassola1994},
we can see that for $n=0$, which is representative of the ``UV'' class, $-1<n<1$, we
obtain a finite number of shocks per unit length with very few high and low-mass
objects. In contrast, for $n=-2$, which is representative of the ``IR'' class, $-3<n<-1$,
we observe a proliferation of small shocks which appear to fill all of space (up to the
resolution of the simulation). This agrees with theoretical results, which show that
shocks are isolated and in finite number per unit length for $n=0$ 
\cite{AvellanedaE1995,Frachebourg2000,Valageas2009c},
whereas they are dense in Eulerian space for $n=-2$ 
\cite{Sinai1992,Bertoin1998,Valageas2009a}.
We obtain similar figures for other indices $n$ in both characteristic classes, $-1<n<1$
and $-3<n<-1$.

\subsection{Shock mass function}
\label{Shock-mass-function}

By averaging over many realizations, and over several output times for each realization 
(thanks to the self-similarity of the dynamics), we can measure the shock mass function
$N(M)\dd M$, defined as the mean number of shocks of mass within $[M,M+\dd M[$
over a unit-length interval. We have shown our results for several values of the index $n$
in Fig.~\ref{figmNm} (to avoid having a huge vertical range we actually plot the
product $M\times N(M)$). We can clearly see the power-law regime at low mass and
the exponential cutoff at high mass, with a strong dependence on $n$. We can check
that our numerical results agree with the exact analytical results that have been
obtained for both cases $n=0$ 
\cite{Frachebourg2000,Valageas2009c},
\beqa
n=0 : \;\; N(M) & = & 2M \inta \frac{\dd s_1}{2\pi\ii} \, \frac{e^{-s_1M}}{\Ai(s_1)^2}
\nonumber \\
&& \times \inta \frac{\dd s_2}{2\pi\ii} \, \frac{e^{s_2M}\Ai'(s_2)}{\Ai(s_2)} ,
\label{NMn0}
\eeqa
and $n=-2$ \cite{Bertoin1998,Valageas2009a},
\beq 
n=-2 : \;\; N(M) = \frac{1}{\sqrt{\pi}} \, M^{-3/2} \, e^{-M} .
\label{NMn-2}
\eeq
In agreement with Fig.~\ref{figldensity}, the shock mass function grows more slowly
than $1/M$ at low mass for $-1<n<1$, which leads to a finite number of shocks per
unit length, whereas it grows faster than $1/M$ for $-3<n<-1$, which leads to an
infinite number of shocks per unit length because of a divergent number of small
shocks (while the total mass remains unity).

We have checked in Fig.~\ref{figlNmp} that the high-mass tail of the shock mass
function agrees with the analytical predictions (\ref{NMp})-(\ref{NMp-n0}).
For $n=-2$ we can also check that our numerical
result agrees reasonably well with the exact derivative obtained from Eq.(\ref{NMn-2}).
For $n=-1.5$, using the normalization given in Eq.(\ref{NMp-sd})
(i.e. $I_{-1.5}$) appears to provide a reasonable approximation to the high-mass
asymptote. This means that for $-2<n\leq -1.5$ shocks have not significantly
modified the quantitative profile of the saddle point.
It is interesting to note that the rate of convergence to the asymptotic regime
(\ref{NMp}) decreases with $n$. This is also due to the fact that the exponential
cutoff is smoother for lower $n$, in agreement with the exponent (\ref{NMp})
and Fig.~\ref{figmNm}, so that the rare-event limit associated with these
asymptotic behaviors is reached at higher masses for lower $n$.
Note that the deviations from the asymptotic behavior (\ref{NMp}) are magnified
in Fig.~\ref{figlNmp} and would appear much smaller in Fig.~\ref{figmNm} as the
exponential falloff is already very steep over this mass range and one would not
distinguish in this figure the subdominant effect of power-law prefactors.

\begin{figure}
\begin{center}
\epsfxsize=7 cm \epsfysize=5 cm {\epsfbox{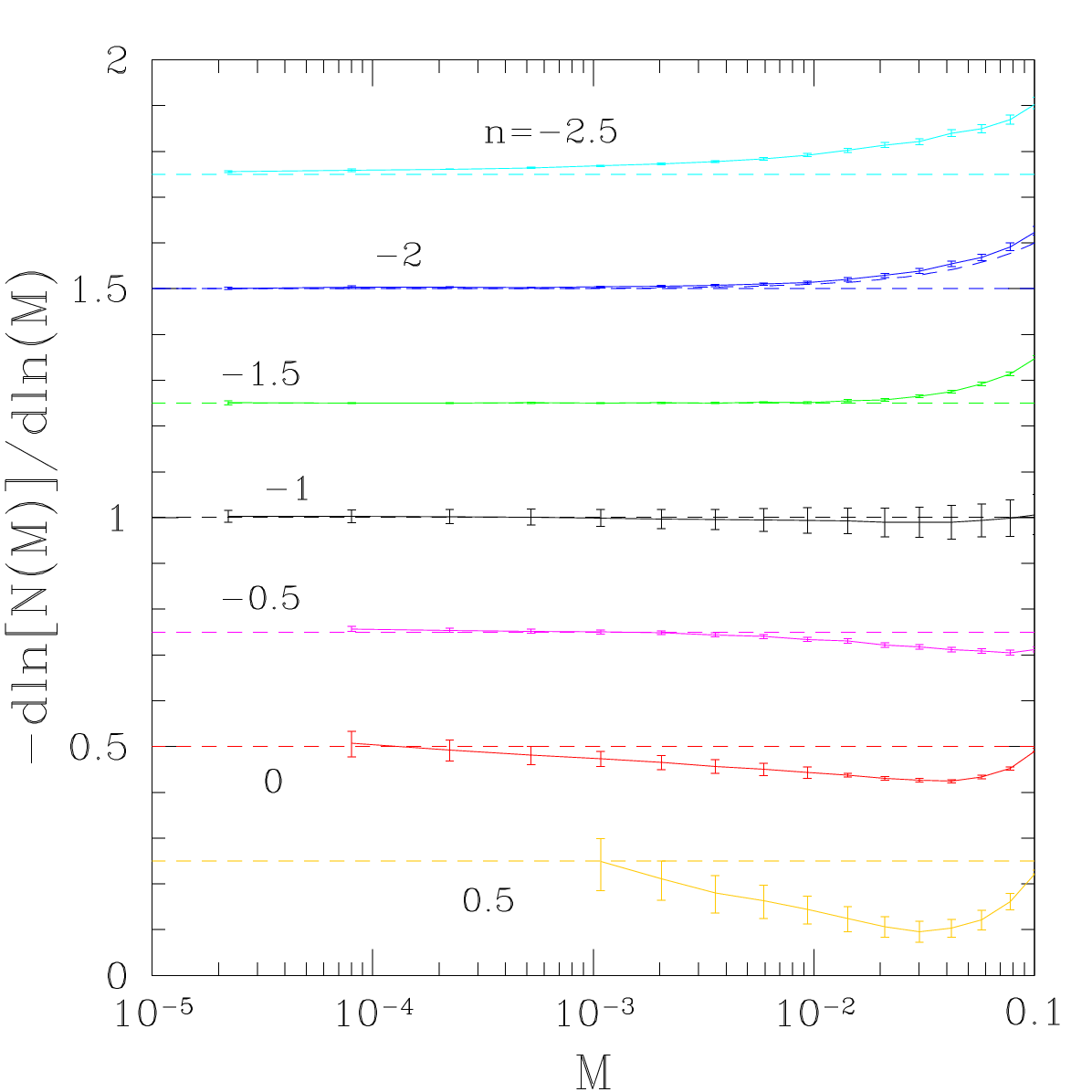}}
\end{center}
\caption{The derivative $-\dd\ln[N(M)]/\dd\ln(M)$ at low mass. The horizontal
dashed lines show the asymptotic behavior (\ref{NMm}). For $n=-2$ the curved
dashed line is the exact derivative obtained from Eq.(\ref{NMn-2}).}
\label{figlNmm}
\end{figure}

At low mass, previous numerical simulations and heuristic arguments
\cite{She1992,Vergassola1994} suggest the power-law tail
\beq
-3<n<1 , \;\; M \rightarrow 0 : \;\; N(M) \sim M^{(n-1)/2} ,
\label{NMm}
\eeq
which has only been proved rigorously for the white-noise case $n=0$
\cite{Frachebourg2000,Valageas2009c} and the Brownian case $n=-2$
\cite{Bertoin1998,Valageas2009a}.
As seen in Fig.~\ref{figlNmm}, where we plot the derivative $-\dd\ln[N(M)]/\dd\ln(M)$,
our numerical results agree with the scalings (\ref{NMm}), and for $n=-2$ with
the full result (\ref{NMn-2}). Contrary to the high-mass tail, the rate of convergence
to this asymptotic behavior is roughly the same for all $n$ over $-3<n<1$.

\subsection{Density distribution}

\begin{figure*}
\begin{center}
\epsfxsize=7 cm \epsfysize=5 cm {\epsfbox{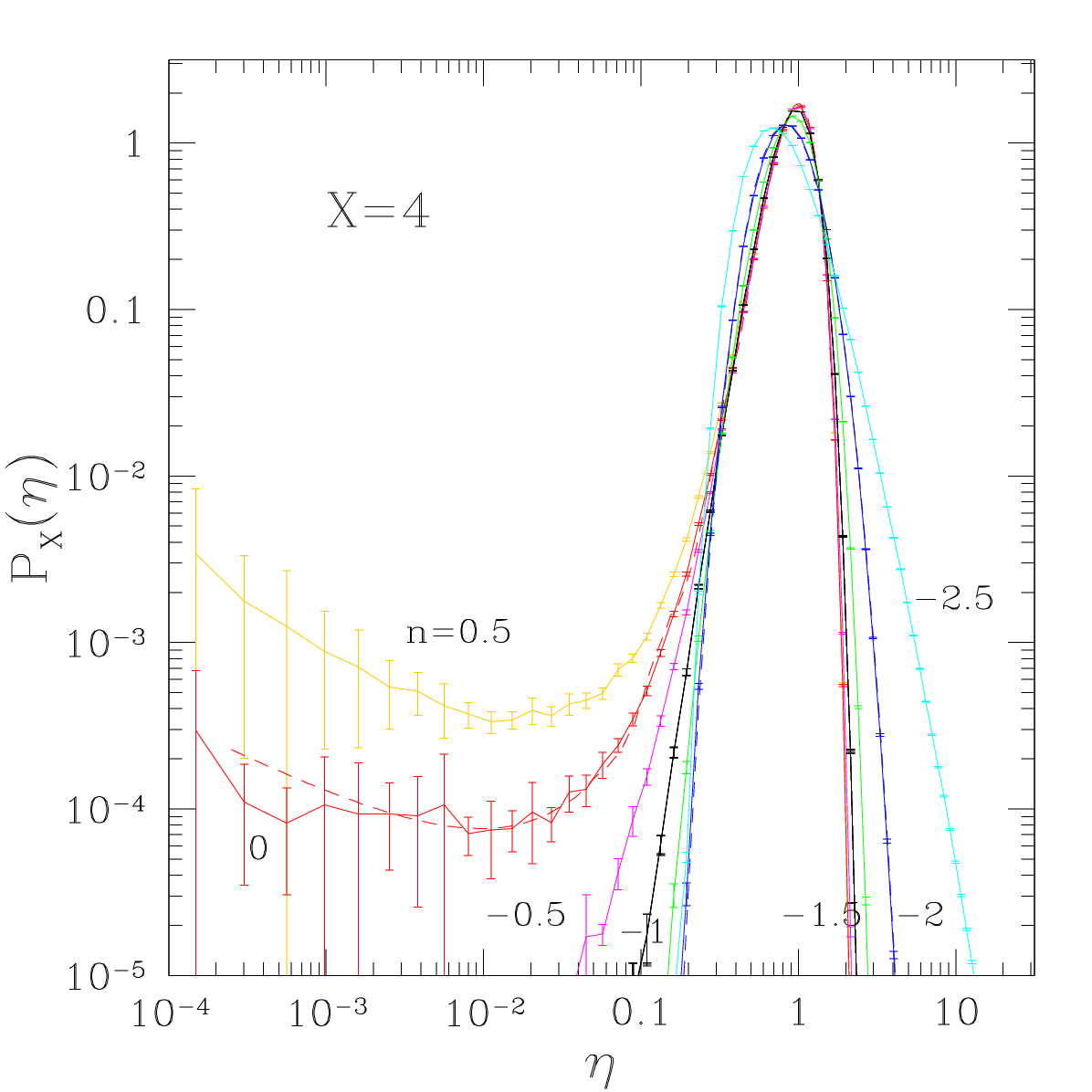}}
\epsfxsize=7 cm \epsfysize=5 cm {\epsfbox{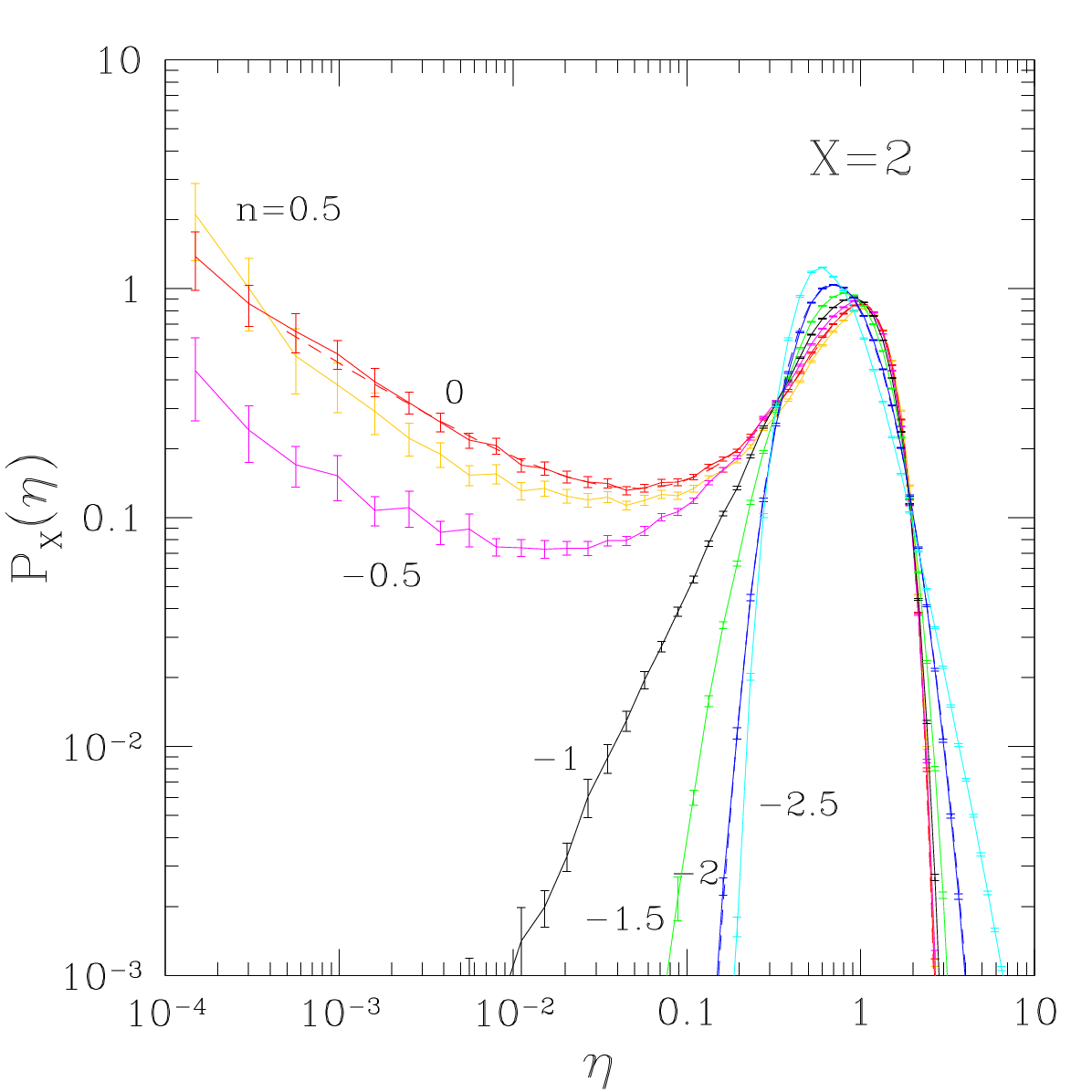}}
\epsfxsize=7 cm \epsfysize=5 cm {\epsfbox{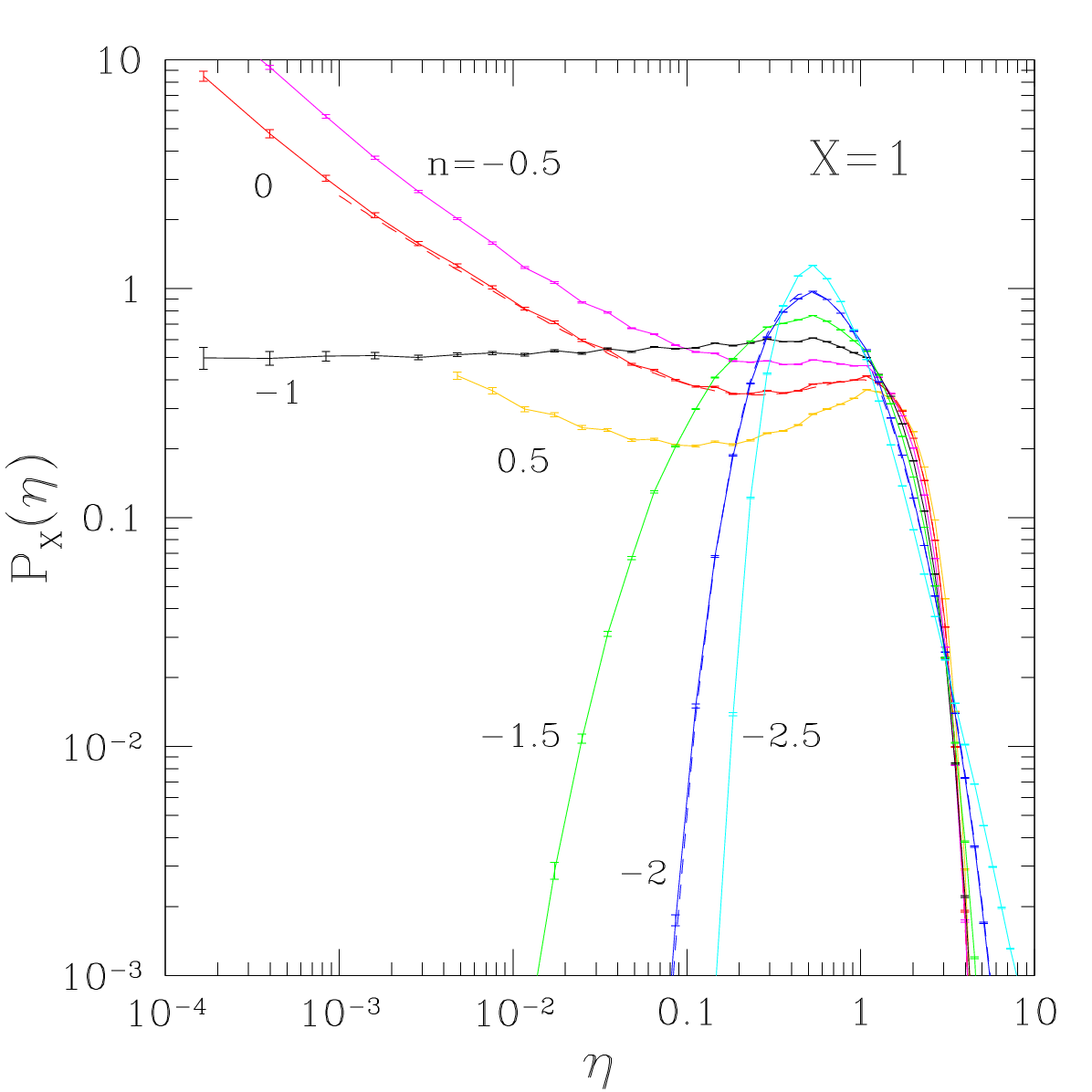}}
\epsfxsize=7 cm \epsfysize=5 cm {\epsfbox{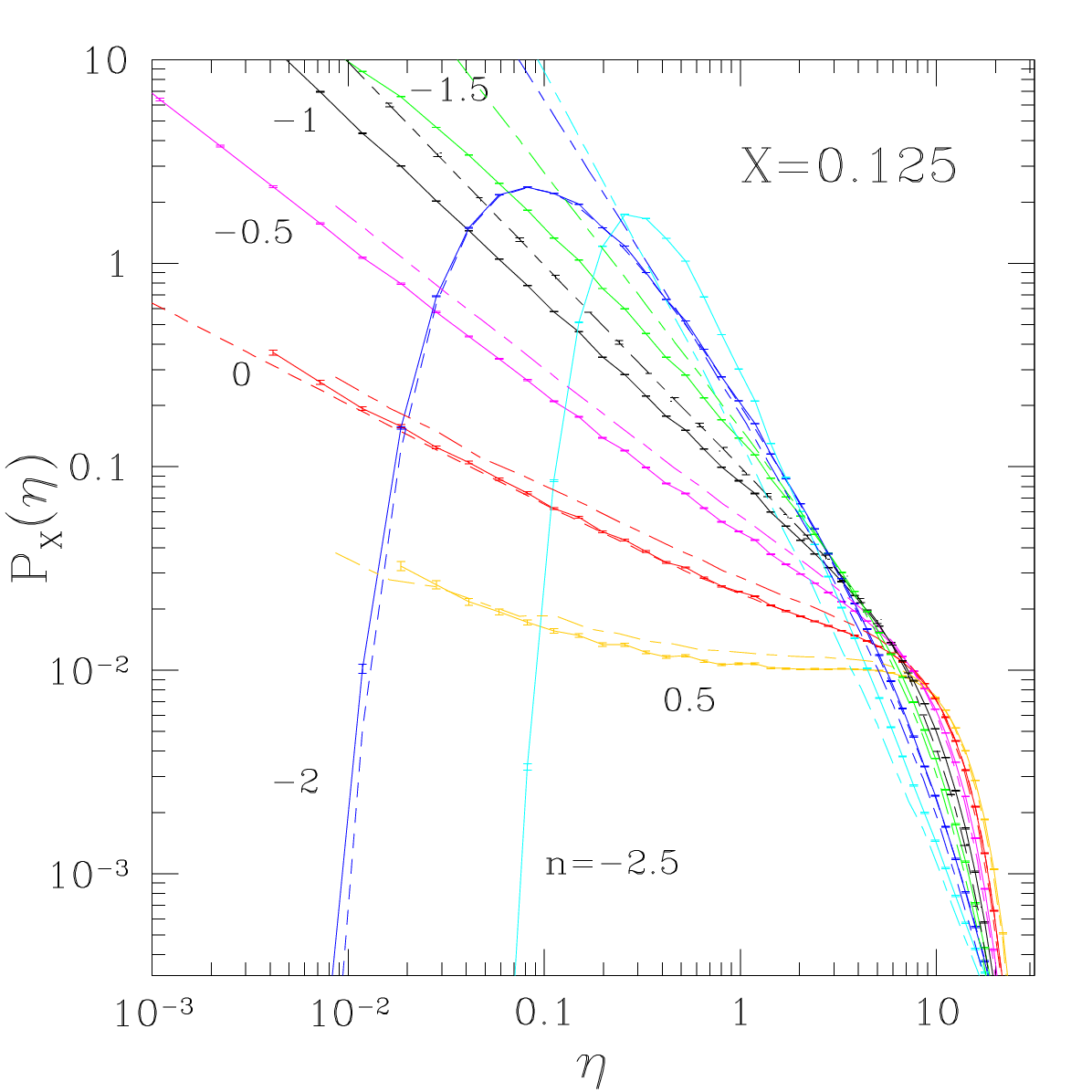}}
\end{center}
\caption{The probability distribution, $P_X(\eta)$, of the overdensity $\eta$
within intervals of length $X$. Smaller $X$ (from top to bottom) probe deeper into
the nonlinear regime. For $n=-2$ and $n=0$ the dashed lines are the exact
analytical results (\ref{PXeta-2}) and (\ref{PXneq}). For $-1<n<1$ there is an additional
Dirac contribution ($\propto \delta_D(\eta)$), associated with empty cells, that does not
appear in the figures. In the last panel ($X=0.125$) the dot-dashed lines are the
asymptotic behavior (\ref{PXeta-NL}).}
\label{figPxrho}
\end{figure*}

\begin{figure*}
\begin{center}
\epsfxsize=7 cm \epsfysize=5 cm {\epsfbox{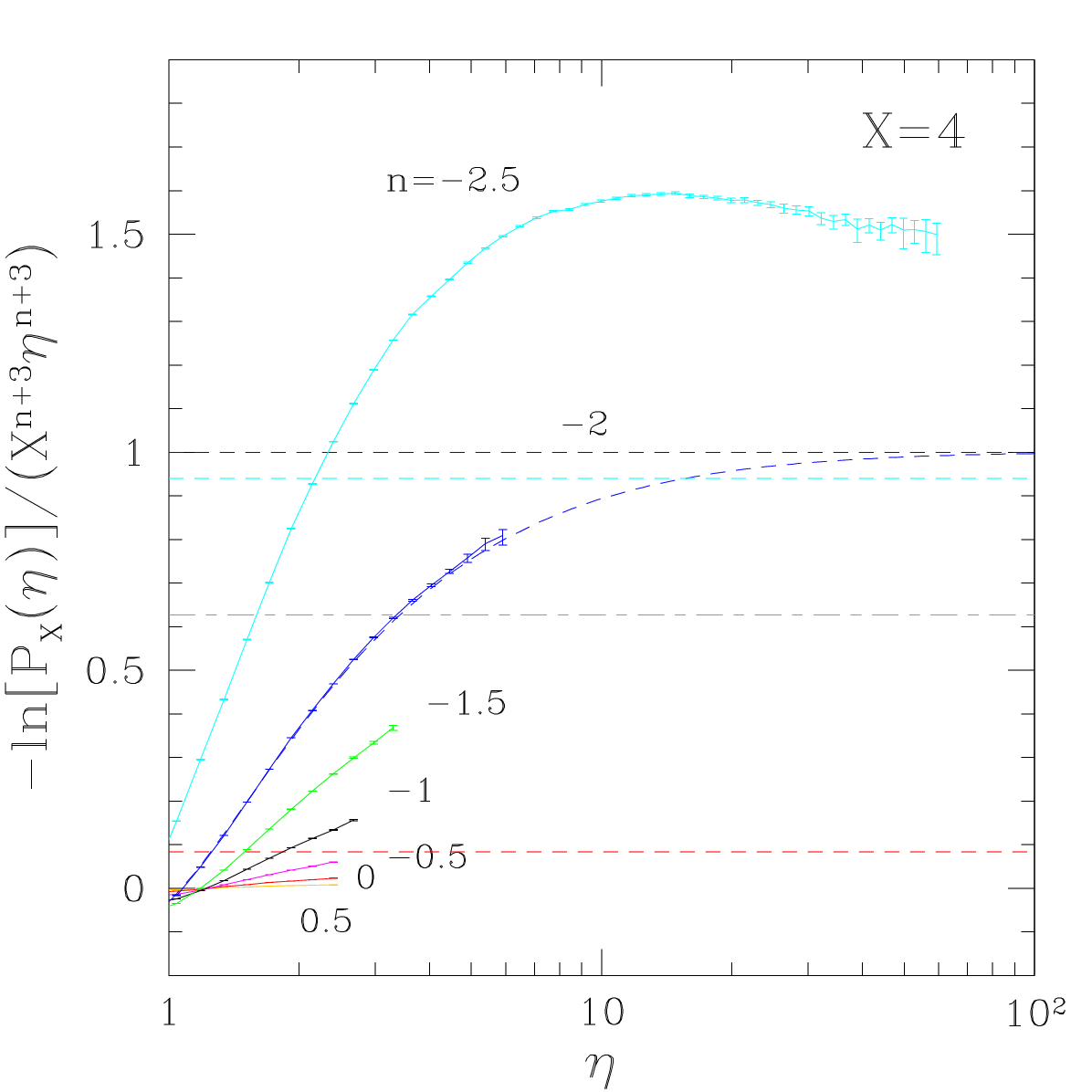}}
\epsfxsize=7 cm \epsfysize=5 cm {\epsfbox{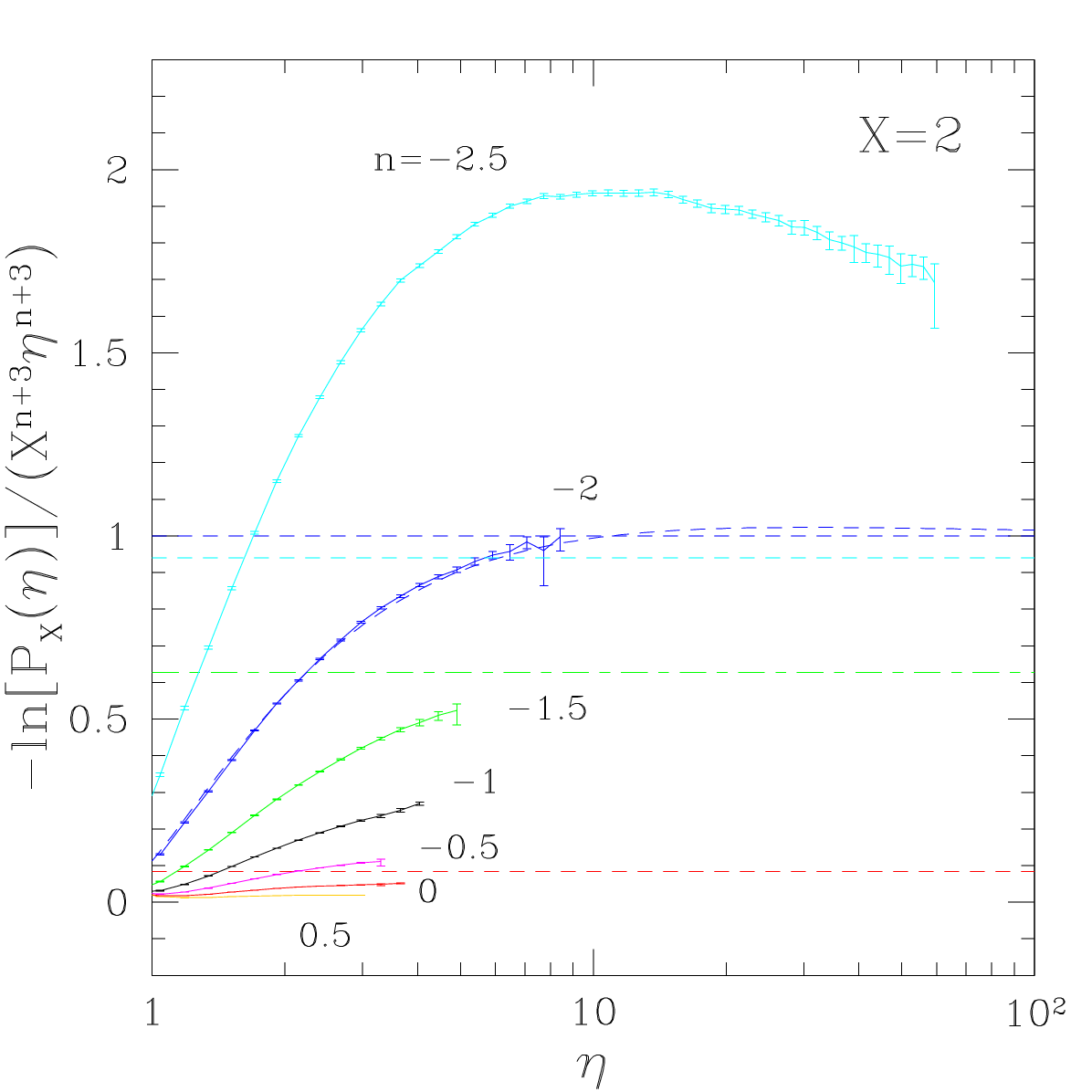}}
\epsfxsize=7 cm \epsfysize=5 cm {\epsfbox{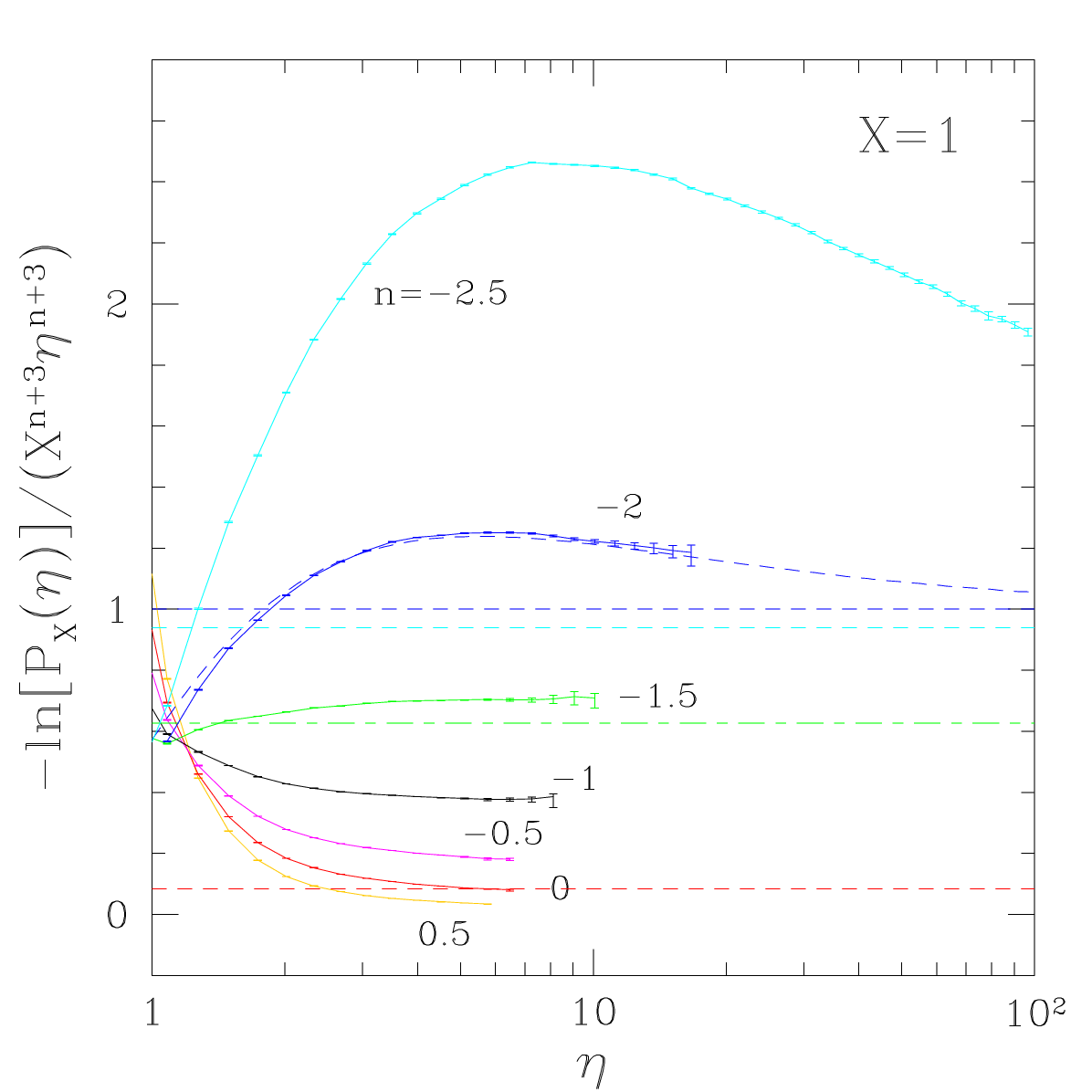}}
\epsfxsize=7 cm \epsfysize=5 cm {\epsfbox{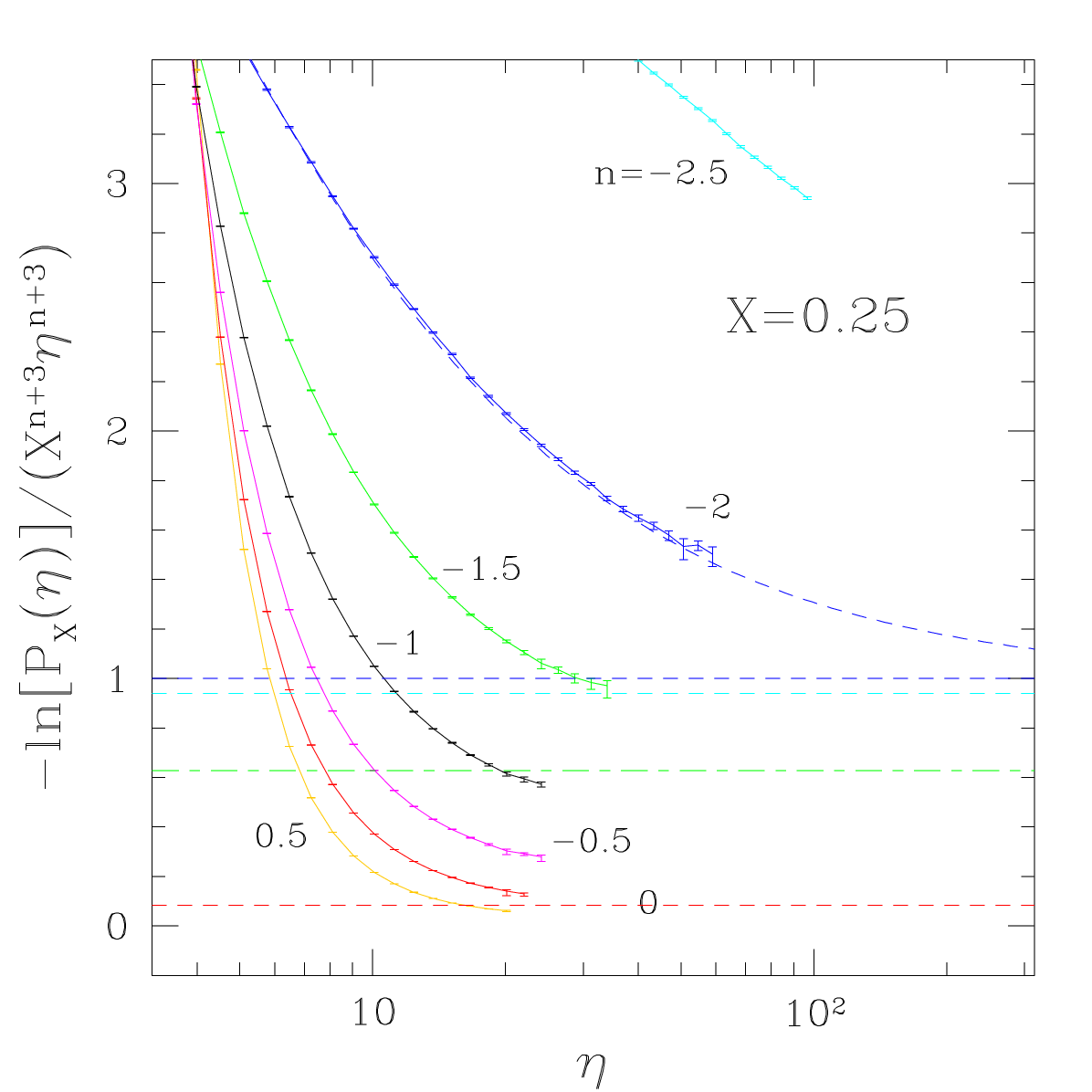}}
\end{center}
\caption{The ratio $-\ln P_X(\eta)/(X^{n+3}\eta^{n+3})$, which characterizes the
high-density tail of the probability distribution $P_X(\eta)$. The horizontal dashed
lines are the exact asymptotic values (\ref{PXeta-large-eta}) and
(\ref{PXeta-large-eta-n0}) for $n=-2.5, -2$, and $0$. For $n=-2$ the curved dashed line
is the exact ratio obtained from Eq.(\ref{PXeta-2}). For $n=-1.5$ the dot-dashed line
is the value obtained from Eq.(\ref{PXeta-large-eta}), which is only approximate
in this case.}
\label{figPxlrhop}
\end{figure*}

We now turn to the statistical properties of the smoothed density field. More
precisely, we study the probability distribution function, $P_X(\eta)$, of the
overdensity $\eta$ within an interval of length $x$,
\beq
\eta = \frac{m}{\rho_0 x} = \frac{M}{X} .
\label{eta}
\eeq
By conservation of matter we have $\lag \eta\rag=1$.
The exact expression of $P_X(\eta)$ is again explicitly known for the two cases
$n=-2$ \cite{Valageas2009a},
\beq
n=-2 : \;\; P_X(\eta) = \sqrt{\frac{X}{\pi}} \, e^{2X} \, \eta^{-3/2} \, e^{-X(\eta+1/\eta)} ,
\label{PXeta-2}
\eeq
and $n=0$ \cite{Valageas2009c},
\beq
n=0 : \;\; P_X(\eta) = P_X^0 \, \delta_D(\eta) + P_X^{\neq}(\eta) ,
\label{PXeta0}
\eeq
with
\beq
P_X^0 = \sqrt{\frac{\pi}{X}} \, e^{-\frac{X^3}{12}} \! 
\inta\frac{\dd s_1\dd s_2}{(2\pi\ii)^2} 
\frac{e^{(s_1+s_2)X/2+(s_1-s_2)^2/(4X)}}{\Ai(s_1)\Ai(s_2)}
\label{PX0}
\eeq
and
\beqa
P_X^{\neq}(\eta) & = & 2 \sqrt{\pi X^3} \, e^{-X^3/12} \inta\frac{\dd s\dd s_1\dd s_2}
{(2\pi\ii)^3} \nonumber \\
&& \times \frac{e^{sX(\eta-1)+(s_1+s_2)X/2+(s_1-s_2)^2/(4X)}}
{\Ai(s_1)\Ai(s_2)\Ai(s_1-s)\Ai(s_2-s)} \nonumber \\
&& \times \int_0^{\infty} \dd r \, e^{Xr} \Ai(r+s_1)\Ai(r+s_2) .
\label{PXneq}
\eeqa
In Eq.(\ref{PXeta0}) the Dirac term is associated with the nonzero probability,
$P_X^0$, to have an empty interval, in agreement with
section~\ref{Distribution-of-shocks} and Fig.~\ref{figldensity}. The second term
$P_X^{\neq}(\eta)$ is the regular part associated with nonempty intervals.
There is no Dirac term in Eq.(\ref{PXeta-2}) since shocks are dense for $n=-2$,
as recalled in section~\ref{Distribution-of-shocks} and seen in Fig.~\ref{figldensity},
so that the probability to have an empty interval is zero.

We show in Fig.~\ref{figPxrho} the evolution of $P_X(\eta)$ as we go from large
scales or early times (top) to small scales or late times (bottom), that is, from the
quasi-linear regime to the highly nonlinear regime. 
We can check that our numerical results agree with the exact results (\ref{PXeta-2})
and (\ref{PXeta0}) obtained for $n=-2$ and $n=0$.
At larger scales we recover a probability distribution that is increasingly peaked
around the mean, $\lag \eta\rag=1$, whereas at smaller scales an intermediate
power-law regime develops. This is similar to the behavior observed in cosmology
for the density field built by the gravitational dynamics
\cite{Balian1989,Bouchet1991,Colombi1994,Valageas2000},
starting from Gaussian initial conditions such as (\ref{ndef}).
For $-3<n\leq -2$, at large scales and finite $\eta$ one goes to a quasi-linear regime
governed by a regular saddle-point \cite{Valageas2009b} with
\beqa
\lefteqn{-3 <n \leq -2 , \;\; X\rightarrow \infty : } \nonumber \\
&& \ln P_X(\eta) \sim 
- \left[\eta^{(n+1)/2}-\eta^{(n+3)/2}\right]^2/(2\sigma^2(X/2)) \nonumber \\
&& \hspace{1.5cm} \sim - \frac{X^{n+3}}{I_n} \, 
\left[\eta^{(n+1)/2}-\eta^{(n+3)/2}\right]^2 .
\label{PXeta-QL}
\eeqa
For $n>-2$ shocks appear as soon as $t>0$ and modify the numerical factor $I_n$
in Eq.(\ref{PXeta-QL}) but not the main exponents. In particular, for $n=0$ one has
\cite{Valageas2009c}
\beqa
\lefteqn{n=0, \;\; X\rightarrow \infty, \;\; |\eta-1| \gg X^{-1} : } \nonumber \\
&& \hspace{2cm} \ln P_X(\eta) \sim - \frac{X^3}{12} |\eta-1|^3 .
\label{PXeta-QL-n0}
\eeqa
Thus we recover the large-$X$ and large-$\eta$ exponents of Eq.(\ref{PXeta-QL}),
but the functional form over $\eta$ has been modified.
Thus, whereas for $-3<n \leq -2$ the probability distribution $P_X(\eta)$ goes to
a Gaussian at large scales or early times, and we recover the Gaussian initial
conditions (i.e. the linear regime), this is no longer the case for $n>-1$.  
Indeed, as seen from Eq.(\ref{PXeta-QL}) for $-3<n\leq -2$, in the limit
$X\rightarrow \infty$ typical density fluctuations have
$|\eta-1|\sim \sigma \propto X^{-(n+3)/2}$, so that we can expand the argument
over $\eta$ around $\eta=1$. This gives
\beqa
\lefteqn{-3 <n \leq -2 , \;\; X\rightarrow \infty , \;\; |\eta-1| \ll X^{-(n+3)/3} \; : }
\nonumber \\
&& \hspace{2cm} P_X(\eta) \sim e^{-(\eta-1)^2/(2\sigma^2(X/2))} ,
\label{PXeta-G}
\eeqa
which coincides with the Gaussian associated with the linear density contrast
$\delta_L$.
For $-2<n<-1$ shocks have a modest effect on the relevant saddle point
\cite{Valageas2009b} and we expect to recover the Gaussian (\ref{PXeta-G})
at large scales, but the asymptotic behavior (\ref{PXeta-QL}) is no longer valid:
shocks modify the dependence on $\eta$, whence the value of the exponential
cutoff for any finite $\eta$. For $-1<n<1$, where the linear density variance
(\ref{sigmadef}) diverges, clearly one cannot recover a Gaussian such as
(\ref{PXeta-G}) at large scales: shocks govern the dynamics at all scales and the
probability distribution $P_X(\eta)$ is always strongly non-Gaussian,
as explicitly shown by Eq.(\ref{PXeta-QL-n0}) for the case $n=0$.
%For finite $X$, in addition to the Dirac term (\ref{PX0}) associated with empty cells
%the regular part (\ref{PXneq}) shows an inverse square-root low-density tail
%\cite{Valageas2009c}
%\beqa
%\lefteqn{n=0, \;\; X\rightarrow \infty, \;\; \eta \ll X^{-3} :  } \nonumber \\
%&& \hspace{0.8cm} P_X(\eta) \sim \frac{X}{\Ai'(-\omega_1)^2} \,
%e^{-\omega_1 X-X^3/12} \, \eta^{-1/2} ,
%\label{PXeta-QL-n0-tail}
%\eeqa
%where $-\omega_1\simeq -2.338$ is the first zero of the Airy function $\Ai(x)$.
We can check in Fig.~\ref{figPxrho} that the features associated with either
case $n=0$ and $n=-2$ (such as the power-law tail/exponential cutoff
at low densities, the nonzero/zero probability of empty cells) extend to the
classes $-1<n<1$ and $-3<n<1$ respectively.

At small scales the probability distribution $P_X(\eta)$ is governed by the shock
mass function, since it is dominated by the probability to have a shock of mass
$M=\eta X$ within the cell of size $X$, which gives \cite{Valageas2009c}
\beq
-3<n<1, \;\; X\rightarrow 0 : \;\; P_X(\eta) \sim X^2 \, N(\eta X) .
\label{PXeta-NL}
\eeq
The asymptotic behavior (\ref{PXeta-NL}) holds at fixed $\eta$, and it does not describe
the low-density exponential cutoff that is always present for $-3<n<-1$ (but is repelled
to $\eta\rightarrow 0$ as $X$ goes to zero).
One can explicitly check on the exact expressions obtained for $N(M)$ and 
$P_X(\eta)$ in both cases $n=0$ and $n=-2$ that they agree with (\ref{PXeta-NL}).
Since the analytical expression of the mass function $N(M)$ is only known for these
two cases, $n=0$ and $n=-2$, we plot in the lower panel of Fig.~\ref{figPxrho}
the asymptotic quantity $X^2 \, N(\eta X)$ (dot-dashed lines) obtained using 
the mass functions measured from our numerical computations and shown
in Fig.~\ref{figmNm}. We can see that the behavior (\ref{PXeta-NL})
can already be clearly seen at $X \sim 0.125$, especially for $-1<n<1$.
For $-3<n<-1$ the low-density falloff has not been repelled to very low $\eta$ yet
so that the convergence to (\ref{PXeta-NL}) only appears for $\eta > 1$.

%The scaling (\ref{PXeta-NL}) gives a weight that decreases as $X$ for the
%contribution of this regular part above any fixed low-density threshold
%$\eta_{\rm th}$, $\int_{\eta_{\rm th}}^{\infty} \dd\eta \, P_X(\eta) \sim X$
%for $X\rightarrow 0$.
%For $-1<n<1$ this expresses the fact that shocks are isolated and in finite number
%per unit length, so that the probability to have at least one shock decreases as
%$X$ at low scales, whereas the probability to have an empty cell, associated
%with a Dirac term of the form $P_X^0 \delta_D(\eta)$, goes to unity.
%This decrease of the probability to have a nonzero density is clearly seen in
%Fig.~\ref{figPxrho} for $n=0$ and $n=0.5$.
%For $-3<n<-1$, where shocks are dense and there are no empty cells, the smooth
%probability distribution $P_X(\eta)$ is always normalized to unity.
%This remains consistent with the scaling (\ref{PXeta-NL}) because the latter does not
%extend down to $\eta=0$. There is a high peak at a low density $\eta_-(X)$, which
%goes to zero at small scales, before the very low density exponential cutoff.
%Thus, as seen in Fig.~\ref{figPxrho}, at small highly nonlinear scales, most cells
%have a low density of order $\eta_-(X)$, with $\eta_-\rightarrow 0$ for
%$X \rightarrow 0$, which is beyond the reach of the asymptotic behavior
%(\ref{PXeta-NL}).

%

%

In order to see more clearly the high-density tail of the probability distribution
$P_X(\eta)$ we show in Fig.~\ref{figPxlrhop} the quantity
$-\ln P_X(\eta)/(X^{n+3}\eta^{n+3})$. Indeed, in this rare-event limit one can
still apply a saddle-point approach, which yields \cite{Valageas2009b}
\beqa
\lefteqn{ -3 < n \leq -2 , \;\; \eta \rightarrow \infty : } \nonumber \\
&& \ln P_X(\eta) \sim - \frac{\eta^{n+3}}{2\sigma^2(X/2)} 
= - \frac{X^{n+3}}{I_n} \, \eta^{n+3} .
\label{PXeta-large-eta}
\eeqa
Indeed, this is governed by the same saddle point as the one associated with
(\ref{PXeta-QL}), even though we now consider the limit of large $\eta$ at fixed
$X$, instead of large $X$ at fixed $\eta$.
For $n>-2$ the saddle point develops shocks, which modify the numerical factor
in (\ref{PXeta-large-eta}) but not the exponents. In particular, for $n=0$
this gives
\beq
n=0, \;\; \eta \rightarrow \infty : \;\; \ln P_X(\eta) \sim - \frac{X^3}{12} \, \eta^3 .
\label{PXeta-large-eta-n0}
\eeq
Of course, one can check that the asymptotic behaviors (\ref{PXeta-large-eta}) and
(\ref{PXeta-large-eta-n0}) agree with the full expressions (\ref{PXeta-2}) and
(\ref{PXneq}) obtained for $n=-2$ and $n=0$.
We can see in Fig.~\ref{figPxlrhop} that our numerical results reach a constant
asymptote at high density, in agreement with the general scaling
(\ref{PXeta-large-eta}), and for $n\leq -2$ and $n=0$ they are consistent with the
theoretical values (\ref{PXeta-large-eta}) and (\ref{PXeta-large-eta-n0}).
For $n=-1.5$, using the normalization given in Eq.(\ref{PXeta-large-eta})
(i.e. $I_{-1.5}$) again appears to provide a reasonable approximation to the high-density
asymptote (albeit slightly lower).
In agreement with Fig.~\ref{figlNmp}, this means that for $-2<n\leq -1.5$
shocks have not significantly modified the quantitative profile of the high-density
saddle point.
As for the high-mass tail of the mass functions displayed in Fig.~\ref{figlNmp},
the convergence to the asymptotic behavior (\ref{PXeta-large-eta}) is slower
for lower $n$. We can also note that the shape of the function
$-\ln P_X(\eta)/(X^{n+3}\eta^{n+3})$ depends on scale, as it typically reaches its
asymptote from below at large $X$ and from above at low $X$.
As shown by the exact ratio obtained from Eq.(\ref{PXeta-2}) for the case $n=-2$,
which agrees with our numerical computations, this is not a numerical artifact.
Again, note that Fig.~\ref{figPxlrhop} magnifies the deviations from
(\ref{PXeta-large-eta}), due for instance to subdominant power-law prefactors,
which would not be easily distinguished in Fig.~\ref{figPxrho} as the exponential
falloffs are already very steep over this density range.

\subsection{Low-order density cumulants}
\label{Low-order-density-cumulants}

\begin{figure}
\begin{center}
\epsfxsize=7 cm \epsfysize=5 cm {\epsfbox{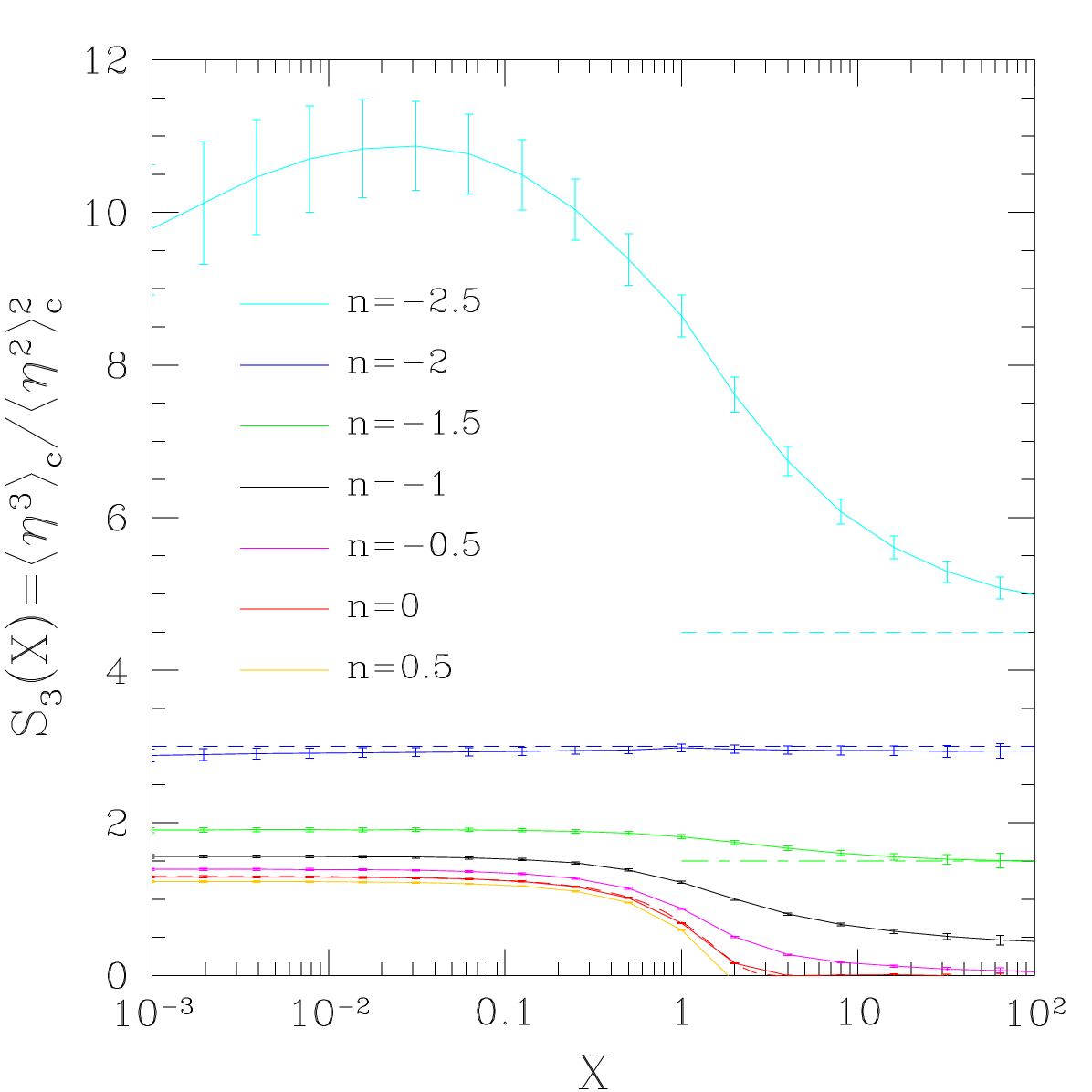}}
\epsfxsize=7 cm \epsfysize=5 cm {\epsfbox{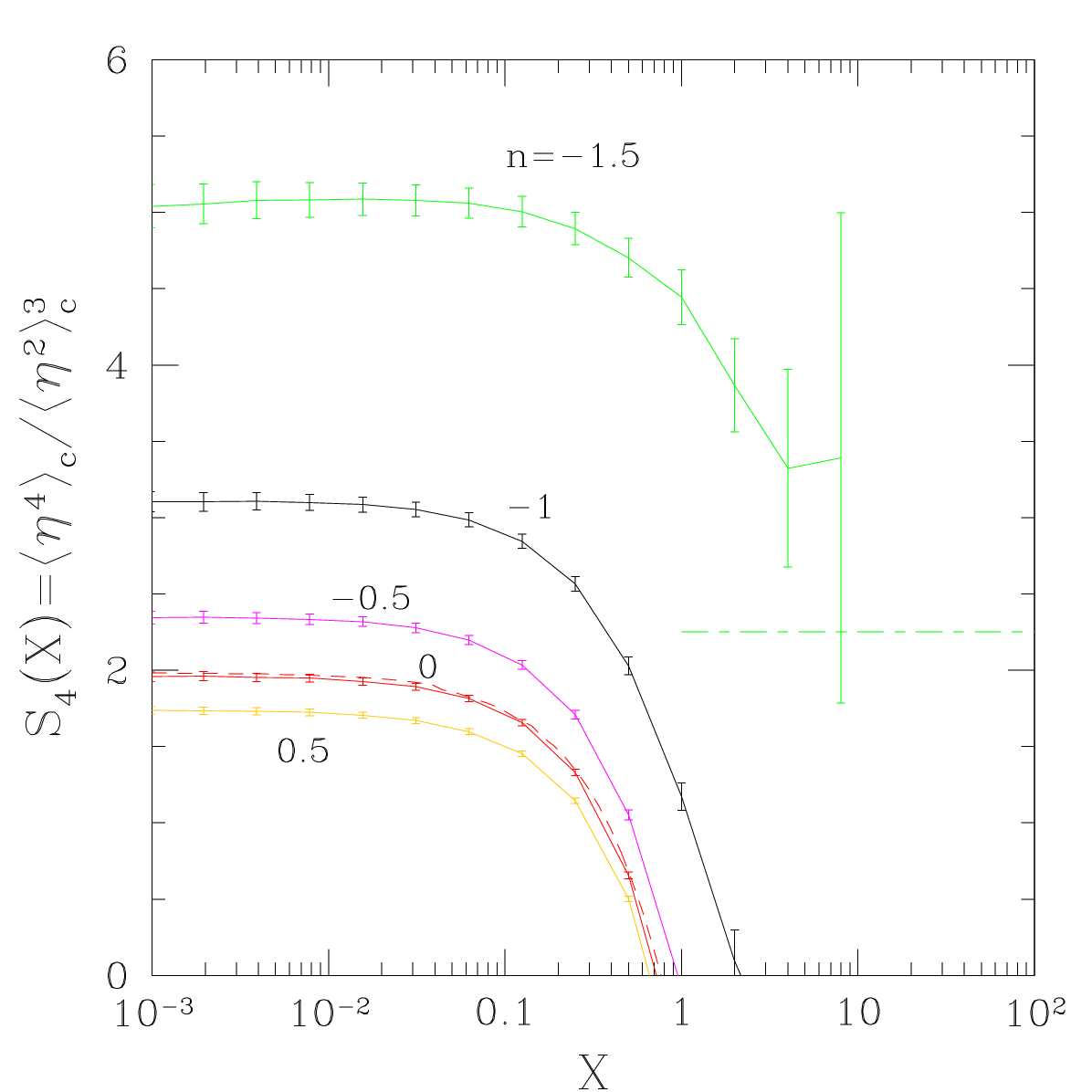}}
\end{center}
\caption{The low-order coefficients $S_3$ (upper panel) and $S_4$ (lower
panel) defined in Eq.(\ref{Spdef}). For $n=0$ and $n=-2$ the dashed curves are
the exact analytical results. Both $S_3$ and $S_4$ are constant in the case $n=-2$.
For $n=-2.5$ the dashed line at $X>1$ shows the large-scale asymptote (\ref{S34-QL}).
For $n=-1.5$ the dot-dashed line at $X>1$ is the value obtained from (\ref{S34-QL}),
which is only approximate in this case.}
\label{figSrho}
\end{figure}

We finally test our results with the use of the low-order cumulants defined, $S_p$, as
\beq
S_p= \frac{\lag\eta^p\rag_c}{\lag\eta^2\rag_c^{p-1}} .
\label{Spdef}
\eeq
They are known to reach constant values at large scales, and those values
can be exactly computed in both the exact dynamics (see \cite{2002PhR...367....1B} and references therein) and for the Burgers equations\footnote{In this case, it corresponds to the Zel'dovich approximation, see \cite{1995ApJ...443..479B}}. Thus, for the 1D case, those parameters reach a constant asymptote at large scales when $-3<n<-1$, with
\cite{Valageas2009b}
\beqa
\lefteqn{-3<n\leq -2 , \; X\rightarrow \infty : \;\; S_3 \rightarrow -3(n+1) ,}
\label{S3-d1} \\
&&  \hspace{2.5cm} S_4 \rightarrow 3(9+16n+7n^2) .
\label{S34-QL}
\eeqa
For $-2<n<-1$ shocks modify the large-scale asymptotes, while for $-1<n<1$
the coefficients $S_p$ typically vanish for odd $p$ and go to infinity for even $p$,
as can be explicitly checked in the case $n=0$ \cite{Valageas2009c}
where exact results can be derived from Eq.(\ref{PXneq}).
On the other hand, at small scales they reach constant asymptotes, for all $n$
in the range $-3<n<1$, as the density cumulants are governed by the pointlike
masses associated with shocks.
%This directly follows from the scaling (\ref{PXeta-NL}) of the density 
%probability distribution function, see the discussion in section~7 of
%\cite{Valageas2009c}.
The exact values of these small-scale asymptotes, associated with the highly
nonlinear regime and governed by the shock mass function, are only known for
the two cases $n=0$ \cite{Valageas2009c} and $n=-2$ \cite{Valageas2009a}.
In the case $n=-2$ it happens that the coefficients $S_p$ are actually
scale-independent, so that the quasilinear values (\ref{S34-QL}) hold for all $X$
and we have \cite{Valageas2009a}
\beq
n=-2 : \;\;\;\; S_p = (2p-3)!!
\label{Sp-2}
\eeq
%This means that the 1D case $n=-2$ provides an exact realization of the
%``stable-clustering ansatz'' introduced in \cite{Davis1977,Peebles1980} as
%an approximate model for the density field generated by the gravitational
%dynamics in cosmology.
We can check in Fig.~\ref{figSrho}, where we plot the coefficients $S_3$ and
$S_4$, that our numerical computations agree with the results recalled above.
In particular, we clearly see the small-scale universal constant asymptotes, due
to shocks, except for the case $n=-2.5$. There it is not clear whether the
deviation is due to the finite numerical resolution or to the slow convergence
to the small-scale asymptote. At large scale we can see the same curve approach
the asymptote (\ref{S34-QL}).
As for the high-mass and high-density tails of the shock mass function and of the
density distribution, for the case $n=-1.5$ the value given by Eq.(\ref{S34-QL})
is a very good approximation for $S_3$ at large scales. This again means that
shocks do not have a significant effect for $-2<n\leq -1.5$. For $S_4$ the error bars
are too large to draw any conclusion on the accuracy of (\ref{S34-QL}).

\subsection{The density power spectrum}
\label{Power-spectra}

\begin{figure}
\begin{center}
\epsfxsize=7 cm \epsfysize=5 cm {\epsfbox{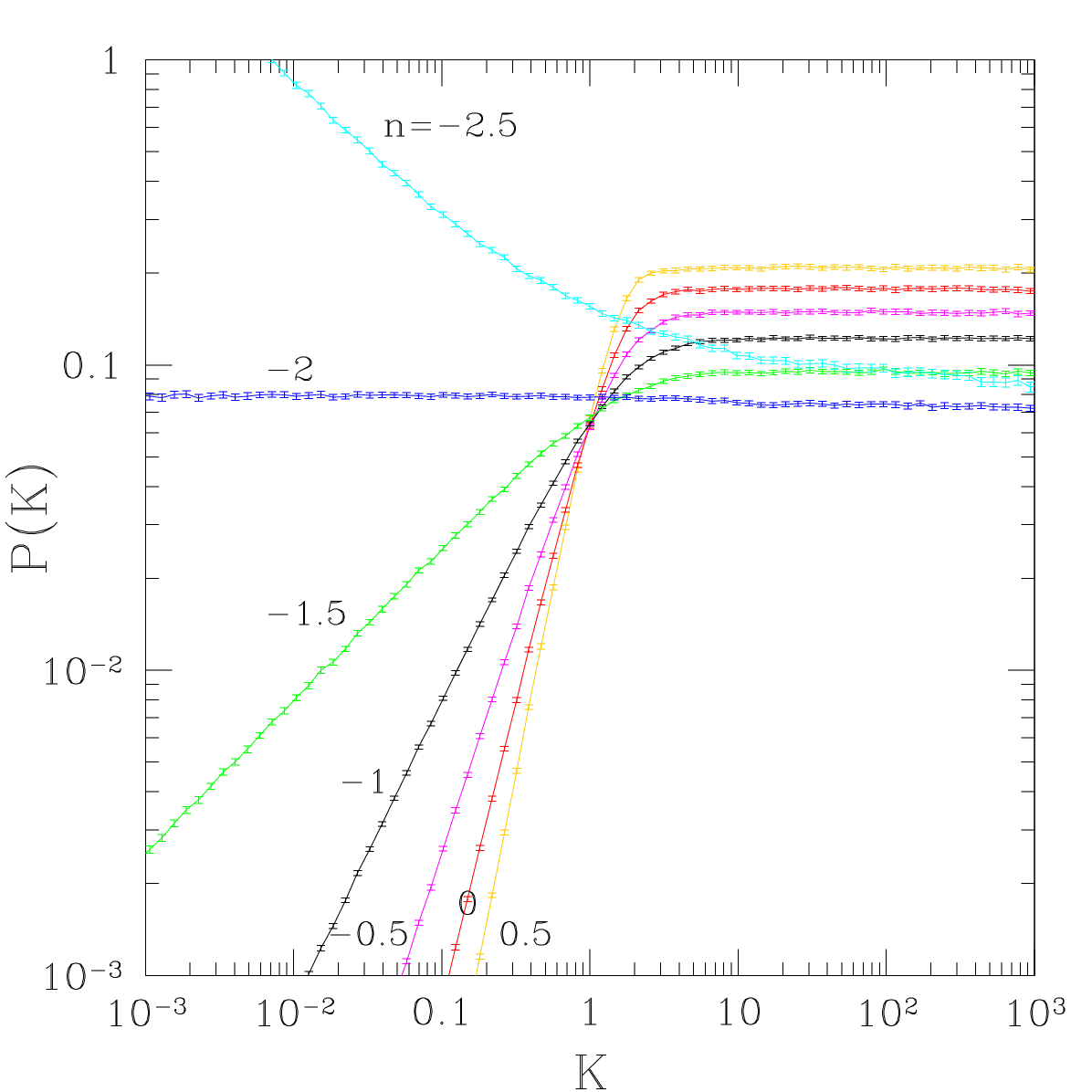}}
\end{center}
\caption{The density power spectrum $P(K)$. For $n=0$ and $n=-2$ the dashed
lines are the exact analytical results, which are almost completely masked by
the numerical results.}
\label{figPk}
\end{figure}

%

%\begin{figure}
%\begin{center}
%\epsfxsize=7 cm \epsfysize=5 cm {\epsfbox{Ek.ps}}
%\end{center}
%\caption{The energy spectrum $E(K)$ (i.e. velocity power spectrum). For $n=0$ and
%$n=-2$ the numerical results almost perfectly match the analytical results 
%shown by the dashed lines.}
%\label{figEk}
%\end{figure}

We show in Fig.~\ref{figPk} the density power spectrum as a function of the wave number $K$. 
At low $K$ in the ``linear
regime'' we recover the initial condition (\ref{PLE0}). Note that this holds for all
$-3<n<1$, even though shocks cannot be ignored even at a qualitative level for
$n>-1$ at all scales (in particular they make the real-space variance
$\lag \eta^2\rag_c$ finite even though the linear variance $\sigma^2$ was infinite).
At high $K$, in the highly nonlinear regime, we recover the universal constant
asymptote, due to shocks (as the haloes that form are pointlike it is easy to see that 
white-noise tails are expected to develop at large $K$.) 
For $n=0$ and $n=-2$ we also plot the analytical results 
\cite{Valageas2009c,Valageas2009a} as dashed lines.They reproduce exactly the numerical results 
showing that the numerical results are free of finite volume effects.

%
% but they cannot be distinguished
%from the numerical results, which allows to check the convergence of numerical
%computations.

%
%Indeed, in the inviscid limit that we consider in this article,
%shocks are infinitesimally thin so that the density field can be seen as a sum of
%Dirac density peaks, see Fig.~\ref{figldensity}, which leads to a ``white-noise''
%tail at high wavenumber for the density power spectrum. This remains valid
%for $-3<n<-1$ where shocks are dense. For $n=-2$, where $P(K)$ also has a constant
%asymptote at low $K$ in the linear regime, it happens that $P(K)$ is truly
%scale-independent \cite{Valageas2009a} (the low and high $K$ asymptotes are equal
%and there is no deviation at intermediate scales).
%Note however that the density field is never a true white-noise at high $k$ 
%as it is not Gaussian and higher-order correlations play a key role, as seen
%in section~\ref{Low-order-density-cumulants}.

%For completeness we show in Fig.~\ref{figEk} the energy spectrum $E(K)$.
%In 1D it is simply related to the density power spectrum by $E(K)=P(K)/K^2$
%and it shows at high $K$ the universal $K^{-2}$ tail due to shocks
%\cite{Gurbatov1997,Valageas2009c}. We can check in Fig.~\ref{figEk} that we recover
%this high-$K$ asymptote, as well as the low-$K$ asymptote (\ref{PLE0}).
%Again, for $n=0$ and $n=-2$ the numerical results match very well the known
%analytical results.

%

%

\section{Algorithms for the 2D Geometrical Adhesion Model}
\label{Algorithms-for-the-2D-GAM}

\subsection{Initial conditions and the Eulerian velocity field}
\label{compute-H(x)}

As for the 1D case, we discretize the system on a regular $N\times N$ grid, with unit
steps and periodic boundary conditions, and we choose $N$ to be a power of $2$,
typically $N=2^{11}=2048$.
To implement the Gaussian initial conditions we
also introduce rescaled coordinates as in (\ref{hx}). It is convenient to use the
velocity potential $\psi_0$, which is obtained from a discrete Fourier transform as
in (\ref{hu0}), with now
\beq
 \lag | \thpsi_{0,\hvk} |^2 \rag = \frac{D}{(2\pi)^2} \, \left(\frac{2\pi}{N}\right)^{n-1}
\, \hk^{n-3} .
\label{varthpsik}
\eeq
This yields the initial velocity field $\vu_0(\vq)$ through Eq.(\ref{thetadef}).

In Eulerian space, the velocity field $\vu(\vx,t)$ and its potential $\psi(\vx,t)$
are again obtained from the Hopf-Cole solution (\ref{Hxphiq}).
Thus, we need to compute the 2D Legendre transform $\cL_{\vx} [ \varphi_L(\vq,t) ]$,
over the regular $N\times N$ grid $\vx_{i_1,i_2}$, from the periodic linear potential
$\varphi_L(\vq,t)$ defined over the regular $N\times N$ grid $\vq_{i_1,i_2}$.
Thanks to the period $N$ of the system the 2D Eulerian coordinates 
$\vx_{i_1,i_2} = (i_1,i_2)$, with $0\leq i \leq N-1$, are associated with 
Lagrangian coordinates $\vq_{j_1,j_2}= (j_1,j_2)$ that obey
$-N/2 \leq j \leq N+N/2-2$. Therefore, we first extend $\varphi_L(\vq,t)$ to the
larger grid $-N/2 \leq j \leq N+N/2-2$ (using periodicity) and next we compute
$\cL_{\vx} [ \varphi_L(\vq,t) ]$, using the well-known property that a 2D Legendre
transform can be obtained from two successive partial 1D Legendre transforms, 
\beqa
\lefteqn{\hspace{-0.5cm} \cL_{x_1,x_2} [ \varphi_L(q_1,q_2) ] = \max_{q_1,q_2} 
\left[ x_1 q_1+x_2 q_2 - \varphi_L(q_1,q_2) \right] } \nonumber \\
& & \hspace{0.5cm} =\max_{q_1} \left[ x_1 q_1 + \max_{q_2} \left[ x_2 q_2
- \varphi_L(q_1,q_2) \right] \right]  \nonumber \\
& & \hspace{0.5cm} = \cL_{x_1} \left[ - \cL_{x_2} (\varphi_L) \right] .
\label{Legendre2D}
\eeqa
Thus, for each 1D Legendre transform in Eq.(\ref{Legendre2D}) we use the
algorithm of Lucet \cite{Lucet1997}, used in appendix~\ref{One-dimension} for the
one-dimensional case and described in 
appendix~\ref{Algorithms-for-the-1D-Burgers-dynamics}, taking advantage
of the fact that all functions are given on regular grids.
This allows to compute the 2D Legendre
transform $H(\vx)$ in linear time over the total number of grid points,
$\Nt=N^2$, which is thus optimal.

\subsection{Direct Lagrangian map}
\label{compute-convex-hull}

In addition to the Eulerian fields $\vu(\vx,t)$ and $\psi(\vx,t)$, the procedure
(\ref{Legendre2D}) yields the inverse Lagrangian map $\vx\mapsto\vq$.
However, contrary to the 1D case this is no longer sufficient to obtain the
direct Lagrangian map $\vq\mapsto\vx$.
In fact, as recalled in section~\ref{Lagrangian-potential}, the latter (and the associated
density field) depends on the precise definition of the inviscid limit.
In this article we consider the procedure described in 
section~\ref{Lagrangian-potential}, where the Lagrangian map $\vx(\vq)$ is defined
from the convex hull $\varphi(\vq)$ by Eq.(\ref{qx-xq}).
As shown in \cite{BernardeauVal2010b}, this corresponds to adding a specific
diffusive term to the right hand side of the equation of continuity,
which vanishes in the inviscid limit $\nu\rightarrow 0^+$ except along shocks.
As described in \cite{BernardeauVal2010b}, see also 
\cite{Gurbatov1991,Vergassola1994,Woyczynski2007},
in two dimensions the convex hull
$\varphi(\vq)$ defines a triangulation of the Lagrangian $\vq$-space
(because the convex hull of $\varphi_L(\vq)$ is made of a set of triangular facets)
which is associated with a Voronoi-like tessellation of the Eulerian $\vx$-space.

From the Legendre duality (\ref{varphicdef}) we can see that within this
prescription the direct Lagrangian map, $\vq\mapsto\vx$, can be obtained from the
Legendre transform of the Eulerian function $H(\vx)$. Thus, the position $\vx$
of the particle of Lagrangian coordinate $\vq$ is given by the point $\vx$ where
the maximum in Eq.(\ref{varphicdef}) is reached. Therefore, since we have already
obtained $H(\vx)$ through a 2D Legendre transform, as explained above,
we could use the same algorithm to apply a second 2D Legendre transform to
$H(\vx)$. This would give $\varphi(\vq)$ on a regular grid, as well as the
direct Lagrangian map, $\vq\mapsto\vx$. As noticed in \cite{Lucet1997}, this
procedure, based on two successive Legendre transforms, provides a very fast
algorithm to compute on a grid the convex hull $\varphi(\vq)$ of any function
$\varphi_L(\vq)$, since it scales linearly with the total number of points $\Nt$
of the grid (as we have recalled above for the computation of $H(\vx)$).

In contrast, it is known that the explicit computation of the 3D convex hull scales at
least as $\Nt \ln \Nt$. The reason for this longer execution time 
is that by ``explicit computation of the 3D convex hull'' we mean that, given the
initial function $\varphi_L(\vq)$ on a grid of $\Nt$ points $\vq_j$,
which defines a set of 3D points $(q_1,q_2,\varphi_L)_j$, 
we want to obtain the subset of $\Nv$ vertices that belong to the lower convex hull
as well as its $\Nf$ triangular facets (which specifies how the vertices are gathered into
triplets, to form these facets; note that each vertex can be a summit of several facets).
Clearly, this involves more information than the mere knowledge of the values
of $\varphi(\vq)$ on a grid, which explains the different scalings with $\Nt$
of these two problems (in particular, once we know the facets of the convex hull
it is immediate to compute $\varphi(\vq)$ on any grid, while the converse is not
true).
Note that this is a truly three-dimensional problem.

In spite of the explicit expression (\ref{varphicdef}), which gives the direct
Lagrangian map $\vx(\vq)$ through the Legendre transform of $H(\vx)$, we use
in this article the explicit computation of the 3D convex hull (i.e. we compute
the list of its triangular facets) to obtain the direct Lagrangian map, $\vq\mapsto\vx$.
This is necessary to obtain the Lagrangian and Eulerian-space tessellations and to
follow the intricate dynamics of shock nodes, which undergo both merging and
fragmentation events. These processes are described in detail in 
\cite{BernardeauVal2010b}, where we used for numerical computations the algorithm
that we describe in
appendix~\ref{Algorithms-for-the-2D-Burgers-dynamics} in the present paper.
If we only require snapshots of the Lagrangian map and of the density field, as in the
present article, the faster Legendre transform (\ref{varphicdef}) would be sufficient as
noticed above.
However, in practice it introduces an additional source of error in numerical
computations. Indeed, the function $H(\vx)$ being defined as the Legendre transform
(\ref{Hxphiq}) and $\varphi_L(\vq)$ being defined on a set of discrete points, it is
piecewise affine. In fact, for the self-similar initial conditions (\ref{ndef}) this is not a
numerical artifact due to the discretization and the planar facets of $H(\vx)$, which
define the Eulerian-space Voronoi-like tessellation, correspond to voids (i.e. empty
regions) in the Eulerian density field. Moreover, their typical size scales with time as
the characteristic scale $L(t)$ defined in Eq.(\ref{Lt}).
However, if we compute $H(\vx)$ on a grid, using the Legendre transform algorithm
described above, it is clear that because of the finite precision of real numbers in
computers such planar facets will show small wrinkles. Then, when we apply a second
Legendre transform to $H(\vx)$ to compute $\varphi(\vq)$, we artificially split large
voids into smaller voids and introduce spurious matter concentrations (associated with
the contact points of these wrinkles with their convex envelope)\footnote{Such an
effect can be easily checked numerically by noticing that if we repeatedly apply such
2D Legendre transforms we obtain an increasing number of shocks nodes, instead of
a stable result as implied by the property $\cL^{2n}(\varphi_L)=\varphi$ for 
any $n\geq 1$.}.
This is not necessarily a serious problem if one is not interested in the properties of the
Lagrangian and Eulerian-space tessellations, as long as one makes sure to discard these
spurious low-mass shock nodes. However, to be fully consistent with our previous work 
\cite{BernardeauVal2010b} and to avoid introducing unnecessary sources of numerical
error we prefer not to use this simple approach and to explicitly compute the 3D
lower convex hull $\varphi$ as well as the Lagrangian and Eulerian-space
tessellations.

Therefore, to obtain $\varphi$ (more precisely, the list of its triangular facets) from
the values $\varphi_L(\vq_j)$ of the linear Lagrangian potential on the 2D grid of
$\Nt$ points $\vq_j$, with $\Nt=N^2$, we need a 3D algorithm, which computes
the convex hull of a finite set of points in three dimensions.
A minor simplification is that we only require its lower part, since $\varphi$ is the
lower convex envelope of $\varphi_L$.
A brute-force method, testing each triple of points as a possible facet, takes a running
time $O(\Nt^4)$, whereas a slight modification improves this to $O(\Nt^3)$
by testing each pair of points as a possible edge \cite{ORourke1998}.
Gift-wrapping algorithms \cite{Chand1970,Preparata1985}
scale as $O(\Nt \Nf)$ (where $\Nf$ is the number of output facets) by generating
facets one at a time via implicit searches. 
Incremental methods gradually update the convex hull as initial points are inserted
one after the other and can achieve an optimal $O(\Nt\ln\Nt)$ expected running time
\cite{Clarkson1989} by randomizing the insertion order (so as to beat the $O(\Nt^2)$
worst-case performance).
Finally, the divide-and-conquer method, proposed in \cite{Shamos1975} for 2D Voronoi
diagrams, and in \cite{Preparata1977} for 3D convex hulls was the earliest algorithm
to achieve optimal $O(\Nt\ln\Nt)$ running time. As its name suggests, this algorithm
divides the point set into two halves by a vertical plane, recursively computes the hull
of each half, and merges the two hulls into one.
As usual, it is the bisection by two and the recursivity that bring down a factor $\Nt$
to $\ln\Nt$ in the running time.
Since we need to compute many convex hulls $\varphi$, with typically
$\Nt=N^2=2^{22}=4194304$, in order to accumulate a sufficiently large number of
realizations and output times to measure the statistical properties of the dynamics over
several regimes, it is necessary to use a fast algorithm that scales as $O(\Nt\ln\Nt)$.

We choose the 3D divide-and-conquer algorithm as implemented by Chan
\cite{Chan2003}.
This provides a transparent interpretation of the method which is well suited to our
case, where the initial points are on a regular grid and we only need the lower part of
the 3D convex hull. We describe this recursive algorithm in
appendix~\ref{Algorithms-for-the-2D-Burgers-dynamics}.
This gives the triangular facets of $\varphi$ as well as the Lagrangian-space
triangulation at any time $t$. Moreover, the slope $(x_1,x_2)$ of
each facet gives the Eulerian-space position $\vx$ of the shock node which contains all
the matter associated with this Lagrangian-space triangle, with a mass equal to 
the triangle area (setting again $\rho_0=1$).
Then, listing for instance in clockwise order the facets that have a common vertex
$\vq_c$ we obtain the Voronoi-like cell associated with $\vq_c$, each of these facets
giving a summit $\vx_j$ of this Eulerian-space cell. These summits are shock nodes
whereas the cell itself is a void (i.e. with zero matter density) and the cell boundaries
are zero-mass shock lines.

\subsection{Comparison with previous numerical studies}
\label{Numerical-comparison}

To compare with previous works, let us first note that some previous numerical studies
\cite{Weinberg1990,Nusser1990,Melott1994} of the ``adhesion model'' are not based
on the Legendre transforms of Sec.~\ref{Lagrangian-potential} but on the standard
continuity equation.
Thus, keeping the viscosity $\nu$ small but finite, they first compute the
velocity field at all times through the Hopf-Cole solution (\ref{Hopf1}) and
next integrate particle trajectories over time using this velocity field.
As we have recalled in Sec.~\ref{Lagrangian-potential}, the dynamics obtained in
this fashion is in fact {\it not} identical to the one studied in this article (and
some other works), as in the inviscid limit the behaviors of particles on the shock
manifold are different.

Other numerical works 
\cite{Kofman1990,Kofman1992,Sahni1994}
have taken advantage of the geometrical interpretation (\ref{Paraboladef}), in terms
of first-contact paraboloids, of the Hopf-Cole solution in the inviscid limit to avoid
integrations over time. Thus, by spanning the Lagrangian $\vq-$space
with paraboloids (\ref{Paraboladef}) of Gaussian curvature $1/t^2$ at the apex,
one can separate Lagrangian coordinates into ``stuck'' and ``free'' particles.
``Free'' particles are such that the paraboloid that makes contact with the
surface $\psi_0(\vq)$ at position $\vq$ has no other intersections with
the initial velocity potential $\psi_0(\vq)$. Therefore, such particles have not
experienced any shock yet, and their Eulerian location $\vx$ is given by the
apex of this first-contact  paraboloid. ``Stuck'' particles are such that this
paraboloid has other intersections with $\psi_0$, which means that they have
already experienced shocks (and their Eulerian location is no longer given by
the apex of this paraboloid).
As recalled in Sec.~\ref{Lagrangian-potential}, this geometrical construction
in terms of paraboloids, which only relies on the Hopf-Cole solution (\ref{psixpsi0q}),
applies to all prescriptions (i.e. using the standard continuity equation as well
as using the ``geometrical model'' (\ref{qx-xq})). It is sufficient to describe
regular regions (i.e. outside of shocks) associated with ``free'' particles, where there
is a one-to-one mapping $\vq\leftrightarrow \vx$. 
In our case, for the power-law initial conditions (\ref{ndef}), this gives the 
Voronoi-like diagrams associated with empty regions and their boundaries,
see for instance Fig.~7 in \cite{BernardeauVal2010b} and
\cite{Kofman1990,Gurbatov1991,Kofman1992,Sahni1994,Woyczynski2007}.
Next, scanning the Eulerian $\vx-$space with paraboloids, one obtains the
Eulerian location of the particles that form the boundaries of the ``stuck'' Lagrangian
domains. Moreover, considering the paraboloids that have three simultaneous
contact points all the particles located within the associated Lagrangian triangle are
set to the Eulerian location given by the apex of the paraboloid.
In this fashion one reconstructs the Lagrangian-space triangulation associated with
the ``geometrical model'' described in Sec.~\ref{Lagrangian-potential}, without
performing Legendre transforms and convex hull constructions.
However, this procedure is rather intricate and the successive scans of the
Lagrangian and Eulerian grids by paraboloids introduce some numerical
inaccuracies.

Finally, the use of Legendre transforms and convex hull constructions was
introduced in \cite{Noullez1994,Vergassola1994}, from the definition
of the ``geometrical model'' described in Sec.~\ref{Lagrangian-potential}.
This provides a very elegant method to obtain the direct Lagrangian
map, $\vq\mapsto\vx$, and the associated matter distribution, without looking
for first-contact paraboloids and trying to invert the $\vx\mapsto\vq$ map.
Then, \cite{Noullez1994,Vergassola1994} first compute the function $H(\vx)$
through the Legendre transform (\ref{Hxphiq}), which gives the inverse map,
$\vx\mapsto\vq$, and the Eulerian velocity field $\vu(\vx)$. Next, they obtain
the direct map, $\vq\mapsto\vx$, and the associated density field, from
the second Eq.(\ref{qx-xq}), by computing the nonlinear Lagrangian potential
$\varphi(\vq)$ from $H(\vx)$ through a second Legendre transform (\ref{varphicdef}).

As explained above, our procedure differs in two respects. For the first step,
we also compute $H(\vx)$ and the velocity field from the Legendre transform
(\ref{Hxphiq}), but as described in Sec.~\ref{compute-H(x)}
we use a faster algorithm that scales as $O(N^2)$, instead of $O[(N\ln N)^2]$ in
\cite{Noullez1994,Vergassola1994}. For the second step, as described in
Sec.~\ref{compute-convex-hull}, we do not compute $\varphi(\vq)$ through the
second Legendre transform (\ref{varphicdef}), to avoid inaccuracies associated with
numerical wrinkles in the planar facets of $H(\vx)$. Rather, we directly compute
$\varphi$ as the convex hull of the linear Lagrangian potential $\varphi_L$,
from Eq.(\ref{phi-convex}), without reference to an intermediate Eulerian
grid. Note that this is a standard problem of computational geometry,
that we solve exactly using a convex hull algorithm.
Therefore, once the initial velocity potential $\psi_0(\vq)$ is given on a grid,
whence the linear Lagrangian potential $\varphi_L$, no further sources of error are 
introduced by the use of an Eulerian grid, since we compute the exact convex
hull $\varphi$ under the form of a list of triangular facets whose slopes give the
exact Eulerian locations of the shock nodes (without reference to an Eulerian
grid). Thus, both the Eulerian-space Voronoi-like tessellation and the Lagrangian-space
Delaunay-like triangulation are exactly computed, without introducing further
inaccuracies (up to the precision with which real numbers are represented by
the computer). In particular, this removes any source of ambiguities.
As noticed in Sec.~\ref{compute-convex-hull}, although using the second Legendre
transform (\ref{varphicdef}) would provide a faster algorithm that scales as
$O(N^2)$, the exact convex hull algorithm that we use in this article
scales as $O(N^2\ln N)$, which is still faster than the method used in previous
works.

\subsection{Computation of the 3D lower convex hull $\varphi$}
\label{Algorithms-for-the-2D-Burgers-dynamics}

\begin{figure}
\begin{center}
\epsfxsize=7 cm \epsfysize=4.5 cm {\epsfbox{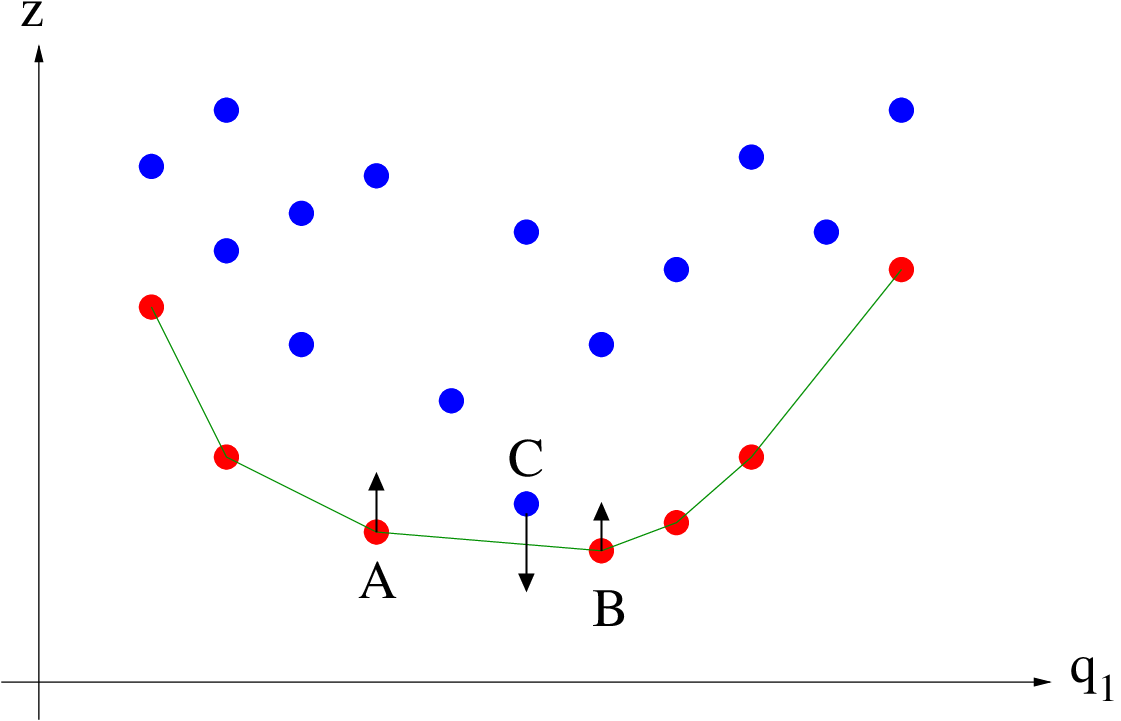}}
\end{center}
\caption{The projection of the 3D points $\{(q_1,q_2,\varphi_L)_j\}$ onto the
$(q_1,z)$ plane. The convex hull of these points gives the vertices of the 3D
convex hull $\varphi$ observed from planes of a given slope $x_2$ along direction
$q_2$. As $x_2$ increases the points move in the plane $(q_1,z)$ along verticals,
with a ``speed'' equal to $-q_2$, shown by the arrows for points $A,B$ and $C$.
For a slightly larger value of $x_2$ the point $C$ will move below the segment
$[AB]$ and appear as a new vertex in the convex envelope $z_c(q_1)$. This means
that points $\{A,B,C\}$ are a triangular facet of the 3D convex hull $\varphi$.}
\label{figconv2D}
\end{figure}

We describe here the 3D divide-and-conquer algorithm that we use to compute the
convex hull $\varphi$ (more precisely, the list of its triangular facets) following
the implementation introduced by Chan \cite{Chan2003}.
The main point is to transform the 3D problem into a ``kinetic'' 2D problem, which is
easier to visualize. This proceeds by scanning the convex hull $\varphi$ in order of
increasing slope $x_2$ along direction $2$, so that one only needs to study the
evolution with ``time'' $x_2$  of a curve $z(q_1)$, as explained below.

The input is the set of 3D points
$\{(q_1,q_2,\varphi_L)_j\}$ with $j=1,..,\Nt$. Let us choose a given slope $x_2$ along
the direction $q_2$ and consider the points $\{(q_1,z)_j\}$ in the 2D plane $(q_1,z)$
with $z\equiv \varphi_L-x_2 q_2$.
Thus, $z_j$ is the signed vertical distance between the 3D point $(j)$ and the plane
$P_{x_2}$ of equation $\varphi_L=x_2 q_2$. Then, the convex envelope of the set of
2D points $\{(q_1,z)_j\}$ gives the vertices of $\varphi$ that come into contact with
planes of slopes $(x_1,x_2)$ (i.e. of equation $\varphi_L=x_1 q_1+x_2 q_2+c$). 
As one goes from left to right, i.e. in order of increasing $q_1$, the slope $x_1$ along
the axis $q_1$ also increases (as for the 1D case studied in section~\ref{algorithm-1d}).
This is shown in Fig.~\ref{figconv2D}.
The $N$ data points with the same coordinate $q_1$ on the initial $N\times N$ grid
appear on the same vertical line in the $(q_1,z)$ plane.
By going from $x_2=-\infty$ up to $x_2=+\infty$ one scans in this fashion all vertices
of the lower convex hull $\varphi$. 
Therefore, one obtains a movie, running with ``time'' $x_2$, where points of fixed
abscissa $q_1$ and varying height $z=\varphi_L-x_2 q_2$ move along verticals at
different speeds $-q_2$, so that the convex envelope in the $(q_1,z)$ plane evolves
with time, as its vertices can appear and disappear.

In addition, one obtains the triangular facets by recording the
{\it insertion} and {\it deletion} events. In an {\it insertion} event, a new vertex $C$
of abscissa $q_1^C$ appears at a ``time'' $x_2^*$ in the convex hull $z_c(q_1)$,
in-between vertices $A$ and $B$ (this occurs when the point $C$ crosses from above
the segment $[AB]$ in the plane $(q_1,z)$). Then, the triplet $\{A,B,C\}$ is a planar
facet of $\varphi$ (with a slope $x_2^*$ along direction $2$).
In a {\it deletion} event, a vertex $C$ located between vertices $A$ and $B$ of the
convex hull $z_c(q_1)$ disappears, and this again means that the triplet
$\{A,B,C\}$ is a planar facet of $\varphi$. 
In our case, where the initial points are located on a regular $N\times N$ grid over the
$(q_1,q_2)$ plane, we also have {\it exchange} events, as a new vertex $C$ can replace
an older vertex $C'$ that has the same coordinate $q_1$. At the crossing time $x_2^*$
these vertices are located between two vertices $A$ and $B$ and we obtain two facets,
$\{A,C,C'\}$ and $\{B,C,C'\}$, of the convex hull $\varphi$. 
We show in Fig.~\ref{figconv2D} a configuration just before an {\it insertion} event,
as the vertex $C$ will soon move below the segment $[AB]$ (the arrows show the
``velocities'' $-q_2$ of these points).
By computing analytically the successive crossing times $x_2^*$ we move from
one event to the next one. Therefore, we do not need any discretization over $x_2$
and we obtain the exact list of the facets of the convex hull $\varphi$, in order
of increasing slope $x_2$.
This fully defines $\varphi$.

\begin{figure}
\begin{center}
\epsfxsize=7 cm \epsfysize=4.5 cm {\epsfbox{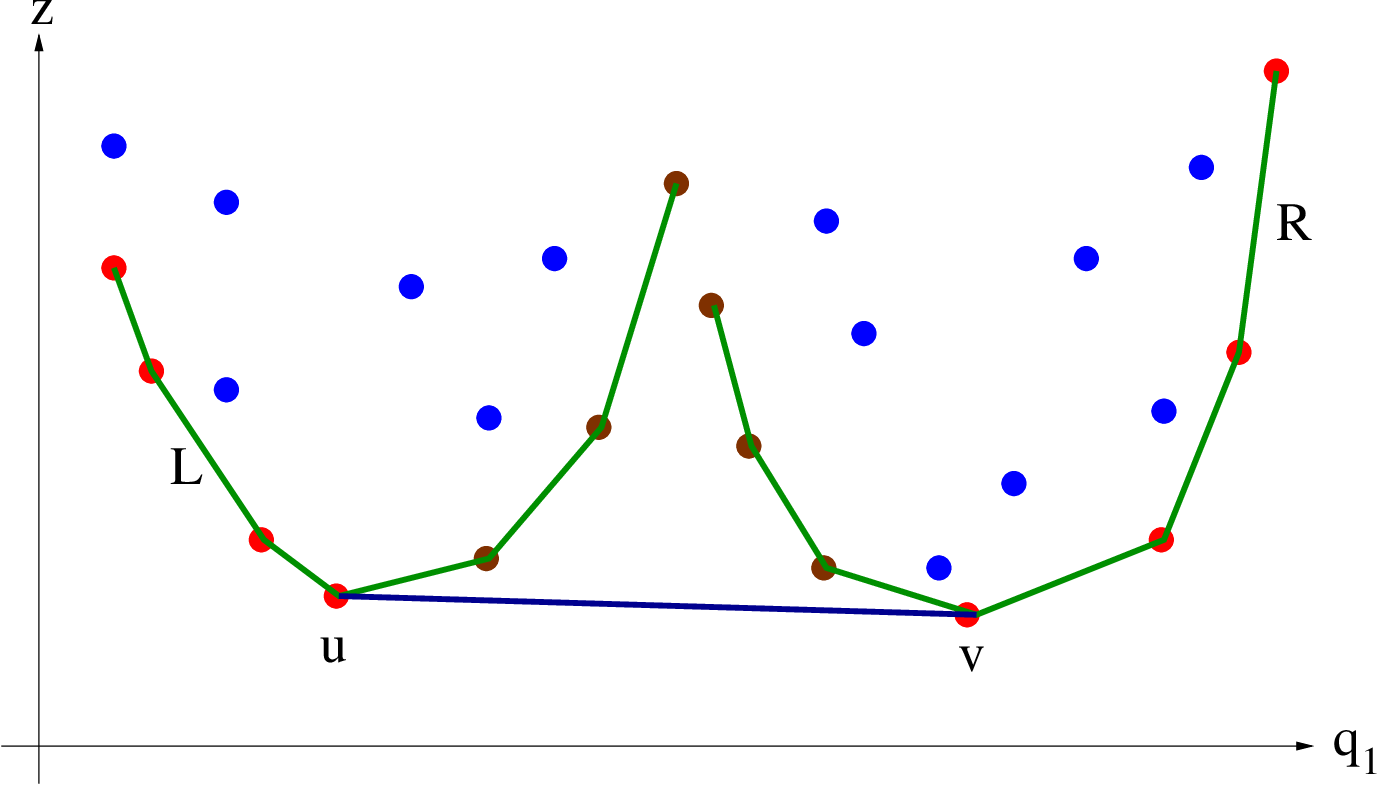}}
\end{center}
\caption{Merging of the 2D convex hulls $L$ and $R$ associated with the left and
right halves of the points in the $(q_1,z)$ plane. The convex envelope of all the
points is obtained by determining the bridge between vertices $u$ and $v$ of
the left and right hulls. As $x_2$ increases, changes to $L$ and $R$ on outer sides
of the bridge are recorded whereas changes to $L$ and $R$ within $]u,v[$ do not
contribute. In addition, as neighbors of $u$ and $v$ cross the line $(uv)$ the
bridge is modified.}
\label{figbridge2D}
\end{figure}

The algorithm as described so far takes a running time $O(\Nt^2)$. In order to
achieve a faster $O(\Nt\ln\Nt)$ performance, we adapt the divide-and-conquer method
introduced in \cite{Chan2003}.
Since the data points are given on a regular grid over the $(q_1,q_2)$
plane, the columns along $q_2$ at fixed $q_1$ (which appear as vertical
columns in the $(q_1,z)$ plane as in Fig.~\ref{figconv2D}) are stored in increasing
order of $q_1$.
Then, we recursively create a movie for the left lower hull $L$ of the $N/2$ left columns
and a movie for the right lower hull $R$ of the $N/2$ right columns. Next, we build
a movie for the lower hull of all columns by a merging process, described in
Fig.~\ref{figbridge2D}.
This is done by identifying the common tangent $uv$, called the bridge, and removing
the part of $L$ to the right of $u$ and the part of $R$ to the left of $v$.
Then, starting from $x_2=-\infty$, as ``time'' progresses events that take place on
either side of the bridge are recorded, whereas events that take place in-between
vertices $u$ and $v$ do not contribute. In addition, by paying attention to the
neighbors of $u$ and $v$ we update the position of the bridge.

As for the 2D Legendre transform (\ref{Hxphiq}) used to compute $H(\vx)$ and the
velocity field, we first extend $\varphi_L(\vq)$ to the larger grid
$-N/2\leq j \leq N+N/2-2$ (using periodicity) and we compute the convex hull
associated with this set of points. This ensures that boundary effects are eliminated
for the points in the range $0\leq j \leq N-1$ we are interested in.
We obtain in this fashion the triangular facets of $\varphi$ over this domain, which
defines the Lagrangian-space triangulation at a given time $t$. Moreover, the slope
$(x_1,x_2)$ of each facet gives the Eulerian-space position $\vx$ of the shock node
which contains all the matter associated with this Lagrangian-space triangle, with a
mass equal to $\rho_0$ times the triangle area.
Thus, from the list of the triangular facets of $\varphi$ we obtain at once
both the Lagrangian and Eulerian-space tessellations, which fully define the
distribution of matter at a given time, see \cite{BernardeauVal2010b} for
detailed descriptions.

\section{Separable case in $d$ dimensions}
\label{Separable}

We describe here the factorizable solutions of the dynamics presented in
Sec.~\ref{Separable-text}.

\subsection{General index $n$}
\label{General-index-n}

For the factorizable initial conditions presented in Sec.~\ref{Separable-text},
defined by Eq.(\ref{psi0_sep}) with independent 1D Gaussian fields along the
different directions, the mass $M$ of a shock node is the product of the
``1D masses'' $M_i$ along directions $i$ (since all directions are described by
the same index $n$ we can work with the dimensionless scaling variables as
defined in (\ref{QXU}), with a unique characteristic length $L(t)$ given by
Eq.(\ref{Lt})). Then, the shock mass function writes as
\beq
N(M) = \int_0^{\infty} \prod_{i=1}^d \dd M_i \, N^{(i)}(M_i)  
\, \delta_D(M-\prod _i M_i) ,
\label{NMprod}
\eeq
where $N^{(i)}(M)$ is the 1D shock mass function along direction $i$.
In our case all 1D mass functions are identical, $N^{(i)}(M)=N^{(1)}(M)$, where
$N^{(1)}(M)$ is the 1D mass function associated with the index $n$ studied
in section~\ref{Shock-mass-function}.
Introducing the Mellin transform $\hN(s)$ of the shock mass function,
\beqa
\hN(s) & = & \int_0^{\infty} \dd M \, M^{s-1} \, N(M) , \label{MellinNs} \\
N(M) & = & \int_{c-\ii\infty}^{c+\infty} \frac{\dd s}{2\pi\ii} \, M^{-s} \, \hN(s) ,
\label{MellinNM}
\eeqa
(where $c$ is an arbitrary real number within the fundamental strip of $\hN(s)$),
we obtain at once
\beq
\hN(s) = \left(\hN^{(1)}(s)\right)^d .
\label{Ns-N1s}
\eeq
Assuming that the 1D mass functions show the low-mass power-law tails
(\ref{NMm}) without logarithmic prefactors (which has only been proved for
$n=0$ and $n=-2$ but is consistent with numerical simulations for other values
of $n$, see Figs.~\ref{figmNm} and \ref{figlfnu}), we obtain
\beqa
s\rightarrow -\frac{n-1}{2} & : & \hN^{(1)}(s) \sim \frac{1}{s+(n-1)/2} , \\
&& \hN(s) \sim \left(s+\frac{n-1}{2}\right)^{-d} .
\label{Ns-small}
\eeqa
From standard properties of the Mellin transform this yields the low-mass asymptotic
behavior (\ref{Md-Mm}).
Thus, we obtain for all dimensions $d$ a low-mass power-law tail, with the same
exponent $(n-1)/2$, but with a logarithmic prefactor to the power $(d-1)$.
The 1D high-mass cutoff (\ref{NMp}) gives the large-$s$ behaviors (keeping only
the leading-order term)
\beq
s\rightarrow\infty : \;\; \hN^{(1)}(s) \sim s^{s/(n+3)} , \;\;
\hN(s) \sim s^{d s/(n+3)} ,
\label{Ns-large}
\eeq
whence the asymptotic tail (\ref{Md-Mp}).
Thus, we obtain for all dimensions $d$ a high-mass modified-exponential cutoff,
but with an exponent $(n+3)/d$ that decreases at higher $d$.

The same analysis applies to the probability distribution, $P_X(\eta)$, of the
overdensity within cubic cells of size $X$. Indeed, we again have
$\eta=\prod_{i=1}^d \eta_i$, where $\eta_i$ is the ''1D overdensity'' along direction
$i$, and the density probability distribution function writes as
\beq
P_X(\eta) = \int_0^{\infty} \prod_{i=1}^d \dd\eta_i \, P_X^{(i)}(\eta_i) \,
\delta_D(\eta-\prod_i\eta_i) .
\label{PXeta-prod}
\eeq
This yields the Mellin transform
\beq
\hP_X(s) = \left(\hP_X^{(1)}(s)\right)^d .
\label{PXs}
\eeq
As for the shock mass function this leads to the high-density cutoff
\beq
\eta\rightarrow\infty : \;\; \ln P_X(\eta) \sim - X^{n+3} \, \eta^{(n+3)/d} .
\label{PXd-Mp}
\eeq
Again we recover the characteristic exponents (\ref{PReta-large-eta-2d}) of the
isotropic case.

\subsection{Index $n=-2$}
\label{n=-2}

\begin{figure}
\begin{center}
\epsfxsize=7 cm \epsfysize=5 cm {\epsfbox{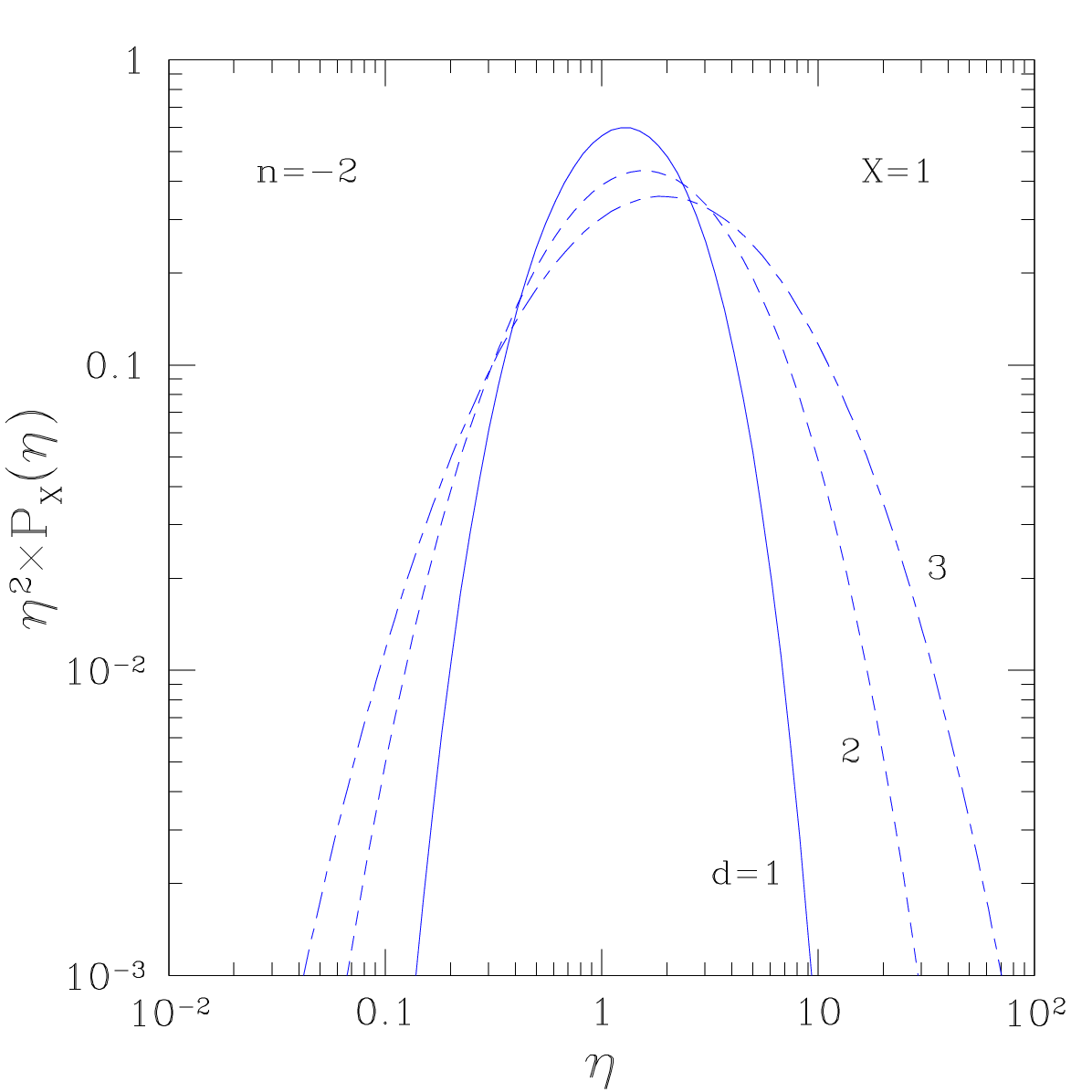}}
\epsfxsize=7 cm \epsfysize=5 cm {\epsfbox{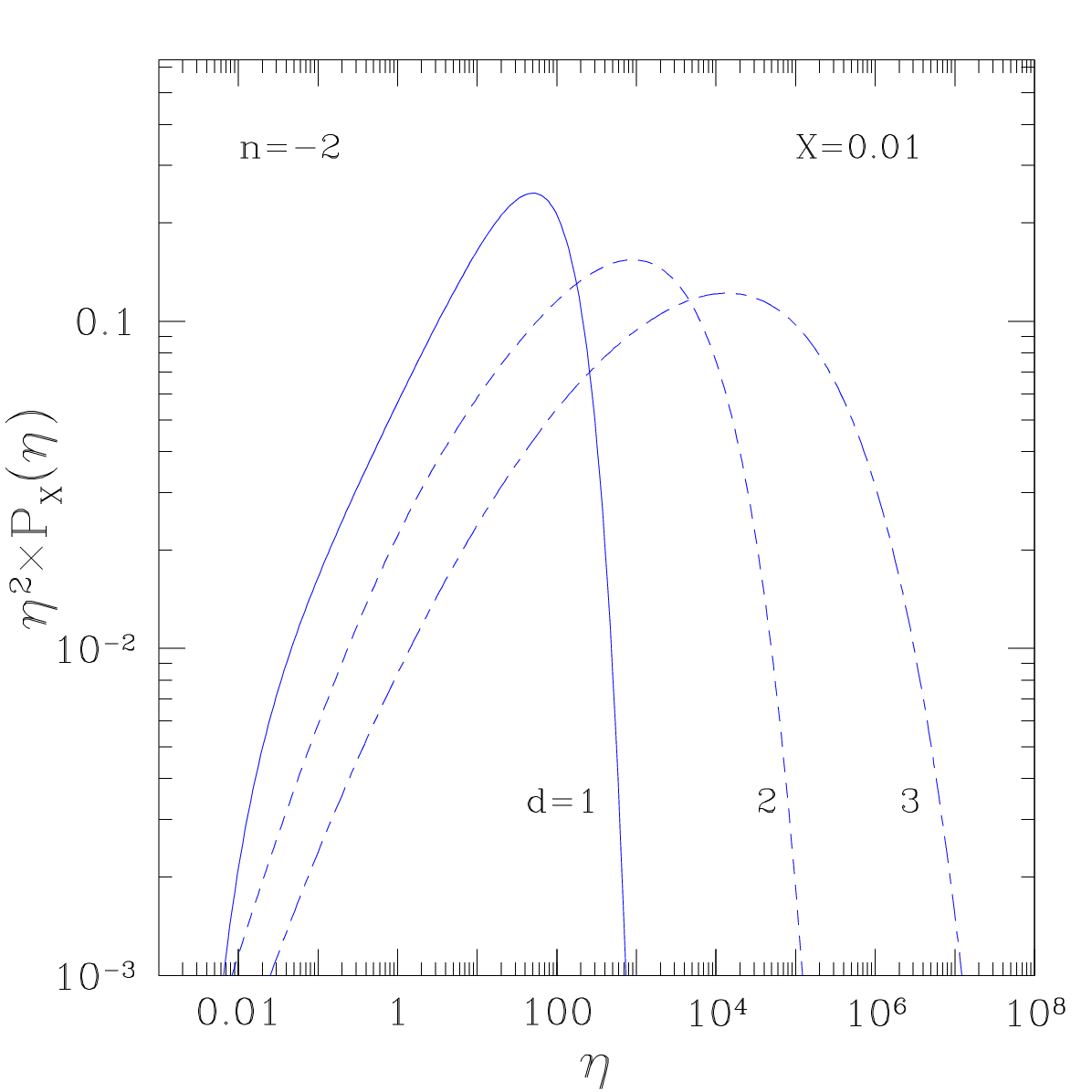}}
\end{center}
\caption{The product $\eta^2 P_X(\eta)$, where $P_X(\eta)$ is the probability
distribution function of the overdensity $\eta$ within cubic cells of size $X$,
for the separable dynamics with $n=-2$. Dimensions $d=1$ (solid line),
$d=2$ (dashed line), and $d=3$ (dot-dashed line), are shown for two
values of the cell size $X$.}
\label{figrho2Pxrho_n-2}
\end{figure}

For $n=-2$, where the 1D mass function $N^{(1)}(M)$ and density probability
distribution $P_X^{(1)}(\eta)$ are given by the simple expressions (\ref{NMn-2})
and (\ref{PXeta-2}), we can derive explicit expressions. Thus, the 1D Mellin
transforms write as
\beqa
n=-2 & : & \hN^{(1)}(s) = \frac{1}{\sqrt{\pi}} \, \Gamma\left(s-\frac{3}{2}\right) , \\
&& \hP_X^{(1)}(s) = 2 \sqrt{\frac{X}{\pi}} \, e^{2X} \, K_{s-3/2}(2X) ,
\eeqa
where $K_{\nu}$ is the modified Bessel function of the second kind of order $\nu$.
This gives the shock mass function and the density probability distribution for all
$d$ through Eqs.(\ref{Ns-N1s}) and (\ref{PXs}).
At low dimensions it is simpler to use the integrals (\ref{NMprod}) and
(\ref{PXeta-prod}), which gives for instance in 2D the expressions
(\ref{NM-n-2-sep})-(\ref{PXn-2-sep}).
This yields the asymptotic behaviors
\beqa
\hspace{-0.6cm} d=2, \; M\rightarrow 0 & : & \; N(M) \sim \frac{-\ln M}{\pi}
\, M^{-3/2} , \label{NM0n-2-sep} \\
M\rightarrow \infty & : & \; N(M) \sim \frac{1}{\sqrt{\pi}} \, M^{-7/4}
\, e^{-2\sqrt{M}} , \label{NMinfn-2-sep}
\eeqa
and
\beqa
d=2,  \; \eta\rightarrow 0 & : & \; \ln P_X(\eta) 
\sim - 2X/\sqrt{\eta} ,  \label{PX0n-2sep} \\
\eta\rightarrow \infty & : & \; \ln P_X(\eta) \sim - 2X\sqrt{\eta} , \label{PXinfn-2sep} 
\eeqa
which agree with the general results (\ref{Md-Mm}), (\ref{Md-Mp}), and 
(\ref{PXd-Mp}).

We show our results in Figs.~\ref{figm2Nm_n-2} and \ref{figrho2Pxrho_n-2}
for the shock mass function $N(M)$ and the probability distribution $P_X(\eta)$,
in dimensions $d=1,2$ and $3$ and for the index $n=-2$.
To see more clearly the intermediate power-law regime we plot the product
$\eta^2 P_X(\eta)$ in Fig.~\ref{figrho2Pxrho_n-2}. In agreement
with Eq.(\ref{PXd-Mp}), for higher $d$ the density probability distribution
shows smoother high- and low-density cutoffs and a broader peak, which again
can be understood as the result of a multiplicative process.
As for the isotropic case, at smaller scales an intermediate power-law regime
develops, as can be seen explicitly in Eq.(\ref{PXn-2-sep}) in 2D. However,
because of logarithmic prefactors it is more difficult to see it in the figure
for higher $d$.

\begin{acknowledgments}
This work is supported in part by the French Agence Nationale de la Recherche
under grant ANR-07-BLAN-0132 (BLAN07-1-212615).
\end{acknowledgments}

\bibliography{ref}   % name your BibTeX data base

\end{document}